\documentclass[acmsmall,screen,nonacm]{acmart}
\usepackage[T1]{fontenc}
\usepackage[utf8]{inputenc}
\usepackage{amsthm}
\usepackage{amsmath}
\usepackage{amssymb}
\usepackage{subcaption}
\usepackage{environ}
\usepackage[capitalize, noabbrev]{cleveref}
\usepackage{xifthen}
\usepackage{expl3}
\usepackage{xparse}
\usepackage{newunicodechar}
\usepackage{hhline}
\usepackage[tikz]{bclogo}
\usepackage[normalem]{ulem}
\makeatletter
\def\uwave{\bgroup \markoverwith{\lower3.5\p@\hbox{\sixly \textcolor{red}{\char58}}}\ULon}
\font\sixly=lasy6 
\makeatother

\newunicodechar{⥶}{\ensuremath{\raisebox{0.2em}{$\scriptstyle<$}\hspace*{-0.55em}\raisebox{-0.12em}{$\scriptscriptstyle\leftarrow$}}}
\usepackage[supertabular]{ottalt}

\newcommand{\ottdrule}[4][]{{\displaystyle\frac{\begin{array}{l}#2\end{array}}{#3}\quad\ottdrulename{#4}}}
\newcommand{\ottusedrule}[1]{\[#1\]}
\newcommand{\ottpremise}[1]{ #1 \\}
\newenvironment{ottdefnblock}[3][]{ \framebox{\mbox{#2}} \quad #3 \\[0pt]}{}

\newcommand{\ottnt}[1]{\mathit{#1}}
\newcommand{\ottmv}[1]{\mathit{#1}}
\newcommand{\ottkw}[1]{\mathbf{#1}}
\newcommand{\ottsym}[1]{#1}
\newcommand{\ottcom}[1]{\text{#1}}
\newcommand{\ottdrulename}[1]{\textsc{#1}}

\newcommand{\ottmetavartabular}[1]{\begin{supertabular}{ll}#1\end{supertabular}}

\newcommand{\ottafterlastrule}{\\}
\newcommand{\ottmetavars}{
\ottmetavartabular{
 $ \ottmv{var} ,\, \ottmv{x} ,\, \ottmv{y} ,\, \ottmv{d} ,\, \ottmv{dd} ,\, \ottmv{un} ,\, \ottmv{xs} ,\, \ottmv{ys} ,\, \ottmv{ex} ,\, \ottmv{st} ,\, \ottmv{tree} ,\, \ottmv{tl} ,\, \ottmv{tr} ,\, \ottmv{dtree} ,\, \ottmv{f} ,\, \ottmv{dh} ,\, \ottmv{dt} ,\, \ottmv{dx} ,\, \ottmv{dy} ,\, \ottmv{dxs} ,\, \ottmv{dys} ,\, \ottmv{dv} ,\, \ottmv{dtlr} ,\, \ottmv{dtl} ,\, \ottmv{dtr} ,\, \ottmv{q} ,\, \ottmv{tok} ,\, \ottmv{z} ,\, \ottmv{front} ,\, \ottmv{back} ,\, \ottmv{l} ,\, \ottmv{dl} $ & \ottcom{Variable names} \\
 $ \ottmv{k} $ & \ottcom{Index for ranges} \\
}}

\newcommand{\ottdruleTyXXvalXXHole}[1]{\ottdrule[#1]{%
}{
 \ottshname{\hboxed{ \ottshname{h} } }:\!_{\!   \ottsmode{1}  \hspace{-0.15ex}  \ottsmode{\nu}   } \ottstype{T}   \!\!\pmb{\phantom{a}^{\scriptscriptstyle \mathrm{v} }\!\!\vdash}\,   \ottshname{\hboxed{ \ottshname{h} } }   \pmb{:}  \ottstype{T}}{%
{\ottdrulename{Ty\_val\_Hole}}{\text{\CHole}}%
}}

\newcommand{\ottdruleTyXXvalXXDest}[1]{\ottdrule[#1]{%
\ottpremise{  \ottsmode{1}  \hspace{-0.15ex}  \ottsmode{\nu}  \texttt{ ⥶ }\ottsmode{m}\texttt{ }}%
}{
 \ottshname{\destminus} \ottshname{h} :\!_{\! \ottsmode{m} }\ottstype{\lfloor}\,\!_{ \ottsmode{n} } \ottstype{T} \ottstype{\rfloor}   \!\!\pmb{\phantom{a}^{\scriptscriptstyle \mathrm{v} }\!\!\vdash}\,   \ottshname{\destminus} \ottshname{h}   \pmb{:}   \ottstype{\lfloor}\,\!_{\mydestm{ \ottsmode{n} } } \ottstype{T} \ottstype{\rfloor} }{%
{\ottdrulename{Ty\_val\_Dest}}{\text{\CDest}}%
}}

\newcommand{\ottdruleTyXXvalXXUnit}[1]{\ottdrule[#1]{%
}{
 \smallbullet   \!\!\pmb{\phantom{a}^{\scriptscriptstyle \mathrm{v} }\!\!\vdash}\,  \ottsctor{()}  \pmb{:}   \ottstype{1} }{%
{\ottdrulename{Ty\_val\_Unit}}{\text{\CUnit}}%
}}

\newcommand{\ottdruleTyXXvalXXFun}[1]{\ottdrule[#1]{%
\ottpremise{ \Delta ,~  \ottmv{x} :\!_{\! \ottsmode{m} } \ottstype{T}    \,\pmb{\vdash}\,  \ottnt{u}  \pmb{:}  \ottstype{U}}%
}{
\Delta  \!\!\pmb{\phantom{a}^{\scriptscriptstyle \mathrm{v} }\!\!\vdash}\,   \lamvnt{ \ottmv{x} }{ \ottsmode{m} }{ \ottnt{u} }   \pmb{:}   \ottstype{T} \,_{\myfuntm{ \ottsmode{m} } }\!\ottstype{\multimap}\, \ottstype{U} }{%
{\ottdrulename{Ty\_val\_Fun}}{\text{\CFun}}%
}}

\newcommand{\ottdruleTyXXvalXXLeft}[1]{\ottdrule[#1]{%
\ottpremise{\Theta  \!\!\pmb{\phantom{a}^{\scriptscriptstyle \mathrm{v} }\!\!\vdash}\,  \ottnt{v_{{\mathrm{1}}}}  \pmb{:}  \ottstype{T_{{\mathrm{1}}}}}%
}{
\Theta  \!\!\pmb{\phantom{a}^{\scriptscriptstyle \mathrm{v} }\!\!\vdash}\,  \ottsctor{Inl} \, \ottnt{v_{{\mathrm{1}}}}  \pmb{:}   \ottstype{T_{{\mathrm{1}}}} \ottstype{\oplus} \ottstype{T_{{\mathrm{2}}}} }{%
{\ottdrulename{Ty\_val\_Left}}{\text{\CLeft}}%
}}

\newcommand{\ottdruleTyXXvalXXRight}[1]{\ottdrule[#1]{%
\ottpremise{\Theta  \!\!\pmb{\phantom{a}^{\scriptscriptstyle \mathrm{v} }\!\!\vdash}\,  \ottnt{v_{{\mathrm{2}}}}  \pmb{:}  \ottstype{T_{{\mathrm{2}}}}}%
}{
\Theta  \!\!\pmb{\phantom{a}^{\scriptscriptstyle \mathrm{v} }\!\!\vdash}\,  \ottsctor{Inr} \, \ottnt{v_{{\mathrm{2}}}}  \pmb{:}   \ottstype{T_{{\mathrm{1}}}} \ottstype{\oplus} \ottstype{T_{{\mathrm{2}}}} }{%
{\ottdrulename{Ty\_val\_Right}}{\text{\CRight}}%
}}

\newcommand{\ottdruleTyXXvalXXProd}[1]{\ottdrule[#1]{%
\ottpremise{\Theta_{{\mathrm{1}}}  \!\!\pmb{\phantom{a}^{\scriptscriptstyle \mathrm{v} }\!\!\vdash}\,  \ottnt{v_{{\mathrm{1}}}}  \pmb{:}  \ottstype{T_{{\mathrm{1}}}}}%
\ottpremise{\Theta_{{\mathrm{2}}}  \!\!\pmb{\phantom{a}^{\scriptscriptstyle \mathrm{v} }\!\!\vdash}\,  \ottnt{v_{{\mathrm{2}}}}  \pmb{:}  \ottstype{T_{{\mathrm{2}}}}}%
}{
\Theta_{{\mathrm{1}}}  +  \Theta_{{\mathrm{2}}}  \!\!\pmb{\phantom{a}^{\scriptscriptstyle \mathrm{v} }\!\!\vdash}\,   \ottsctor{(} \ottnt{v_{{\mathrm{1}}}} \,\ottsctor{,}~ \ottnt{v_{{\mathrm{2}}}} \ottsctor{)}   \pmb{:}   \ottstype{T_{{\mathrm{1}}}} \ottstype{\otimes} \ottstype{T_{{\mathrm{2}}}} }{%
{\ottdrulename{Ty\_val\_Prod}}{\text{\CProd}}%
}}

\newcommand{\ottdruleTyXXvalXXExp}[1]{\ottdrule[#1]{%
\ottpremise{\Theta  \!\!\pmb{\phantom{a}^{\scriptscriptstyle \mathrm{v} }\!\!\vdash}\,  \ottnt{v'}  \pmb{:}  \ottstype{T}}%
}{
 \ottsmode{n} \hspace{-0.3ex}\cdot\hspace{-0.3ex} \Theta   \!\!\pmb{\phantom{a}^{\scriptscriptstyle \mathrm{v} }\!\!\vdash}\,   \expcons{ \ottsmode{n} } \ottnt{v'}   \pmb{:}   \ottstype{!}_{ \ottsmode{n} } \ottstype{T} }{%
{\ottdrulename{Ty\_val\_Exp}}{\text{\CExp}}%
}}

\newcommand{\ottdruleTyXXvalXXAmpar}[1]{\ottdrule[#1]{%
\ottpremise{    \ottsmode{1}  \hspace{-0.15ex}  \ottsmode{\uparrow}   \hspace{-0.3ex}\cdot\hspace{-0.3ex} \Delta_{{\mathrm{1}}}  ,~ \Delta_{{\mathrm{3}}}   \!\!\pmb{\phantom{a}^{\scriptscriptstyle \mathrm{v} }\!\!\vdash}\,  \ottnt{v_{{\mathrm{1}}}}  \pmb{:}  \ottstype{T}}%
\ottpremise{ \Delta_{{\mathrm{2}}} ,~  \ottshname{\destminus^{\scriptscriptstyle\text{-}1} } \Delta_{{\mathrm{3}}}    \!\!\pmb{\phantom{a}^{\scriptscriptstyle \mathrm{v} }\!\!\vdash}\,  \ottnt{v_{{\mathrm{2}}}}  \pmb{:}  \ottstype{U}}%
}{
 \Delta_{{\mathrm{1}}} ,~ \Delta_{{\mathrm{2}}}   \!\!\pmb{\phantom{a}^{\scriptscriptstyle \mathrm{v} }\!\!\vdash}\,   _{  \ottshname{\mathsfbf{hnames}(} \Delta_{{\mathrm{3}}} \ottshname{)}  \!}\ottsctor{\langle} \ottnt{v_{{\mathrm{2}}}} \,\ottsctor{\bbcomma}~ \ottnt{v_{{\mathrm{1}}}} \ottsctor{\rangle}   \pmb{:}   \ottstype{U} \,\ottstype{\ltimes}\, \ottstype{T} }{%
{\ottdrulename{Ty\_val\_Ampar}}{\text{\CAmpar}}%
}}

\newcommand{\ottdefnTyXXval}[1]{\begin{ottdefnblock}[#1]{$\Theta  \!\!\pmb{\phantom{a}^{\scriptscriptstyle \mathrm{v} }\!\!\vdash}\,  \ottnt{v}  \pmb{:}  \ottstype{T}$}{\ottcom{Typing judgment for values}}
\ottusedrule{\ottdruleTyXXvalXXHole{}}
\ottusedrule{\ottdruleTyXXvalXXDest{}}
\ottusedrule{\ottdruleTyXXvalXXUnit{}}
\ottusedrule{\ottdruleTyXXvalXXFun{}}
\ottusedrule{\ottdruleTyXXvalXXLeft{}}
\ottusedrule{\ottdruleTyXXvalXXRight{}}
\ottusedrule{\ottdruleTyXXvalXXProd{}}
\ottusedrule{\ottdruleTyXXvalXXExp{}}
\ottusedrule{\ottdruleTyXXvalXXAmpar{}}
\end{ottdefnblock}}

\newcommand{\ottdruleTyXXtermXXVal}[1]{\ottdrule[#1]{%
\ottpremise{\texttt{DisposableOnly }\Gamma\texttt{ }}%
\ottpremise{\Delta  \!\!\pmb{\phantom{a}^{\scriptscriptstyle \mathrm{v} }\!\!\vdash}\,  \ottnt{v}  \pmb{:}  \ottstype{T}}%
}{
 \Gamma ,~ \Delta   \,\pmb{\vdash}\,  \ottnt{v}  \pmb{:}  \ottstype{T}}{%
{\ottdrulename{Ty\_term\_Val}}{\text{\CVal}}%
}}

\newcommand{\ottdruleTyXXtermXXVar}[1]{\ottdrule[#1]{%
\ottpremise{\texttt{DisposableOnly }\Gamma\texttt{ }}%
\ottpremise{  \ottsmode{1}  \hspace{-0.15ex}  \ottsmode{\nu}  \texttt{ ⥶ }\ottsmode{m}\texttt{ }}%
}{
 \Gamma ,~  \ottmv{x} :\!_{\! \ottsmode{m} } \ottstype{T}    \,\pmb{\vdash}\,  \ottmv{x}  \pmb{:}  \ottstype{T}}{%
{\ottdrulename{Ty\_term\_Var}}{\text{\CVar}}%
}}

\newcommand{\ottdruleTyXXtermXXApp}[1]{\ottdrule[#1]{%
\ottpremise{\Gamma_{{\mathrm{1}}}  \,\pmb{\vdash}\,  \ottnt{t}  \pmb{:}  \ottstype{T}}%
\ottpremise{\Gamma_{{\mathrm{2}}}  \,\pmb{\vdash}\,  \ottnt{t'}  \pmb{:}   \ottstype{T} \,_{\myfuntm{ \ottsmode{m} } }\!\ottstype{\multimap}\, \ottstype{U} }%
}{
 \ottsmode{m} \hspace{-0.3ex}\cdot\hspace{-0.3ex} \Gamma_{{\mathrm{1}}}   +  \Gamma_{{\mathrm{2}}}  \,\pmb{\vdash}\,   \ottnt{t'} ~ \ottnt{t}   \pmb{:}  \ottstype{U}}{%
{\ottdrulename{Ty\_term\_App}}{\text{\CApp}}%
}}

\newcommand{\ottdruleTyXXtermXXPatU}[1]{\ottdrule[#1]{%
\ottpremise{\Gamma_{{\mathrm{1}}}  \,\pmb{\vdash}\,  \ottnt{t}  \pmb{:}   \ottstype{1} }%
\ottpremise{\Gamma_{{\mathrm{2}}}  \,\pmb{\vdash}\,  \ottnt{u}  \pmb{:}  \ottstype{U}}%
}{
\Gamma_{{\mathrm{1}}}  +  \Gamma_{{\mathrm{2}}}  \,\pmb{\vdash}\,   \ottnt{t} \patu \ottnt{u}   \pmb{:}  \ottstype{U}}{%
{\ottdrulename{Ty\_term\_PatU}}{\text{\CPatU}}%
}}

\newcommand{\ottdruleTyXXtermXXPatS}[1]{\ottdrule[#1]{%
\ottpremise{\Gamma_{{\mathrm{1}}}  \,\pmb{\vdash}\,  \ottnt{t}  \pmb{:}   \ottstype{T_{{\mathrm{1}}}} \ottstype{\oplus} \ottstype{T_{{\mathrm{2}}}} }%
\ottpremise{ \Gamma_{{\mathrm{2}}} ,~  \ottmv{x_{{\mathrm{1}}}} :\!_{\! \ottsmode{m} } \ottstype{T_{{\mathrm{1}}}}    \,\pmb{\vdash}\,  \ottnt{u_{{\mathrm{1}}}}  \pmb{:}  \ottstype{U}}%
\ottpremise{ \Gamma_{{\mathrm{2}}} ,~  \ottmv{x_{{\mathrm{2}}}} :\!_{\! \ottsmode{m} } \ottstype{T_{{\mathrm{2}}}}    \,\pmb{\vdash}\,  \ottnt{u_{{\mathrm{2}}}}  \pmb{:}  \ottstype{U}}%
}{
 \ottsmode{m} \hspace{-0.3ex}\cdot\hspace{-0.3ex} \Gamma_{{\mathrm{1}}}   +  \Gamma_{{\mathrm{2}}}  \,\pmb{\vdash}\,   \ottkw{case}_{\mycasem{ \ottsmode{m} } }\, \ottnt{t} ~\ottkw{of}~\{\ottsctor{Inl}\, \ottmv{x_{{\mathrm{1}}}} \pmb{\mapsto} \ottnt{u_{{\mathrm{1}}}} \,,~\ottsctor{Inr}\, \ottmv{x_{{\mathrm{2}}}} \pmb{\mapsto} \ottnt{u_{{\mathrm{2}}}} \}   \pmb{:}  \ottstype{U}}{%
{\ottdrulename{Ty\_term\_PatS}}{\text{\CPatS}}%
}}

\newcommand{\ottdruleTyXXtermXXPatP}[1]{\ottdrule[#1]{%
\ottpremise{\Gamma_{{\mathrm{1}}}  \,\pmb{\vdash}\,  \ottnt{t}  \pmb{:}   \ottstype{T_{{\mathrm{1}}}} \ottstype{\otimes} \ottstype{T_{{\mathrm{2}}}} }%
\ottpremise{  \Gamma_{{\mathrm{2}}} ,~  \ottmv{x_{{\mathrm{1}}}} :\!_{\! \ottsmode{m} } \ottstype{T_{{\mathrm{1}}}}   ,~  \ottmv{x_{{\mathrm{2}}}} :\!_{\! \ottsmode{m} } \ottstype{T_{{\mathrm{2}}}}    \,\pmb{\vdash}\,  \ottnt{u}  \pmb{:}  \ottstype{U}}%
}{
 \ottsmode{m} \hspace{-0.3ex}\cdot\hspace{-0.3ex} \Gamma_{{\mathrm{1}}}   +  \Gamma_{{\mathrm{2}}}  \,\pmb{\vdash}\,   \ottkw{case}_{\mycasem{ \ottsmode{m} } }\, \ottnt{t} ~\ottkw{of}~\ottsctor{(} \ottmv{x_{{\mathrm{1}}}} \,\ottsctor{,}~ \ottmv{x_{{\mathrm{2}}}} \ottsctor{)} \pmb{\mapsto} \ottnt{u}   \pmb{:}  \ottstype{U}}{%
{\ottdrulename{Ty\_term\_PatP}}{\text{\CPatP}}%
}}

\newcommand{\ottdruleTyXXtermXXPatE}[1]{\ottdrule[#1]{%
\ottpremise{\Gamma_{{\mathrm{1}}}  \,\pmb{\vdash}\,  \ottnt{t}  \pmb{:}   \ottstype{!}_{ \ottsmode{n} } \ottstype{T} }%
\ottpremise{ \Gamma_{{\mathrm{2}}} ,~  \ottmv{x} :\!_{\! \ottsmode{m}  \ottsmode{\hspace{-0.1ex}\cdot\hspace{-0.1ex} }  \ottsmode{n} } \ottstype{T}    \,\pmb{\vdash}\,  \ottnt{u}  \pmb{:}  \ottstype{U}}%
}{
 \ottsmode{m} \hspace{-0.3ex}\cdot\hspace{-0.3ex} \Gamma_{{\mathrm{1}}}   +  \Gamma_{{\mathrm{2}}}  \,\pmb{\vdash}\,   \ottkw{case}_{\mycasem{ \ottsmode{m} } }\, \ottnt{t} ~\ottkw{of}~\expcons{ \ottsmode{n} } \ottmv{x} \, \pmb{\mapsto} \ottnt{u}   \pmb{:}  \ottstype{U}}{%
{\ottdrulename{Ty\_term\_PatE}}{\text{\CPatE}}%
}}

\newcommand{\ottdruleTyXXtermXXUpdA}[1]{\ottdrule[#1]{%
\ottpremise{\Gamma_{{\mathrm{1}}}  \,\pmb{\vdash}\,  \ottnt{t}  \pmb{:}   \ottstype{U} \,\ottstype{\ltimes}\, \ottstype{T} }%
\ottpremise{    \ottsmode{1}  \hspace{-0.15ex}  \ottsmode{\uparrow}   \hspace{-0.3ex}\cdot\hspace{-0.3ex} \Gamma_{{\mathrm{2}}}  ,~  \ottmv{x} :\!_{\!   \ottsmode{1}  \hspace{-0.15ex}  \ottsmode{\nu}   } \ottstype{T}    \,\pmb{\vdash}\,  \ottnt{t'}  \pmb{:}  \ottstype{T'}}%
}{
\Gamma_{{\mathrm{1}}}  +  \Gamma_{{\mathrm{2}}}  \,\pmb{\vdash}\,   \ottkw{upd}_{\ottkw{\ltimes} }\, \ottnt{t} ~\ottkw{with}~ \ottmv{x} \, \pmb{\mapsto} \ottnt{t'}   \pmb{:}   \ottstype{U} \,\ottstype{\ltimes}\, \ottstype{T'} }{%
{\ottdrulename{Ty\_term\_UpdA}}{\text{\CUpdA}}%
}}

\newcommand{\ottdruleTyXXtermXXToA}[1]{\ottdrule[#1]{%
\ottpremise{\Gamma  \,\pmb{\vdash}\,  \ottnt{u}  \pmb{:}  \ottstype{U}}%
}{
\Gamma  \,\pmb{\vdash}\,   \ottkw{to}_{\ottkw{\ltimes} }\, \ottnt{u}   \pmb{:}   \ottstype{U} \,\ottstype{\ltimes}\,  \ottstype{1}  }{%
{\ottdrulename{Ty\_term\_ToA}}{\text{\CToA}}%
}}

\newcommand{\ottdruleTyXXtermXXFromA}[1]{\ottdrule[#1]{%
\ottpremise{\Gamma  \,\pmb{\vdash}\,  \ottnt{t}  \pmb{:}   \ottstype{U} \,\ottstype{\ltimes}\,  \ottstype{(}  \ottstype{!}_{   \ottsmode{1}  \hspace{-0.15ex}  \ottsmode{\infty}   } \ottstype{T}  \ottstype{)}  }%
}{
\Gamma  \,\pmb{\vdash}\,   \ottkw{from}_{\ottkw{\ltimes} }\, \ottnt{t}   \pmb{:}   \ottstype{U} \ottstype{\otimes}  \ottstype{(}  \ottstype{!}_{   \ottsmode{1}  \hspace{-0.15ex}  \ottsmode{\infty}   } \ottstype{T}  \ottstype{)}  }{%
{\ottdrulename{Ty\_term\_FromA}}{\text{\CFromA}}%
}}

\newcommand{\ottdruleTyXXtermXXNewA}[1]{\ottdrule[#1]{%
\ottpremise{\texttt{DisposableOnly }\Gamma\texttt{ }}%
}{
\Gamma  \,\pmb{\vdash}\,   \ottkw{new}_{\ottkw{\ltimes} }   \pmb{:}   \ottstype{T} \,\ottstype{\ltimes}\,  \ottstype{\lfloor}\,\!_{\mydestm{   \ottsmode{1}  \hspace{-0.15ex}  \ottsmode{\nu}   } } \ottstype{T} \ottstype{\rfloor}  }{%
{\ottdrulename{Ty\_term\_NewA}}{\text{\CNewA}}%
}}

\newcommand{\ottdruleTyXXtermXXFillU}[1]{\ottdrule[#1]{%
\ottpremise{\Gamma  \,\pmb{\vdash}\,  \ottnt{t}  \pmb{:}   \ottstype{\lfloor}\,\!_{\mydestm{ \ottsmode{n} } }  \ottstype{1}  \ottstype{\rfloor} }%
}{
\Gamma  \,\pmb{\vdash}\,  \ottnt{t}  \triangleleft  \ottsctor{()}  \pmb{:}   \ottstype{1} }{%
{\ottdrulename{Ty\_term\_FillU}}{\text{\CFillU}}%
}}

\newcommand{\ottdruleTyXXtermXXFillL}[1]{\ottdrule[#1]{%
\ottpremise{\Gamma  \,\pmb{\vdash}\,  \ottnt{t}  \pmb{:}   \ottstype{\lfloor}\,\!_{\mydestm{ \ottsmode{n} } }  \ottstype{T_{{\mathrm{1}}}} \ottstype{\oplus} \ottstype{T_{{\mathrm{2}}}}  \ottstype{\rfloor} }%
}{
\Gamma  \,\pmb{\vdash}\,  \ottnt{t}  \triangleleft \, \ottsctor{Inl}  \pmb{:}   \ottstype{\lfloor}\,\!_{\mydestm{ \ottsmode{n} } } \ottstype{T_{{\mathrm{1}}}} \ottstype{\rfloor} }{%
{\ottdrulename{Ty\_term\_FillL}}{\text{\CFillL}}%
}}

\newcommand{\ottdruleTyXXtermXXFillR}[1]{\ottdrule[#1]{%
\ottpremise{\Gamma  \,\pmb{\vdash}\,  \ottnt{t}  \pmb{:}   \ottstype{\lfloor}\,\!_{\mydestm{ \ottsmode{n} } }  \ottstype{T_{{\mathrm{1}}}} \ottstype{\oplus} \ottstype{T_{{\mathrm{2}}}}  \ottstype{\rfloor} }%
}{
\Gamma  \,\pmb{\vdash}\,  \ottnt{t}  \triangleleft \, \ottsctor{Inr}  \pmb{:}   \ottstype{\lfloor}\,\!_{\mydestm{ \ottsmode{n} } } \ottstype{T_{{\mathrm{2}}}} \ottstype{\rfloor} }{%
{\ottdrulename{Ty\_term\_FillR}}{\text{\CFillR}}%
}}

\newcommand{\ottdruleTyXXtermXXFillP}[1]{\ottdrule[#1]{%
\ottpremise{\Gamma  \,\pmb{\vdash}\,  \ottnt{t}  \pmb{:}   \ottstype{\lfloor}\,\!_{\mydestm{ \ottsmode{n} } }  \ottstype{T_{{\mathrm{1}}}} \ottstype{\otimes} \ottstype{T_{{\mathrm{2}}}}  \ottstype{\rfloor} }%
}{
\Gamma  \,\pmb{\vdash}\,  \ottnt{t}  \triangleleft  \ottsctor{({,})}  \pmb{:}    \ottstype{\lfloor}\,\!_{\mydestm{ \ottsmode{n} } } \ottstype{T_{{\mathrm{1}}}} \ottstype{\rfloor}  \ottstype{\otimes}  \ottstype{\lfloor}\,\!_{\mydestm{ \ottsmode{n} } } \ottstype{T_{{\mathrm{2}}}} \ottstype{\rfloor}  }{%
{\ottdrulename{Ty\_term\_FillP}}{\text{\CFillP}}%
}}

\newcommand{\ottdruleTyXXtermXXFillE}[1]{\ottdrule[#1]{%
\ottpremise{\Gamma  \,\pmb{\vdash}\,  \ottnt{t}  \pmb{:}   \ottstype{\lfloor}\,\!_{\mydestm{ \ottsmode{n} } }  \ottstype{!}_{ \ottsmode{n'} } \ottstype{T}  \ottstype{\rfloor} }%
}{
\Gamma  \,\pmb{\vdash}\,   \ottnt{t} \triangleleft \,\expcons{ \ottsmode{n'} }   \pmb{:}   \ottstype{\lfloor}\,\!_{\mydestm{ \ottsmode{n'}  \ottsmode{\hspace{-0.1ex}\cdot\hspace{-0.1ex} }  \ottsmode{n} } } \ottstype{T} \ottstype{\rfloor} }{%
{\ottdrulename{Ty\_term\_FillE}}{\text{\CFillE}}%
}}

\newcommand{\ottdruleTyXXtermXXFillF}[1]{\ottdrule[#1]{%
\ottpremise{\Gamma_{{\mathrm{1}}}  \,\pmb{\vdash}\,  \ottnt{t}  \pmb{:}   \ottstype{\lfloor}\,\!_{\mydestm{ \ottsmode{n} } }  \ottstype{T} \,_{\myfuntm{ \ottsmode{m} } }\!\ottstype{\multimap}\, \ottstype{U}  \ottstype{\rfloor} }%
\ottpremise{ \Gamma_{{\mathrm{2}}} ,~  \ottmv{x} :\!_{\! \ottsmode{m} } \ottstype{T}    \,\pmb{\vdash}\,  \ottnt{u}  \pmb{:}  \ottstype{U}}%
}{
\Gamma_{{\mathrm{1}}}  +    \ottsmode{(}   \ottsmode{1}  \hspace{-0.15ex}  \ottsmode{\uparrow}    \ottsmode{\hspace{-0.1ex}\cdot\hspace{-0.1ex} }  \ottsmode{n} \ottsmode{)}  \hspace{-0.3ex}\cdot\hspace{-0.3ex} \Gamma_{{\mathrm{2}}}   \,\pmb{\vdash}\,   \ottnt{t} \triangleleft (\lamnt{ \ottmv{x} }{ \ottsmode{m} }{ \ottnt{u} })   \pmb{:}   \ottstype{1} }{%
{\ottdrulename{Ty\_term\_FillF}}{\text{\CFillF}}%
}}

\newcommand{\ottdruleTyXXtermXXFillComp}[1]{\ottdrule[#1]{%
\ottpremise{\Gamma_{{\mathrm{1}}}  \,\pmb{\vdash}\,  \ottnt{t}  \pmb{:}   \ottstype{\lfloor}\,\!_{\mydestm{   \ottsmode{1}  \hspace{-0.15ex}  \ottsmode{\nu}   } } \ottstype{U} \ottstype{\rfloor} }%
\ottpremise{\Gamma_{{\mathrm{2}}}  \,\pmb{\vdash}\,  \ottnt{t'}  \pmb{:}   \ottstype{U} \,\ottstype{\ltimes}\, \ottstype{T} }%
}{
\Gamma_{{\mathrm{1}}}  +     \ottsmode{1}  \hspace{-0.15ex}  \ottsmode{\uparrow}   \hspace{-0.3ex}\cdot\hspace{-0.3ex} \Gamma_{{\mathrm{2}}}   \,\pmb{\vdash}\,   \ottnt{t} \mathop{\triangleleft\mycirc} \ottnt{t'}   \pmb{:}  \ottstype{T}}{%
{\ottdrulename{Ty\_term\_FillComp}}{\text{\CFillComp}}%
}}

\newcommand{\ottdruleTyXXtermXXFillLeaf}[1]{\ottdrule[#1]{%
\ottpremise{\Gamma_{{\mathrm{1}}}  \,\pmb{\vdash}\,  \ottnt{t}  \pmb{:}   \ottstype{\lfloor}\,\!_{\mydestm{ \ottsmode{n} } } \ottstype{T} \ottstype{\rfloor} }%
\ottpremise{\Gamma_{{\mathrm{2}}}  \,\pmb{\vdash}\,  \ottnt{t'}  \pmb{:}  \ottstype{T}}%
}{
\Gamma_{{\mathrm{1}}}  +    \ottsmode{(}   \ottsmode{1}  \hspace{-0.15ex}  \ottsmode{\uparrow}    \ottsmode{\hspace{-0.1ex}\cdot\hspace{-0.1ex} }  \ottsmode{n} \ottsmode{)}  \hspace{-0.3ex}\cdot\hspace{-0.3ex} \Gamma_{{\mathrm{2}}}   \,\pmb{\vdash}\,   \ottnt{t} \blacktriangleleft \ottnt{t'}   \pmb{:}   \ottstype{1} }{%
{\ottdrulename{Ty\_term\_FillLeaf}}{\text{\CFillLeaf}}%
}}

\newcommand{\ottdefnTyXXterm}[1]{\begin{ottdefnblock}[#1]{$\Gamma  \,\pmb{\vdash}\,  \ottnt{t}  \pmb{:}  \ottstype{T}$}{\ottcom{Typing judgment for terms}}
\ottusedrule{\ottdruleTyXXtermXXVal{}}
\ottusedrule{\ottdruleTyXXtermXXVar{}}
\ottusedrule{\ottdruleTyXXtermXXApp{}}
\ottusedrule{\ottdruleTyXXtermXXPatU{}}
\ottusedrule{\ottdruleTyXXtermXXPatS{}}
\ottusedrule{\ottdruleTyXXtermXXPatP{}}
\ottusedrule{\ottdruleTyXXtermXXPatE{}}
\ottusedrule{\ottdruleTyXXtermXXUpdA{}}
\ottusedrule{\ottdruleTyXXtermXXToA{}}
\ottusedrule{\ottdruleTyXXtermXXFromA{}}
\ottusedrule{\ottdruleTyXXtermXXNewA{}}
\ottusedrule{\ottdruleTyXXtermXXFillU{}}
\ottusedrule{\ottdruleTyXXtermXXFillL{}}
\ottusedrule{\ottdruleTyXXtermXXFillR{}}
\ottusedrule{\ottdruleTyXXtermXXFillP{}}
\ottusedrule{\ottdruleTyXXtermXXFillE{}}
\ottusedrule{\ottdruleTyXXtermXXFillF{}}
\ottusedrule{\ottdruleTyXXtermXXFillComp{}}
\ottusedrule{\ottdruleTyXXtermXXFillLeaf{}}
\end{ottdefnblock}}

\newcommand{\ottdruleTyXXstermXXFromAPP}[1]{\ottdrule[#1]{%
\ottpremise{\Gamma  \,\pmb{\vdash}\,  \ottnt{t}  \pmb{:}   \ottstype{T} \,\ottstype{\ltimes}\,  \ottstype{1}  }%
}{
\Gamma  \,\pmb{\vdash}\,   \ottkw{from}_{\ottkw{\ltimes} }'  \ottnt{t}   \pmb{:}  \ottstype{T}}{%
{\ottdrulename{Ty\_sterm\_FromA'}}{\text{\CFromA'}}%
}}

\newcommand{\ottdruleTyXXstermXXUnit}[1]{\ottdrule[#1]{%
\ottpremise{\texttt{DisposableOnly }\Gamma\texttt{ }}%
}{
\Gamma  \,\pmb{\vdash}\,   \ottsctor{()}   \pmb{:}   \ottstype{1} }{%
{\ottdrulename{Ty\_sterm\_Unit}}{\text{\CUnit}}%
}}

\newcommand{\ottdruleTyXXstermXXFun}[1]{\ottdrule[#1]{%
\ottpremise{ \Gamma_{{\mathrm{2}}} ,~  \ottmv{x} :\!_{\! \ottsmode{m} } \ottstype{T}    \,\pmb{\vdash}\,  \ottnt{u}  \pmb{:}  \ottstype{U}}%
}{
\Gamma_{{\mathrm{2}}}  \,\pmb{\vdash}\,   \lamnt{ \ottmv{x} }{ \ottsmode{m} }{ \ottnt{u} }   \pmb{:}   \ottstype{T} \,_{\myfuntm{ \ottsmode{m} } }\!\ottstype{\multimap}\, \ottstype{U} }{%
{\ottdrulename{Ty\_sterm\_Fun}}{\text{\CFun}}%
}}

\newcommand{\ottdruleTyXXstermXXLeft}[1]{\ottdrule[#1]{%
\ottpremise{\Gamma_{{\mathrm{2}}}  \,\pmb{\vdash}\,  \ottnt{t}  \pmb{:}  \ottstype{T_{{\mathrm{1}}}}}%
}{
\Gamma_{{\mathrm{2}}}  \,\pmb{\vdash}\,   \ottsctor{Inl}\, \ottnt{t}   \pmb{:}   \ottstype{T_{{\mathrm{1}}}} \ottstype{\oplus} \ottstype{T_{{\mathrm{2}}}} }{%
{\ottdrulename{Ty\_sterm\_Left}}{\text{\CLeft}}%
}}

\newcommand{\ottdruleTyXXstermXXRight}[1]{\ottdrule[#1]{%
\ottpremise{\Gamma_{{\mathrm{2}}}  \,\pmb{\vdash}\,  \ottnt{t}  \pmb{:}  \ottstype{T_{{\mathrm{2}}}}}%
}{
\Gamma_{{\mathrm{2}}}  \,\pmb{\vdash}\,   \ottsctor{Inr}\, \ottnt{t}   \pmb{:}   \ottstype{T_{{\mathrm{1}}}} \ottstype{\oplus} \ottstype{T_{{\mathrm{2}}}} }{%
{\ottdrulename{Ty\_sterm\_Right}}{\text{\CRight}}%
}}

\newcommand{\ottdruleTyXXstermXXExp}[1]{\ottdrule[#1]{%
\ottpremise{\Gamma_{{\mathrm{2}}}  \,\pmb{\vdash}\,  \ottnt{t}  \pmb{:}  \ottstype{T}}%
}{
 \ottsmode{m} \hspace{-0.3ex}\cdot\hspace{-0.3ex} \Gamma_{{\mathrm{2}}}   \,\pmb{\vdash}\,   \expcons{ \ottsmode{m} } \ottnt{t}   \pmb{:}   \ottstype{!}_{ \ottsmode{m} } \ottstype{T} }{%
{\ottdrulename{Ty\_sterm\_Exp}}{\text{\CExp}}%
}}

\newcommand{\ottdruleTyXXstermXXProd}[1]{\ottdrule[#1]{%
\ottpremise{\Gamma_{{\mathrm{21}}}  \,\pmb{\vdash}\,  \ottnt{t_{{\mathrm{1}}}}  \pmb{:}  \ottstype{T_{{\mathrm{1}}}}}%
\ottpremise{\Gamma_{{\mathrm{22}}}  \,\pmb{\vdash}\,  \ottnt{t_{{\mathrm{2}}}}  \pmb{:}  \ottstype{T_{{\mathrm{2}}}}}%
}{
\Gamma_{{\mathrm{21}}}  +  \Gamma_{{\mathrm{22}}}  \,\pmb{\vdash}\,   \ottsctor{(} \ottnt{t_{{\mathrm{1}}}} \,\ottsctor{,}~ \ottnt{t_{{\mathrm{2}}}} \ottsctor{)}   \pmb{:}   \ottstype{T_{{\mathrm{1}}}} \ottstype{\otimes} \ottstype{T_{{\mathrm{2}}}} }{%
{\ottdrulename{Ty\_sterm\_Prod}}{\text{\CProd}}%
}}

\newcommand{\ottdefnTyXXsterm}[1]{\begin{ottdefnblock}[#1]{$\Gamma  \,\pmb{\vdash}\,  \ottnt{t}  \pmb{:}  \ottstype{T}$}{\ottcom{Derived typing judgment for syntactic sugar forms}}
\ottusedrule{\ottdruleTyXXstermXXFromAPP{}}
\ottusedrule{\ottdruleTyXXstermXXUnit{}}
\ottusedrule{\ottdruleTyXXstermXXFun{}}
\ottusedrule{\ottdruleTyXXstermXXLeft{}}
\ottusedrule{\ottdruleTyXXstermXXRight{}}
\ottusedrule{\ottdruleTyXXstermXXExp{}}
\ottusedrule{\ottdruleTyXXstermXXProd{}}
\end{ottdefnblock}}

\newcommand{\ottdruleTyXXectxsXXId}[1]{\ottdrule[#1]{%
}{
  \smallbullet  \,\pmb{\dashv}\, \raisebox{0.075em}{$\scriptstyle []$} \pmb{:} \ottstype{U_{{\mathrm{0}}}} \ottstype{\rightarrowtail} \ottstype{U_{{\mathrm{0}}}} }{%
{\ottdrulename{Ty\_ectxs\_Id}}{\text{\CId}}%
}}

\newcommand{\ottdruleTyXXectxsXXAppOne}[1]{\ottdrule[#1]{%
\ottpremise{   \ottsmode{m} \hspace{-0.3ex}\cdot\hspace{-0.3ex} \Delta_{{\mathrm{1}}}  ,~ \Delta_{{\mathrm{2}}}  \,\pmb{\dashv}\, \ottnt{E} \pmb{:} \ottstype{U} \ottstype{\rightarrowtail} \ottstype{U_{{\mathrm{0}}}} }%
\ottpremise{\Delta_{{\mathrm{2}}}  \,\pmb{\vdash}\,  \ottnt{t'}  \pmb{:}   \ottstype{T} \,_{\myfuntm{ \ottsmode{m} } }\!\ottstype{\multimap}\, \ottstype{U} }%
}{
 \Delta_{{\mathrm{1}}} \,\pmb{\dashv}\,  \ottnt{E} \hspace*{0.4em}\circ\hspace*{0.4em}   \ottnt{t'} ~ \raisebox{0.075em}{$\scriptstyle []$}    \pmb{:} \ottstype{T} \ottstype{\rightarrowtail} \ottstype{U_{{\mathrm{0}}}} }{%
{\ottdrulename{Ty\_ectxs\_App1}}{\text{\CApp\textsubscript{1}}}%
}}

\newcommand{\ottdruleTyXXectxsXXAppTwo}[1]{\ottdrule[#1]{%
\ottpremise{   \ottsmode{m} \hspace{-0.3ex}\cdot\hspace{-0.3ex} \Delta_{{\mathrm{1}}}  ,~ \Delta_{{\mathrm{2}}}  \,\pmb{\dashv}\, \ottnt{E} \pmb{:} \ottstype{U} \ottstype{\rightarrowtail} \ottstype{U_{{\mathrm{0}}}} }%
\ottpremise{\Delta_{{\mathrm{1}}}  \,\pmb{\vdash}\,  \ottnt{v}  \pmb{:}  \ottstype{T}}%
}{
 \Delta_{{\mathrm{2}}} \,\pmb{\dashv}\,  \ottnt{E} \hspace*{0.4em}\circ\hspace*{0.4em}   \raisebox{0.075em}{$\scriptstyle []$} ~ \ottnt{v}    \pmb{:}  \ottstype{(}  \ottstype{T} \,_{\myfuntm{ \ottsmode{m} } }\!\ottstype{\multimap}\, \ottstype{U}  \ottstype{)}  \ottstype{\rightarrowtail} \ottstype{U_{{\mathrm{0}}}} }{%
{\ottdrulename{Ty\_ectxs\_App2}}{\text{\CApp\textsubscript{2}}}%
}}

\newcommand{\ottdruleTyXXectxsXXPatU}[1]{\ottdrule[#1]{%
\ottpremise{  \Delta_{{\mathrm{1}}} ,~ \Delta_{{\mathrm{2}}}  \,\pmb{\dashv}\, \ottnt{E} \pmb{:} \ottstype{U} \ottstype{\rightarrowtail} \ottstype{U_{{\mathrm{0}}}} }%
\ottpremise{\Delta_{{\mathrm{2}}}  \,\pmb{\vdash}\,  \ottnt{u}  \pmb{:}  \ottstype{U}}%
}{
 \Delta_{{\mathrm{1}}} \,\pmb{\dashv}\,  \ottnt{E} \hspace*{0.4em}\circ\hspace*{0.4em}   \raisebox{0.075em}{$\scriptstyle []$} \patu \ottnt{u}    \pmb{:}  \ottstype{1}  \ottstype{\rightarrowtail} \ottstype{U_{{\mathrm{0}}}} }{%
{\ottdrulename{Ty\_ectxs\_PatU}}{\text{\CPatU}}%
}}

\newcommand{\ottdruleTyXXectxsXXPatS}[1]{\ottdrule[#1]{%
\ottpremise{   \ottsmode{m} \hspace{-0.3ex}\cdot\hspace{-0.3ex} \Delta_{{\mathrm{1}}}  ,~ \Delta_{{\mathrm{2}}}  \,\pmb{\dashv}\, \ottnt{E} \pmb{:} \ottstype{U} \ottstype{\rightarrowtail} \ottstype{U_{{\mathrm{0}}}} }%
\ottpremise{ \Delta_{{\mathrm{2}}} ,~  \ottmv{x_{{\mathrm{1}}}} :\!_{\! \ottsmode{m} } \ottstype{T_{{\mathrm{1}}}}    \,\pmb{\vdash}\,  \ottnt{u_{{\mathrm{1}}}}  \pmb{:}  \ottstype{U}}%
\ottpremise{ \Delta_{{\mathrm{2}}} ,~  \ottmv{x_{{\mathrm{2}}}} :\!_{\! \ottsmode{m} } \ottstype{T_{{\mathrm{2}}}}    \,\pmb{\vdash}\,  \ottnt{u_{{\mathrm{2}}}}  \pmb{:}  \ottstype{U}}%
}{
 \Delta_{{\mathrm{1}}} \,\pmb{\dashv}\,  \ottnt{E} \hspace*{0.4em}\circ\hspace*{0.4em}   \ottkw{case}_{ \ottsmode{m} }~ \raisebox{0.075em}{$\scriptstyle []$} ~\ottkw{of}~\{\ottsctor{Inl}\, \ottmv{x_{{\mathrm{1}}}} \pmb{\mapsto} \ottnt{u_{{\mathrm{1}}}} \,,~\ottsctor{Inr}\, \ottmv{x_{{\mathrm{2}}}} \pmb{\mapsto} \ottnt{u_{{\mathrm{2}}}} \}    \pmb{:}  \ottstype{(}  \ottstype{T_{{\mathrm{1}}}} \ottstype{\oplus} \ottstype{T_{{\mathrm{2}}}}  \ottstype{)}  \ottstype{\rightarrowtail} \ottstype{U_{{\mathrm{0}}}} }{%
{\ottdrulename{Ty\_ectxs\_PatS}}{\text{\CPatS}}%
}}

\newcommand{\ottdruleTyXXectxsXXPatP}[1]{\ottdrule[#1]{%
\ottpremise{   \ottsmode{m} \hspace{-0.3ex}\cdot\hspace{-0.3ex} \Delta_{{\mathrm{1}}}  ,~ \Delta_{{\mathrm{2}}}  \,\pmb{\dashv}\, \ottnt{E} \pmb{:} \ottstype{U} \ottstype{\rightarrowtail} \ottstype{U_{{\mathrm{0}}}} }%
\ottpremise{  \Delta_{{\mathrm{2}}} ,~  \ottmv{x_{{\mathrm{1}}}} :\!_{\! \ottsmode{m} } \ottstype{T_{{\mathrm{1}}}}   ,~  \ottmv{x_{{\mathrm{2}}}} :\!_{\! \ottsmode{m} } \ottstype{T_{{\mathrm{2}}}}    \,\pmb{\vdash}\,  \ottnt{u}  \pmb{:}  \ottstype{U}}%
}{
 \Delta_{{\mathrm{1}}} \,\pmb{\dashv}\,  \ottnt{E} \hspace*{0.4em}\circ\hspace*{0.4em}   \ottkw{case}_{ \ottsmode{m} }~ \raisebox{0.075em}{$\scriptstyle []$} ~\ottkw{of}~\ottsctor{(} \ottmv{x_{{\mathrm{1}}}} \,\ottsctor{,}~ \ottmv{x_{{\mathrm{2}}}} \ottsctor{)} \pmb{\mapsto} \ottnt{u}    \pmb{:}  \ottstype{(}  \ottstype{T_{{\mathrm{1}}}} \ottstype{\otimes} \ottstype{T_{{\mathrm{2}}}}  \ottstype{)}  \ottstype{\rightarrowtail} \ottstype{U_{{\mathrm{0}}}} }{%
{\ottdrulename{Ty\_ectxs\_PatP}}{\text{\CPatP}}%
}}

\newcommand{\ottdruleTyXXectxsXXPatE}[1]{\ottdrule[#1]{%
\ottpremise{   \ottsmode{m} \hspace{-0.3ex}\cdot\hspace{-0.3ex} \Delta_{{\mathrm{1}}}  ,~ \Delta_{{\mathrm{2}}}  \,\pmb{\dashv}\, \ottnt{E} \pmb{:} \ottstype{U} \ottstype{\rightarrowtail} \ottstype{U_{{\mathrm{0}}}} }%
\ottpremise{ \Delta_{{\mathrm{2}}} ,~  \ottmv{x} :\!_{\! \ottsmode{m}  \ottsmode{\hspace{-0.1ex}\cdot\hspace{-0.1ex} }  \ottsmode{m'} } \ottstype{T}    \,\pmb{\vdash}\,  \ottnt{u}  \pmb{:}  \ottstype{U}}%
}{
 \Delta_{{\mathrm{1}}} \,\pmb{\dashv}\,  \ottnt{E} \hspace*{0.4em}\circ\hspace*{0.4em}   \ottkw{case}_{ \ottsmode{m} }~ \raisebox{0.075em}{$\scriptstyle []$} ~\ottkw{of}~\expcons{ \ottsmode{m'} } \ottmv{x} \pmb{\mapsto} \ottnt{u}    \pmb{:}  \ottstype{!}_{ \ottsmode{m'} } \ottstype{T}  \ottstype{\rightarrowtail} \ottstype{U_{{\mathrm{0}}}} }{%
{\ottdrulename{Ty\_ectxs\_PatE}}{\text{\CPatE}}%
}}

\newcommand{\ottdruleTyXXectxsXXUpdA}[1]{\ottdrule[#1]{%
\ottpremise{  \Delta_{{\mathrm{1}}} ,~ \Delta_{{\mathrm{2}}}  \,\pmb{\dashv}\, \ottnt{E} \pmb{:}  \ottstype{U} \,\ottstype{\ltimes}\, \ottstype{T'}  \ottstype{\rightarrowtail} \ottstype{U_{{\mathrm{0}}}} }%
\ottpremise{    \ottsmode{1}  \hspace{-0.15ex}  \ottsmode{\uparrow}   \hspace{-0.3ex}\cdot\hspace{-0.3ex} \Delta_{{\mathrm{2}}}  ,~  \ottmv{x} :\!_{\!   \ottsmode{1}  \hspace{-0.15ex}  \ottsmode{\nu}   } \ottstype{T}    \,\pmb{\vdash}\,  \ottnt{t'}  \pmb{:}  \ottstype{T'}}%
}{
 \Delta_{{\mathrm{1}}} \,\pmb{\dashv}\,  \ottnt{E} \hspace*{0.4em}\circ\hspace*{0.4em}   \ottkw{upd}_{\ottkw{\ltimes} }\, \raisebox{0.075em}{$\scriptstyle []$} ~\ottkw{with}~ \ottmv{x} \, \pmb{\mapsto} \ottnt{t'}    \pmb{:}  \ottstype{(}  \ottstype{U} \,\ottstype{\ltimes}\, \ottstype{T}  \ottstype{)}  \ottstype{\rightarrowtail} \ottstype{U_{{\mathrm{0}}}} }{%
{\ottdrulename{Ty\_ectxs\_UpdA}}{\text{\CUpdA}}%
}}

\newcommand{\ottdruleTyXXectxsXXToA}[1]{\ottdrule[#1]{%
\ottpremise{ \Delta \,\pmb{\dashv}\, \ottnt{E} \pmb{:}  \ottstype{(}  \ottstype{U} \,\ottstype{\ltimes}\,  \ottstype{1}   \ottstype{)}  \ottstype{\rightarrowtail} \ottstype{U_{{\mathrm{0}}}} }%
}{
 \Delta \,\pmb{\dashv}\,  \ottnt{E} \hspace*{0.4em}\circ\hspace*{0.4em}   \ottkw{to}_{\ottkw{\ltimes} }\, \raisebox{0.075em}{$\scriptstyle []$}    \pmb{:} \ottstype{U} \ottstype{\rightarrowtail} \ottstype{U_{{\mathrm{0}}}} }{%
{\ottdrulename{Ty\_ectxs\_ToA}}{\text{\CToA}}%
}}

\newcommand{\ottdruleTyXXectxsXXFromA}[1]{\ottdrule[#1]{%
\ottpremise{ \Delta \,\pmb{\dashv}\, \ottnt{E} \pmb{:}  \ottstype{(}  \ottstype{U} \ottstype{\otimes}  \ottstype{(}  \ottstype{!}_{   \ottsmode{1}  \hspace{-0.15ex}  \ottsmode{\infty}   } \ottstype{T}  \ottstype{)}   \ottstype{)}  \ottstype{\rightarrowtail} \ottstype{U_{{\mathrm{0}}}} }%
}{
 \Delta \,\pmb{\dashv}\,  \ottnt{E} \hspace*{0.4em}\circ\hspace*{0.4em}   \ottkw{from}_{\ottkw{\ltimes} }\, \raisebox{0.075em}{$\scriptstyle []$}    \pmb{:}  \ottstype{(}  \ottstype{U} \,\ottstype{\ltimes}\,  \ottstype{(}  \ottstype{!}_{   \ottsmode{1}  \hspace{-0.15ex}  \ottsmode{\infty}   } \ottstype{T}  \ottstype{)}   \ottstype{)}  \ottstype{\rightarrowtail} \ottstype{U_{{\mathrm{0}}}} }{%
{\ottdrulename{Ty\_ectxs\_FromA}}{\text{\CFromA}}%
}}

\newcommand{\ottdruleTyXXectxsXXFillU}[1]{\ottdrule[#1]{%
\ottpremise{ \Delta \,\pmb{\dashv}\, \ottnt{E} \pmb{:}  \ottstype{1}  \ottstype{\rightarrowtail} \ottstype{U_{{\mathrm{0}}}} }%
}{
 \Delta \,\pmb{\dashv}\,  \ottnt{E} \hspace*{0.4em}\circ\hspace*{0.4em}  \raisebox{0.075em}{$\scriptstyle []$}  \triangleleft  \ottsctor{()}   \pmb{:}  \ottstype{\lfloor}\,\!_{\mydestm{ \ottsmode{n} } }  \ottstype{1}  \ottstype{\rfloor}  \ottstype{\rightarrowtail} \ottstype{U_{{\mathrm{0}}}} }{%
{\ottdrulename{Ty\_ectxs\_FillU}}{\text{\CFillU}}%
}}

\newcommand{\ottdruleTyXXectxsXXFillL}[1]{\ottdrule[#1]{%
\ottpremise{ \Delta \,\pmb{\dashv}\, \ottnt{E} \pmb{:}  \ottstype{\lfloor}\,\!_{\mydestm{ \ottsmode{n} } } \ottstype{T_{{\mathrm{1}}}} \ottstype{\rfloor}  \ottstype{\rightarrowtail} \ottstype{U_{{\mathrm{0}}}} }%
}{
 \Delta \,\pmb{\dashv}\,  \ottnt{E} \hspace*{0.4em}\circ\hspace*{0.4em}  \raisebox{0.075em}{$\scriptstyle []$}  \triangleleft \, \ottsctor{Inl}   \pmb{:}  \ottstype{\lfloor}\,\!_{\mydestm{ \ottsmode{n} } }  \ottstype{T_{{\mathrm{1}}}} \ottstype{\oplus} \ottstype{T_{{\mathrm{2}}}}  \ottstype{\rfloor}  \ottstype{\rightarrowtail} \ottstype{U_{{\mathrm{0}}}} }{%
{\ottdrulename{Ty\_ectxs\_FillL}}{\text{\CFillL}}%
}}

\newcommand{\ottdruleTyXXectxsXXFillR}[1]{\ottdrule[#1]{%
\ottpremise{ \Delta \,\pmb{\dashv}\, \ottnt{E} \pmb{:}  \ottstype{\lfloor}\,\!_{\mydestm{ \ottsmode{n} } } \ottstype{T_{{\mathrm{2}}}} \ottstype{\rfloor}  \ottstype{\rightarrowtail} \ottstype{U_{{\mathrm{0}}}} }%
}{
 \Delta \,\pmb{\dashv}\,  \ottnt{E} \hspace*{0.4em}\circ\hspace*{0.4em}  \raisebox{0.075em}{$\scriptstyle []$}  \triangleleft \, \ottsctor{Inr}   \pmb{:}  \ottstype{\lfloor}\,\!_{\mydestm{ \ottsmode{n} } }  \ottstype{T_{{\mathrm{1}}}} \ottstype{\oplus} \ottstype{T_{{\mathrm{2}}}}  \ottstype{\rfloor}  \ottstype{\rightarrowtail} \ottstype{U_{{\mathrm{0}}}} }{%
{\ottdrulename{Ty\_ectxs\_FillR}}{\text{\CFillR}}%
}}

\newcommand{\ottdruleTyXXectxsXXFillP}[1]{\ottdrule[#1]{%
\ottpremise{ \Delta \,\pmb{\dashv}\, \ottnt{E} \pmb{:}  \ottstype{(}   \ottstype{\lfloor}\,\!_{\mydestm{ \ottsmode{n} } } \ottstype{T_{{\mathrm{1}}}} \ottstype{\rfloor}  \ottstype{\otimes}  \ottstype{\lfloor}\,\!_{\mydestm{ \ottsmode{n} } } \ottstype{T_{{\mathrm{2}}}} \ottstype{\rfloor}   \ottstype{)}  \ottstype{\rightarrowtail} \ottstype{U_{{\mathrm{0}}}} }%
}{
 \Delta \,\pmb{\dashv}\,  \ottnt{E} \hspace*{0.4em}\circ\hspace*{0.4em}  \raisebox{0.075em}{$\scriptstyle []$}  \triangleleft  \ottsctor{({,})}   \pmb{:}  \ottstype{\lfloor}\,\!_{\mydestm{ \ottsmode{n} } }  \ottstype{T_{{\mathrm{1}}}} \ottstype{\otimes} \ottstype{T_{{\mathrm{2}}}}  \ottstype{\rfloor}  \ottstype{\rightarrowtail} \ottstype{U_{{\mathrm{0}}}} }{%
{\ottdrulename{Ty\_ectxs\_FillP}}{\text{\CFillP}}%
}}

\newcommand{\ottdruleTyXXectxsXXFillE}[1]{\ottdrule[#1]{%
\ottpremise{ \Delta \,\pmb{\dashv}\, \ottnt{E} \pmb{:}  \ottstype{\lfloor}\,\!_{\mydestm{ \ottsmode{m}  \ottsmode{\hspace{-0.1ex}\cdot\hspace{-0.1ex} }  \ottsmode{n} } } \ottstype{T} \ottstype{\rfloor}  \ottstype{\rightarrowtail} \ottstype{U_{{\mathrm{0}}}} }%
}{
 \Delta \,\pmb{\dashv}\,  \ottnt{E} \hspace*{0.4em}\circ\hspace*{0.4em}   \raisebox{0.075em}{$\scriptstyle []$} \triangleleft \,\expcons{ \ottsmode{m} }    \pmb{:}  \ottstype{\lfloor}\,\!_{\mydestm{ \ottsmode{n} } }  \ottstype{!}_{ \ottsmode{m} } \ottstype{T}  \ottstype{\rfloor}  \ottstype{\rightarrowtail} \ottstype{U_{{\mathrm{0}}}} }{%
{\ottdrulename{Ty\_ectxs\_FillE}}{\text{\CFillE}}%
}}

\newcommand{\ottdruleTyXXectxsXXFillF}[1]{\ottdrule[#1]{%
\ottpremise{  \Delta_{{\mathrm{1}}} ,~   \ottsmode{(}   \ottsmode{1}  \hspace{-0.15ex}  \ottsmode{\uparrow}    \ottsmode{\hspace{-0.1ex}\cdot\hspace{-0.1ex} }  \ottsmode{n} \ottsmode{)}  \hspace{-0.3ex}\cdot\hspace{-0.3ex} \Delta_{{\mathrm{2}}}   \,\pmb{\dashv}\, \ottnt{E} \pmb{:}  \ottstype{1}  \ottstype{\rightarrowtail} \ottstype{U_{{\mathrm{0}}}} }%
\ottpremise{ \Delta_{{\mathrm{2}}} ,~  \ottmv{x} :\!_{\! \ottsmode{m} } \ottstype{T}    \,\pmb{\vdash}\,  \ottnt{u}  \pmb{:}  \ottstype{U}}%
}{
 \Delta_{{\mathrm{1}}} \,\pmb{\dashv}\,  \ottnt{E} \hspace*{0.4em}\circ\hspace*{0.4em}   \raisebox{0.075em}{$\scriptstyle []$} \triangleleft (\lamnt{ \ottmv{x} }{ \ottsmode{m} }{ \ottnt{u} })    \pmb{:}  \ottstype{\lfloor}\,\!_{\mydestm{ \ottsmode{n} } }  \ottstype{T} \,_{\myfuntm{ \ottsmode{m} } }\!\ottstype{\multimap}\, \ottstype{U}  \ottstype{\rfloor}  \ottstype{\rightarrowtail} \ottstype{U_{{\mathrm{0}}}} }{%
{\ottdrulename{Ty\_ectxs\_FillF}}{\text{\CFillF}}%
}}

\newcommand{\ottdruleTyXXectxsXXFillCompOne}[1]{\ottdrule[#1]{%
\ottpremise{  \Delta_{{\mathrm{1}}} ,~    \ottsmode{1}  \hspace{-0.15ex}  \ottsmode{\uparrow}   \hspace{-0.3ex}\cdot\hspace{-0.3ex} \Delta_{{\mathrm{2}}}   \,\pmb{\dashv}\, \ottnt{E} \pmb{:} \ottstype{T} \ottstype{\rightarrowtail} \ottstype{U_{{\mathrm{0}}}} }%
\ottpremise{\Delta_{{\mathrm{2}}}  \,\pmb{\vdash}\,  \ottnt{t'}  \pmb{:}   \ottstype{U} \,\ottstype{\ltimes}\, \ottstype{T} }%
}{
 \Delta_{{\mathrm{1}}} \,\pmb{\dashv}\,  \ottnt{E} \hspace*{0.4em}\circ\hspace*{0.4em}   \raisebox{0.075em}{$\scriptstyle []$} \mathop{\triangleleft\mycirc} \ottnt{t'}    \pmb{:}  \ottstype{\lfloor}\,\!_{\mydestm{   \ottsmode{1}  \hspace{-0.15ex}  \ottsmode{\nu}   } } \ottstype{U} \ottstype{\rfloor}  \ottstype{\rightarrowtail} \ottstype{U_{{\mathrm{0}}}} }{%
{\ottdrulename{Ty\_ectxs\_FillComp1}}{\text{\CFillComp\textsubscript{1}}}%
}}

\newcommand{\ottdruleTyXXectxsXXFillCompTwo}[1]{\ottdrule[#1]{%
\ottpremise{  \Delta_{{\mathrm{1}}} ,~    \ottsmode{1}  \hspace{-0.15ex}  \ottsmode{\uparrow}   \hspace{-0.3ex}\cdot\hspace{-0.3ex} \Delta_{{\mathrm{2}}}   \,\pmb{\dashv}\, \ottnt{E} \pmb{:} \ottstype{T} \ottstype{\rightarrowtail} \ottstype{U_{{\mathrm{0}}}} }%
\ottpremise{\Delta_{{\mathrm{1}}}  \,\pmb{\vdash}\,  \ottnt{v}  \pmb{:}   \ottstype{\lfloor}\,\!_{\mydestm{   \ottsmode{1}  \hspace{-0.15ex}  \ottsmode{\nu}   } } \ottstype{U} \ottstype{\rfloor} }%
}{
 \Delta_{{\mathrm{2}}} \,\pmb{\dashv}\,  \ottnt{E} \hspace*{0.4em}\circ\hspace*{0.4em}   \ottnt{v} \mathop{\triangleleft\mycirc} \raisebox{0.075em}{$\scriptstyle []$}    \pmb{:}  \ottstype{U} \,\ottstype{\ltimes}\, \ottstype{T}  \ottstype{\rightarrowtail} \ottstype{U_{{\mathrm{0}}}} }{%
{\ottdrulename{Ty\_ectxs\_FillComp2}}{\text{\CFillComp\textsubscript{2}}}%
}}

\newcommand{\ottdruleTyXXectxsXXFillLeafOne}[1]{\ottdrule[#1]{%
\ottpremise{  \Delta_{{\mathrm{1}}} ,~   \ottsmode{(}   \ottsmode{1}  \hspace{-0.15ex}  \ottsmode{\uparrow}    \ottsmode{\hspace{-0.1ex}\cdot\hspace{-0.1ex} }  \ottsmode{n} \ottsmode{)}  \hspace{-0.3ex}\cdot\hspace{-0.3ex} \Delta_{{\mathrm{2}}}   \,\pmb{\dashv}\, \ottnt{E} \pmb{:}  \ottstype{1}  \ottstype{\rightarrowtail} \ottstype{U_{{\mathrm{0}}}} }%
\ottpremise{\Delta_{{\mathrm{2}}}  \,\pmb{\vdash}\,  \ottnt{t'}  \pmb{:}  \ottstype{T}}%
}{
 \Delta_{{\mathrm{1}}} \,\pmb{\dashv}\,  \ottnt{E} \hspace*{0.4em}\circ\hspace*{0.4em}   \raisebox{0.075em}{$\scriptstyle []$} \blacktriangleleft \ottnt{t'}    \pmb{:}  \ottstype{\lfloor}\,\!_{\mydestm{ \ottsmode{n} } } \ottstype{T} \ottstype{\rfloor}  \ottstype{\rightarrowtail} \ottstype{U_{{\mathrm{0}}}} }{%
{\ottdrulename{Ty\_ectxs\_FillLeaf1}}{\text{\CFillLeaf\textsubscript{1}}}%
}}

\newcommand{\ottdruleTyXXectxsXXFillLeafTwo}[1]{\ottdrule[#1]{%
\ottpremise{  \Delta_{{\mathrm{1}}} ,~   \ottsmode{(}   \ottsmode{1}  \hspace{-0.15ex}  \ottsmode{\uparrow}    \ottsmode{\hspace{-0.1ex}\cdot\hspace{-0.1ex} }  \ottsmode{n} \ottsmode{)}  \hspace{-0.3ex}\cdot\hspace{-0.3ex} \Delta_{{\mathrm{2}}}   \,\pmb{\dashv}\, \ottnt{E} \pmb{:}  \ottstype{1}  \ottstype{\rightarrowtail} \ottstype{U_{{\mathrm{0}}}} }%
\ottpremise{\Delta_{{\mathrm{1}}}  \,\pmb{\vdash}\,  \ottnt{v}  \pmb{:}   \ottstype{\lfloor}\,\!_{\mydestm{ \ottsmode{n} } } \ottstype{T} \ottstype{\rfloor} }%
}{
 \Delta_{{\mathrm{2}}} \,\pmb{\dashv}\,  \ottnt{E} \hspace*{0.4em}\circ\hspace*{0.4em}   \ottnt{v} \blacktriangleleft \raisebox{0.075em}{$\scriptstyle []$}    \pmb{:} \ottstype{T} \ottstype{\rightarrowtail} \ottstype{U_{{\mathrm{0}}}} }{%
{\ottdrulename{Ty\_ectxs\_FillLeaf2}}{\text{\CFillLeaf\textsubscript{2}}}%
}}

\newcommand{\ottdruleTyXXectxsXXOpenAmpar}[1]{\ottdrule[#1]{%
\ottpremise{ \ottshname{\mathsfbf{hnames}(} \ottnt{E} \ottshname{)} \texttt{ \#\# } \ottshname{\mathsfbf{hnames}(} \Delta_{{\mathrm{3}}} \ottshname{)} \texttt{ }}%
\ottpremise{  \Delta_{{\mathrm{1}}} ,~ \Delta_{{\mathrm{2}}}  \,\pmb{\dashv}\, \ottnt{E} \pmb{:}  \ottstype{(}  \ottstype{U} \,\ottstype{\ltimes}\, \ottstype{T'}  \ottstype{)}  \ottstype{\rightarrowtail} \ottstype{U_{{\mathrm{0}}}} }%
\ottpremise{ \Delta_{{\mathrm{2}}} ,~  \ottshname{\destminus^{\scriptscriptstyle\text{-}1} } \Delta_{{\mathrm{3}}}    \!\!\pmb{\phantom{a}^{\scriptscriptstyle \mathrm{v} }\!\!\vdash}\,  \ottnt{v_{{\mathrm{2}}}}  \pmb{:}  \ottstype{U}}%
}{
     \ottsmode{1}  \hspace{-0.15ex}  \ottsmode{\uparrow}   \hspace{-0.3ex}\cdot\hspace{-0.3ex} \Delta_{{\mathrm{1}}}  ,~ \Delta_{{\mathrm{3}}}  \,\pmb{\dashv}\,  \ottnt{E} \hspace*{0.4em}\circ\hspace*{0.4em}   ^{\text{op}\!}_{  \ottshname{\mathsfbf{hnames}(} \Delta_{{\mathrm{3}}} \ottshname{)}  \!}\ottsctor{\langle} \ottnt{v_{{\mathrm{2}}}} \,\ottsctor{\bbcomma}~ \raisebox{0.075em}{$\scriptstyle []$} \ottsctor{\rangle}    \pmb{:} \ottstype{T'} \ottstype{\rightarrowtail} \ottstype{U_{{\mathrm{0}}}} }{%
{\ottdrulename{Ty\_ectxs\_OpenAmpar}}{\text{\COpenAmpar}}%
}}

\newcommand{\ottdefnTyXXectxs}[1]{\begin{ottdefnblock}[#1]{$ \Delta \,\pmb{\dashv}\, \ottnt{E} \pmb{:} \ottstype{T} \ottstype{\rightarrowtail} \ottstype{U_{{\mathrm{0}}}} $}{\ottcom{Typing judgment for evaluation contexts}}
\ottusedrule{\ottdruleTyXXectxsXXId{}}
\ottusedrule{\ottdruleTyXXectxsXXAppOne{}}
\ottusedrule{\ottdruleTyXXectxsXXAppTwo{}}
\ottusedrule{\ottdruleTyXXectxsXXPatU{}}
\ottusedrule{\ottdruleTyXXectxsXXPatS{}}
\ottusedrule{\ottdruleTyXXectxsXXPatP{}}
\ottusedrule{\ottdruleTyXXectxsXXPatE{}}
\ottusedrule{\ottdruleTyXXectxsXXUpdA{}}
\ottusedrule{\ottdruleTyXXectxsXXToA{}}
\ottusedrule{\ottdruleTyXXectxsXXFromA{}}
\ottusedrule{\ottdruleTyXXectxsXXFillU{}}
\ottusedrule{\ottdruleTyXXectxsXXFillL{}}
\ottusedrule{\ottdruleTyXXectxsXXFillR{}}
\ottusedrule{\ottdruleTyXXectxsXXFillP{}}
\ottusedrule{\ottdruleTyXXectxsXXFillE{}}
\ottusedrule{\ottdruleTyXXectxsXXFillF{}}
\ottusedrule{\ottdruleTyXXectxsXXFillCompOne{}}
\ottusedrule{\ottdruleTyXXectxsXXFillCompTwo{}}
\ottusedrule{\ottdruleTyXXectxsXXFillLeafOne{}}
\ottusedrule{\ottdruleTyXXectxsXXFillLeafTwo{}}
\ottusedrule{\ottdruleTyXXectxsXXOpenAmpar{}}
\end{ottdefnblock}}

\newcommand{\ottdruleTyXXcmd}[1]{\ottdrule[#1]{%
\ottpremise{ \Delta \,\pmb{\dashv}\, \ottnt{E} \pmb{:} \ottstype{T} \ottstype{\rightarrowtail} \ottstype{U_{{\mathrm{0}}}} }%
\ottpremise{\Delta  \,\pmb{\vdash}\,  \ottnt{t}  \pmb{:}  \ottstype{T}}%
}{
 \,\pmb{\vdash}\, \ottnt{E} \big[\, \ottnt{t} \,\big] \pmb{:} \ottstype{U_{{\mathrm{0}}}} }{%
{\ottdrulename{Ty\_cmd}}{\text{\CTyCmd}}%
}}

\newcommand{\ottdefnTy}[1]{\begin{ottdefnblock}[#1]{$ \,\pmb{\vdash}\, \ottnt{E} \big[\, \ottnt{t} \,\big] \pmb{:} \ottstype{T} $}{\ottcom{Typing judgment for commands}}
\ottusedrule{\ottdruleTyXXcmd{}}
\end{ottdefnblock}}

  \renewottcommands[ott]

\usepackage{ottstyling}
\patchcmd{\ottmetavars}{$ \ottmv{k} $ & \ottcom{Index for ranges} \\}{}{}{}

\patchcmd{\ottdruleTyXXectxsXXOpenAmpar}{%
\ottpremise{  \Delta_{{\mathrm{1}}} ,~ \Delta_{{\mathrm{2}}}  \,\pmb{\dashv}\, \ottnt{C} \pmb{:}  \ottstype{(}  \ottstype{U} \,\ottstype{\ltimes}\, \ottstype{T'}  \ottstype{)}  \ottstype{\rightarrowtail} \ottstype{U_{{\mathrm{0}}}} }%
\ottpremise{ \Delta_{{\mathrm{2}}} ,~  \ottshname{\destminus^{\scriptscriptstyle\text{-}1} } \Delta_{{\mathrm{3}}}    \!\!\pmb{\phantom{a}^{\scriptscriptstyle \mathrm{v} }\!\!\vdash}\,  \ottnt{v_{{\mathrm{2}}}}  \pmb{:}  \ottstype{U}}%
}{\ottpremise{  \Delta_{{\mathrm{1}}} ,~ \Delta_{{\mathrm{2}}}  \,\pmb{\dashv}\, \ottnt{C} \pmb{:}  \ottstype{(}  \ottstype{U} \,\ottstype{\ltimes}\, \ottstype{T'}  \ottstype{)}  \ottstype{\rightarrowtail} \ottstype{U_{{\mathrm{0}}}}
\qquad
\Delta_{{\mathrm{2}}} ,~  \ottshname{\destminus^{\scriptscriptstyle\text{-}1} } \Delta_{{\mathrm{3}}}    \!\!\pmb{\phantom{a}^{\scriptscriptstyle \mathrm{v} }\!\!\vdash}\,  \ottnt{v_{{\mathrm{2}}}}  \pmb{:}  \ottstype{U}}
}{}{}

\patchcmd{\ottdruleTyXXvalXXAmpar}{%
\ottpremise{   \ottsmode{1}  \hspace{-0.15ex}  \ottsmode{\uparrow}    \ottsmode{\hspace{-0.1ex}\cdot\hspace{-0.1ex} }  \Delta_{{\mathrm{1}}} ,~ \Delta_{{\mathrm{3}}}   \!\!\pmb{\phantom{a}^{\scriptscriptstyle \mathrm{v} }\!\!\vdash}\,  \ottnt{v_{{\mathrm{1}}}}  \pmb{:}  \ottstype{T}}%
\ottpremise{ \Delta_{{\mathrm{2}}} ,~  \ottshname{\destminus^{\scriptscriptstyle\text{-}1} } \Delta_{{\mathrm{3}}}    \!\!\pmb{\phantom{a}^{\scriptscriptstyle \mathrm{v} }\!\!\vdash}\,  \ottnt{v_{{\mathrm{2}}}}  \pmb{:}  \ottstype{U}}%
}{
\ottpremise{   \ottsmode{1}  \hspace{-0.15ex}  \ottsmode{\uparrow}    \ottsmode{\hspace{-0.1ex}\cdot\hspace{-0.1ex} }  \Delta_{{\mathrm{1}}} ,~ \Delta_{{\mathrm{3}}}   \!\!\pmb{\phantom{a}^{\scriptscriptstyle \mathrm{v} }\!\!\vdash}\,  \ottnt{v_{{\mathrm{1}}}}  \pmb{:}  \ottstype{T}
\qquad
\Delta_{{\mathrm{2}}} ,~  \ottshname{\destminus^{\scriptscriptstyle\text{-}1} } \Delta_{{\mathrm{3}}}    \!\!\pmb{\phantom{a}^{\scriptscriptstyle \mathrm{v} }\!\!\vdash}\,  \ottnt{v_{{\mathrm{2}}}}  \pmb{:}  \ottstype{U}}%
}{}{}

\patchcmd{\ottdruleTyXXtermXXPatS}{%
\ottpremise{ \Gamma_{{\mathrm{2}}} ,~  \ottmv{x_{{\mathrm{1}}}} :\!_{\! \ottsmode{m} } \ottstype{T_{{\mathrm{1}}}}    \,\pmb{\vdash}\,  \ottnt{u_{{\mathrm{1}}}}  \pmb{:}  \ottstype{U}}%
\ottpremise{ \Gamma_{{\mathrm{2}}} ,~  \ottmv{x_{{\mathrm{2}}}} :\!_{\! \ottsmode{m} } \ottstype{T_{{\mathrm{2}}}}    \,\pmb{\vdash}\,  \ottnt{u_{{\mathrm{2}}}}  \pmb{:}  \ottstype{U}}%
}{\ottpremise{ \Gamma_{{\mathrm{2}}} ,~  \ottmv{x_{{\mathrm{1}}}} :\!_{\! \ottsmode{m} } \ottstype{T_{{\mathrm{1}}}}    \,\pmb{\vdash}\,  \ottnt{u_{{\mathrm{1}}}}  \pmb{:}  \ottstype{U}%
\qquad
\Gamma_{{\mathrm{2}}} ,~  \ottmv{x_{{\mathrm{2}}}} :\!_{\! \ottsmode{m} } \ottstype{T_{{\mathrm{2}}}}    \,\pmb{\vdash}\,  \ottnt{u_{{\mathrm{2}}}}  \pmb{:}  \ottstype{U}}
}{}{}

\patchcmd{\ottdruleTyXXectxsXXPatS}{%
\ottpremise{\Delta_{{\mathrm{2}}}  +   \ottmv{x_{{\mathrm{1}}}} :\!_{\! \ottsmode{m} } \ottstype{T_{{\mathrm{1}}}}   \,\pmb{\vdash}\,  \ottnt{u_{{\mathrm{1}}}}  \pmb{:}  \ottstype{U}}%
\ottpremise{\Delta_{{\mathrm{2}}}  +   \ottmv{x_{{\mathrm{2}}}} :\!_{\! \ottsmode{m} } \ottstype{T_{{\mathrm{2}}}}   \,\pmb{\vdash}\,  \ottnt{u_{{\mathrm{2}}}}  \pmb{:}  \ottstype{U}}%
}{
  \ottpremise{\Delta_{{\mathrm{2}}}  +   \ottmv{x_{{\mathrm{1}}}} :\!_{\! \ottsmode{m} } \ottstype{T_{{\mathrm{1}}}}   \,\pmb{\vdash}\,  \ottnt{u_{{\mathrm{1}}}}  \pmb{:}  \ottstype{U}
\qquad
  \Delta_{{\mathrm{2}}}  +   \ottmv{x_{{\mathrm{2}}}} :\!_{\! \ottsmode{m} } \ottstype{T_{{\mathrm{2}}}}   \,\pmb{\vdash}\,  \ottnt{u_{{\mathrm{2}}}}  \pmb{:}  \ottstype{U}}
}{}{}

\usepackage{tikzit}
\definecolor{sczcolor}{RGB}{0,0,0}
\definecolor{sczfcolor}{RGB}{128, 128, 128}
\definecolor{scicolor}{RGB}{199, 22, 6}
\definecolor{scifcolor}{RGB}{251, 164, 157}
\definecolor{sciicolor}{RGB}{60, 103, 163}
\definecolor{sciiicolor}{RGB}{97, 5, 94}
\newcommand{\scz}[1]{\textcolor{sczcolor}{#1}}
\newcommand{\sczf}[1]{\textcolor{sczfcolor}{#1}}
\newcommand{\sci}[1]{\textcolor{scicolor}{#1}}
\newcommand{\scif}[1]{\textcolor{scifcolor}{#1}}
\newcommand{\scii}[1]{\textcolor{sciicolor}{#1}}

\tikzstyle{header}=[anchor=mid west, scale=1.0]
\tikzstyle{code}=[anchor=mid west, scale=1.0]
\tikzstyle{scopename}=[anchor=mid west, scale=0.7]
\tikzstyle{smallannot}=[anchor=mid west, scale=0.55]

\tikzstyle{s0s}=[-, draw=sczcolor]
\tikzstyle{s0d}=[dotted, dash pattern=on 2pt off 1pt, draw=sczcolor]
\tikzstyle{s1s}=[scicolor]
\tikzstyle{s1d}=[dotted, dash pattern=on 2pt off 1pt, draw=scicolor]
\tikzstyle{s2s}=[sciicolor]
\tikzstyle{s2d}=[dotted, dash pattern=on 2pt off 1pt, draw=sciicolor]
\tikzstyle{s3s}=[sciiicolor]
\tikzstyle{s3d}=[dotted, dash pattern=on 2pt off 1pt, draw=sciiicolor]
\tikzstyle{a0}=[draw=sczcolor, ->]
\tikzstyle{a1}=[draw=scicolor, ->]
\tikzstyle{a2}=[draw=sciicolor, ->]
\tikzstyle{a3}=[draw=sciiicolor, ->]
\tikzstyle{red}=[densely dotted, draw={rgb,255: red,0; green,0; blue,0}, ->]

\newcommand{\longshort}[2]{#1}
\InputIfFileExists{short-version}{
  \renewcommand{\longshort}[2]{##2}
}

  \InputIfFileExists{no-editing-marks}{
    \def\noeditingmarks{}
  }

  \usepackage{xargs}
  \usepackage[colorinlistoftodos,prependcaption,textsize=tiny]{todonotes}
  \ifx\noeditingmarks\undefined
      \newcommand{\note}[1]{{\color{blue}{\begin{itemize} \item {#1} \end{itemize}}}}
      \newenvironment{alt}{\color{red}}{}

      \newcommand{\TODO}[1]{\textnormal{\textcolor{red}{TODO: #1} } }
      \newcommandx{\unsure}[2][1=]{\todo[linecolor=orange,backgroundcolor=orange!25,bordercolor=orange,#1]{#2}}
      \newcommandx{\info}[2][1=]{\todo[linecolor=green,backgroundcolor=green!25,bordercolor=green,#1]{#2}}
      \newcommandx{\change}[2][1=]{\todo[linecolor=blue,backgroundcolor=blue!25,bordercolor=blue,#1]{#2}}
      \newcommandx{\inconsistent}[2][1=]{\todo[linecolor=red,backgroundcolor=red!25,bordercolor=red,#1]{#2}}
      \newcommandx{\critical}[2][1=]{\todo[linecolor=purple,backgroundcolor=purple!25,bordercolor=purple,#1]{#2}}
      \newcommand{\improvement}[1]{\todo[linecolor=pink,backgroundcolor=pink!25,bordercolor=pink]{#1}}
      \newcommandx{\resolved}[2][1=]{\todo[linecolor=OliveGreen,backgroundcolor=OliveGreen!25,bordercolor=OliveGreen,#1]{#2}} 
  \else
      \newcommand{\TODO}[1]{}

      \newcommand{\note}[1]{}

      \newcommand{\unsure}[2][1=]{}
      \newcommand{\info}[2][1=]{}
      \newcommand{\change}[2]{}
      \newcommand{\inconsistent}[2]{}
      \newcommand{\critical}[2]{}
      \newcommand{\improvement}[1]{}
      \newcommand{\resolved}[2]{}

  \fi

\newcommand{\parr}{\rotatebox[origin=c]{180}{\&}}
\makeatletter
\newcommand{\smallbullet}{} 
\DeclareRobustCommand\smallbullet{%
\mathord{\mathpalette\smallbullet@{0.5}}%
}
\newcommand{\smallbullet@}[2]{%
\vcenter{\hbox{\scalebox{#2}{$\m@th#1\bullet$}}}%
}
\makeatother

\makeatletter
\newcommand{\oset}[3][0ex]{%
\mathrel{\mathop{#3}\limits^{
  \vbox to#1{\kern-2\ex@
  \hbox{$\scriptstyle#2$}\vss}}}}
\makeatother

\def\mycasem#1{\ifthenelse{\equal{#1}{  \ottsmode{1}  \hspace{-0.15ex}  \ottsmode{\nu}  }}{}{#1}}
\def\myfunvm#1{\ifthenelse{\equal{#1}{  \ottsmode{1}  \hspace{-0.15ex}  \ottsmode{\nu}  }}{\,\,}{#1}}
\def\myfuntm#1{\ifthenelse{\equal{#1}{  \ottsmode{1}  \hspace{-0.15ex}  \ottsmode{\nu}  }}{\,}{#1}}
\def\mydestm#1{\ifthenelse{\equal{#1}{  \ottsmode{1}  \hspace{-0.15ex}  \ottsmode{\nu}  }}{}{#1}}
\def\mymul#1{\ifthenelse{\equal{#1}{ \ottsmode{1} }}{}{#1}}

\newcommand{\destcalculus}{\ensuremath{\lambda_d}}

\newcommand\btriangleq{\pmb{\triangleq}}
\newcommand\btriangleqrec{\oset{\mathsf{rec}}{\pmb{\triangleq}}}
\newlength{\interdefskip}
\newcommand{\newtype}[3][]{#2~\ifthenelse{\equal{#1}{}}{\btriangleq}{\btriangleqrec}~#3\\[\interdefskip]}
\newcommand{\newoperator}[5][]{\phantom{a}\!\!\!\!\!\!\begin{array}[t]{l}%
#2 ~\pmb{:}~ #3 \\
#4 ~\ifthenelse{\equal{#1}{}}{\btriangleq}{\btriangleqrec}~ #5
\end{array}\\[\interdefskip]}

\newcommand{\newoperatorb}[5][]{\phantom{a}\!\!\!\!\!\!\begin{array}[t]{l}%
#2 ~\pmb{:}~ #3 \\
#4 ~\ifthenelse{\equal{#1}{}}{\btriangleq}{\btriangleqrec}~\\\myspace{1}#5
\end{array}\\[\interdefskip]}
\newcommand{\figureratio}{0.9}
\newcommand{\codehere}[2][t]{\vspace{-0.05cm}\begin{center}\begin{minipage}[#1]{\figureratio\linewidth}{\small\ensuremath{#2}}\end{minipage}\end{center}\vspace{-0.05cm}}
\NewEnviron{codefig}[2][t]{\begin{figure}[#1]
\codehere{\BODY}#2
\end{figure}}
\NewEnviron{ottfig}[2][t]{\begin{figure}[#1]\visiblespaces
\small\BODY\activespaces #2
\end{figure}}
\newcommand{\sidebysidecodehere}[4]{\begin{center}\begin{minipage}[#1]{\figureratio\linewidth}
\noindent\begin{minipage}[#1]{#2\linewidth-0.02\linewidth}{\small\ensuremath{#3}}\end{minipage}
\hfill
\vrule width 0.5pt 
\hfill
\begin{minipage}[#1]{\linewidth-#2\linewidth-0.02\linewidth}{\small\ensuremath{#4}}\end{minipage}
\end{minipage}\end{center}
}

\newcommand{\sidebysidecodefig}[6][t]{
\begin{figure}[#1]
\sidebysidecodehere{#3}{#4}{#5}{#6}
#2
\end{figure}
}

\newlength{\widthaugment}
\NewEnviron{augmentwidth}[1]
  {\setlength{\widthaugment}{#1}
   \pgfmathsetmacro{\myratio}{\linewidth / (\widthaugment + \linewidth)}
   \scalebox{\myratio}{\begin{minipage}{\linewidth+\widthaugment}\BODY
   \end{minipage}}}

\newcommand{\IfFancyRuleNames}[2]{#2}

\newcommand{\SetPrefix}[1]{\IfFancyRuleNames{\renewcommand{\ottdrulename}[1]{#1}}{}}

\makeatletter

\IfFancyRuleNames{
  \def\CSep{/}
  \def\CHole{id\textsubscript{h}}
  \def\CDest{id\textsubscript{d}}
  \def\CUnit{$\ottstype{1}$I}
  \def\CFun{$\ottstype{\multimap}$I}
  \def\CLeft{$\ottstype{\oplus}$I\textsubscript{1}}
  \def\CRight{$\ottstype{\oplus}$I\textsubscript{2}}
  \def\CProd{$\ottstype{\otimes}$I}
  \def\CExp{$\ottstype{!}$I}
  \def\CAmpar{$\ottstype{\ltimes}$I}
  \def\CVal{fromVal}
  \def\CVar{id\textsubscript{v}}
  \def\CApp{$\ottstype{\multimap}$E}
  \def\CPatU{$\ottstype{1}$E}
  \def\CPatS{$\ottstype{\oplus}$E}
  \def\CPatP{$\ottstype{\otimes}$E}
  \def\CPatE{$\ottstype{!}$E}
  \def\CUpdA{$\ottstype{\ltimes}$upd}
  \def\CToA{$\ottstype{\ltimes}$I\textsubscript{t}}
  \def\CFromA{$\ottstype{\ltimes}$E}
  \def\CNewA{$\ottstype{\ltimes}$I\textsubscript{n}}
  \def\CFillU{$\ottstype{\lfloor 1\rfloor}$E}
  \def\CFillL{$\ottstype{\lfloor \oplus\rfloor}$E\textsubscript{1}}
  \def\CFillR{$\ottstype{\lfloor \oplus\rfloor}$E\textsubscript{2}}
  \def\CFillP{$\ottstype{\lfloor \otimes\rfloor}$E}
  \def\CFillE{$\ottstype{\lfloor !\rfloor}$E}
  \def\CFillF{$\ottstype{\lfloor \multimap\rfloor}$E}
  \def\CFillComp{$\ottstype{\lfloor\smallbullet\rfloor}$E\textsubscript{c}}
  \def\CFillLeaf{$\ottstype{\lfloor\smallbullet\rfloor}$E\textsubscript{l}}
  \def\CId{id}
  \def\COpenAmpar{$\ottstype{\ltimes}$op}
  \def\COpen{op}
  \def\CClose{cl}

  \def\CRed{C}
  \def\CTyTerm{ty-$\ottnt{t}$}
  \def\CTySTerm{ty-$\ottnt{t}{\scriptstyle^{s}}$}
  \def\CTyVal{ty-$\ottnt{v}$}
  \def\CTyEctxs{ty-$\ottnt{E}$}
  \def\CTyCmd{ty-cmd}
  \def\CRed{red}
  \def\rref*#1{\textsc{#1}}
}{
  \renewcommand{\ottdrulename}[1]{\ottalt@replace@cs\ranchor\_-{}#1\\}\renewcommand{\maybecomm}[1]{\bgroup\def\text##1{}#1\egroup}
  \def\CSep{-}
  \def\CHole{Hole}
  \def\CDest{Dest}
  \def\CUnit{Unit}
  \def\CFun{Fun}
  \def\CLeft{Left}
  \def\CRight{Right}
  \def\CProd{Prod}
  \def\CExp{Exp}
  \def\CAmpar{Ampar}
  \def\CVal{Val}
  \def\CVar{Var}
  \def\CApp{App}
  \def\CPatU{PatU}
  \def\CPatS{PatS}
  \def\CPatP{PatP}
  \def\CPatE{PatE}
  \def\CUpdA{UpdA}
  \def\CToA{ToA}
  \def\CFromA{FromA}
  \def\CNewA{NewA}
  \def\CFillU{FillU}
  \def\CFillL{FillL}
  \def\CFillR{FillR}
  \def\CFillP{FillP}
  \def\CFillE{FillE}
  \def\CFillF{FillF}
  \def\CFillComp{FillComp}
  \def\CFillLeaf{FillLeaf}
  \def\CId{Id}
  \def\COpenAmpar{OpenAmpar}
  \def\COpen{Open}
  \def\CClose{Close}

  \def\CRed{Red}
  \def\CTyTerm{Ty-term}
  \def\CTySTerm{Ty-sterm}
  \def\CTyVal{Ty-val}
  \def\CTyEctxs{Ty-ectxs}
  \def\CTyCmd{Ty-cmd}
}
\makeatother

\setcopyright{cc}
\setcctype{by}

\longshort{\settopmatter{printacmref=false}}{

\acmDOI{10.1145/3720423}
\acmYear{2025}
\acmJournal{PACMPL}
\acmVolume{9}
\acmNumber{OOPSLA1}
\acmArticle{89}
\acmMonth{4}
\received{2024-10-15}
\received[accepted]{2025-02-18}

}

\begin{document}


\title[Destination Calculus]{Destination Calculus:\\ A Linear $\lambda-$Calculus for Purely Functional Memory Writes}

\author{Thomas Bagrel}
\orcid{0009-0008-8700-2741}
\affiliation{
\institution{LORIA/Inria}
\department{MOSEL/VERIDIS}
\city{Villers-lès-Nancy}
\country{France}
}
\affiliation{
\institution{Tweag}
\department{OSPO}
\city{Paris}
\country{France}
}
\email{thomas.bagrel@loria.fr}
\email{thomas.bagrel@tweag.io}

\author{Arnaud Spiwack}
\orcid{0000-0002-5985-2086}
\affiliation{
\institution{Tweag}
\department{OSPO}
\position{Director, Research}
\city{Paris}
\country{France}
}
\email{arnaud.spiwack@tweag.io}



\begin{abstract}
  Destination passing ---aka. out parameters--- is taking a parameter to fill rather than returning a result from a function. Due to its apparently imperative nature, destination passing has struggled to find its way to pure functional programming. In this paper, we present a pure functional calculus with destinations at its core. Our calculus subsumes all the similar systems, and can be used to reason about their correctness or extension. In addition, our calculus can express programs that were previously not known to be expressible in a pure language. This is guaranteed by a modal type system where modes are used to manage both linearity and scopes. Type safety of our core calculus was proved formally with the Coq proof assistant.
\end{abstract}

\maketitle



\section{Introduction}\label{sec:intro}

In destination-passing style, a function doesn't return a value: it takes as an argument a location where the output of the function ought to be written. A function of type $ \ottstype{T} \,\ottstype{\to}\, \ottstype{U} $ would, in destination-passing style, have type $  \ottstype{T} \,\ottstype{\to}\,  \ottstype{\lfloor}\,\!_{\mydestm{   \ottsmode{1}  \hspace{-0.15ex}  \ottsmode{\nu}   } } \ottstype{U} \ottstype{\rfloor}   \,\ottstype{\to}\,  \ottstype{1}  $ instead, where $ \ottstype{\lfloor}\,\!_{\mydestm{   \ottsmode{1}  \hspace{-0.15ex}  \ottsmode{\nu}   } } \ottstype{U} \ottstype{\rfloor} $ denotes a destination for value of type $\ottstype{U}$. This style is common in system programming, where destinations $ \ottstype{\lfloor}\,\!_{\mydestm{   \ottsmode{1}  \hspace{-0.15ex}  \ottsmode{\nu}   } } \ottstype{U} \ottstype{\rfloor} $ are more commonly known as “out parameters” (in C, $ \ottstype{\lfloor}\,\!_{\mydestm{   \ottsmode{1}  \hspace{-0.15ex}  \ottsmode{\nu}   } } \ottstype{U} \ottstype{\rfloor} $ would typically be a pointer of type $\ottstype{U*}$).

The reason why system programs rely on destinations so much is that using destinations can save calls to the memory allocator. If a function returns a $\ottstype{U}$, it has to allocate the space for a $\ottstype{U}$. But with destinations, the caller is responsible for finding space for a $\ottstype{U}$. The caller may simply ask the memory allocator for the space, in which case we've saved nothing; but it can also reuse the space of an existing $\ottstype{U}$ that it doesn't need anymore, or space in an array, or even space in a region of memory that the allocator doesn't have access to, like a memory-mapped file.

This does all sound quite imperative, but we argue that the same considerations are relevant for functional programming, albeit to a lesser extent. In fact~\citet{shaikhha_destination-passing_2017} has demonstrated that using destination passing in the intermediate language of a functional array-programming language allowed for significant optimizations. Where destinations truly shine in functional programming, however, is that they increase the expressiveness of the language; destinations as first-class values allow for meaningfully new programs to be written, as first explored in~\cite{bagrel_destination-passing_2024}.

The trouble, of course, is that destinations are imperative; we wouldn't want to sacrifice the immutability of our linked data structures (later on abbreviated \emph{structures}) for the sake of the more situational destinations. The goal here is to extend functional programming just enough to be able to build immutable structures by destination passing without endangering purity and memory safety. This is already what~\cite{bagrel_destination-passing_2024} introduces, using a linear type system to restrict mutation. Destinations become write-once-only references into a structure with holes. Here we follow these leads, but we refine the type system further to allow for even more programs (see \cref{sec:scope-escape-dests}).

There are two key elements to the expressiveness of destination passing:
\begin{itemize}
\item structures can be built in any order. Not only from the leaves to the root, like in ordinary functional programming, but also from the root to the leaves, or any combination thereof. This can be done in ordinary functional programming using function composition in a form of continuation-passing; and destinations act as an optimization. This line of work was pioneered by~\citet{minamide_functional_1998}. While this only increases expressiveness when combined with the next point, the optimization is significant enough that destination passing has been implemented in the Ocaml optimizer to support tail modulo constructor~\cite{bour_tmc_2021};
\item when destinations are first-class values, they can be passed and stored like ordinary values. This is the innovation of~\cite{bagrel_destination-passing_2024} upon which we build. The consequence is that not only the order in which a structure is built is arbitrary, this order can be determined dynamically during the runtime of the program.
\end{itemize}

To support this programming style, we introduce \destcalculus{}. We intend \destcalculus{} to serve as a foundational, theoretical calculus to reason about safe destinations in a functional setting. Indeed \destcalculus{} subsumes all the systems that we've discussed in this section. As such we expect that these systems or their extensions can be justified simply by giving them a translation into \destcalculus{}, in order to get all the safety results and metatheory of \destcalculus{} for free. Even though \destcalculus{} is not really meant to be implemented as a real programming language, we still draft an implementation strategy for it based on efficient mutations, in the line of \cite{bour_tmc_2021,bagrel_destination-passing_2024}.

Our contributions are as follows:
\begin{itemize}
\item \destcalculus{}, a modal, linear, simply typed $\lambda$-calculus with destinations (\cref{sec:syntax-type-system,sec:ectxs-sem}). \destcalculus{} is expressive enough to serve as an encoding for previous calculi with destinations (see~\cref{sec:related-work});
\item a demonstration that \destcalculus{} is more expressive than previous calculi with destinations (\cref{sec:scope-escape-dests,sec:bft}), namely that destinations can be stored in structures with holes. We show how we can improve, in particular, on the breadth-first traversal example of~\cite{bagrel_destination-passing_2024};
\item an implementation strategy for \destcalculus{} which uses mutation without compromising the purity of \destcalculus{} (\cref{sec:implementation});
\item formally-verified proofs, with the Coq proof assistant, of type safety (\cref{sec:formal-proof}).
\end{itemize}

\section{Working with Destinations}\label{sec:working-with-dests}

Let's introduce and get familiar with \destcalculus{}, our simply typed $\lambda$-calculus with destination. The syntax is standard, except that we use linear logic's $ \ottstype{T} \ottstype{\oplus} \ottstype{U} $ and $ \ottstype{T} \ottstype{\otimes} \ottstype{U} $ for sums and products, and linear function arrow $\ottstype{\multimap}$, since \destcalculus{} is linearly typed, even though it isn't a focus in this section.

\subsection{Building up a Vocabulary}\label{ssec:build-up-vocab}

\activespaces

In its simplest form, destination passing, much like continuation passing, is using a location, received as an argument, to return a value. Instead of a linear function with signature $ \ottstype{T} \,_{\myfuntm{   \ottsmode{1}  \hspace{-0.15ex}  \ottsmode{\nu}   } }\!\ottstype{\multimap}\, \ottstype{U} $, in \destcalculus{} you would have $  \ottstype{T} \,_{\myfuntm{   \ottsmode{1}  \hspace{-0.15ex}  \ottsmode{\nu}   } }\!\ottstype{\multimap}\,  \ottstype{\lfloor}\,\!_{\mydestm{   \ottsmode{1}  \hspace{-0.15ex}  \ottsmode{\nu}   } } \ottstype{U} \ottstype{\rfloor}   \,_{\myfuntm{   \ottsmode{1}  \hspace{-0.15ex}  \ottsmode{\nu}   } }\!\ottstype{\multimap}\,  \ottstype{1}  $, where $ \ottstype{\lfloor}\,\!_{\mydestm{   \ottsmode{1}  \hspace{-0.15ex}  \ottsmode{\nu}   } } \ottstype{U} \ottstype{\rfloor} $ is read “destination for type $\ottstype{U}$”.

\longshort{}{
\bigskip 
\bigskip
\bigskip
}

For instance, here is a destination-passing version of the identity function:

\codehere{\newoperator
{\ottkw{dId}}{  \ottstype{T} \,_{\myfuntm{   \ottsmode{1}  \hspace{-0.15ex}  \ottsmode{\nu}   } }\!\ottstype{\multimap}\,  \ottstype{\lfloor}\,\!_{\mydestm{   \ottsmode{1}  \hspace{-0.15ex}  \ottsmode{\nu}   } } \ottstype{T} \ottstype{\rfloor}   \,_{\myfuntm{   \ottsmode{1}  \hspace{-0.15ex}  \ottsmode{\nu}   } }\!\ottstype{\multimap}\,  \ottstype{1}  }
{\ottkw{dId}~\ottmv{x}~\ottmv{d}}{ \ottmv{d} \blacktriangleleft \ottmv{x} }}

We think of a destination as a reference to an uninitialized memory location, and $ \ottmv{d} \blacktriangleleft \ottmv{x} $ (read “fill $\ottmv{d}$ with $\ottmv{x}$”) as writing $\ottmv{x}$ to the memory location.

The form $ \ottmv{d} \blacktriangleleft \ottmv{x} $ is the simplest way to use a destination. But we don't have to fill a destination with a complete value in a single step. Destinations can be filled piecemeal.

\codehere{\newoperator
{\ottkw{fillWithInl}}{  \ottstype{\lfloor}\,\!_{\mydestm{   \ottsmode{1}  \hspace{-0.15ex}  \ottsmode{\nu}   } }  \ottstype{T} \ottstype{\oplus} \ottstype{U}  \ottstype{\rfloor}  \,_{\myfuntm{   \ottsmode{1}  \hspace{-0.15ex}  \ottsmode{\nu}   } }\!\ottstype{\multimap}\,  \ottstype{\lfloor}\,\!_{\mydestm{   \ottsmode{1}  \hspace{-0.15ex}  \ottsmode{\nu}   } } \ottstype{T} \ottstype{\rfloor}  }
{\ottkw{fillWithInl}~\ottmv{d}}{\ottmv{d}  \triangleleft \, \ottsctor{Inl}}}

In this example, we're filling a destination for type $ \ottstype{T} \ottstype{\oplus} \ottstype{U} $ by setting the outermost constructor to left variant $ \ottsctor{Inl} $. We think of $\ottmv{d}  \triangleleft \, \ottsctor{Inl}$ (read “fill $\ottmv{d}$ with $ \ottsctor{Inl} $”) as allocating memory to store a block of the form $ \ottsctor{Inl} ~\holesq$, write the address of that block to the location that $\ottmv{d}$ points to, and return a new destination of type $ \ottstype{\lfloor}\,\!_{\mydestm{   \ottsmode{1}  \hspace{-0.15ex}  \ottsmode{\nu}   } } \ottstype{T} \ottstype{\rfloor} $ pointing to the uninitialized argument of $ \ottsctor{Inl} $. Uninitialized memory, when part of a structure or value, like $\holesq$ in $ \ottsctor{Inl} ~\holesq$, is called a \emph{hole}.

Notice that with $\ottkw{fillWithInl}$ we are constructing the structure from the outermost constructor inward: we've written a value of the form $ \ottsctor{Inl} ~\holesq$ into a hole, but we have yet to describe what goes in the new hole $\holesq$. Such data constructors with uninitialized arguments are called \emph{hollow constructors}\footnote{The full triangle $ \blacktriangleleft $ is used to fill a destination with a fully-formed value, while the \emph{hollow} triangle $ \triangleleft $ is used to fill a destination with a \emph{hollow constructor}.}. This is opposite to how functional programming usually works, where values are built from the innermost constructors outward: first we make a value $\ottnt{v}$ and only then can we use $ \ottsctor{Inl} $ to make $\ottsctor{Inl} \, \ottnt{v}$. This will turn out to be a key ingredient in the expressiveness of destination passing.

Yet, everything we've shown so far could have been done with continuations. So it's worth asking: how are destinations different from continuations? Part of the answer lies in our intention to effectively implement destinations as pointers to uninitialized memory (see~\cref{sec:implementation}). But where destinations really differ from continuations is when one has several destinations at hand. Then they have to fill \emph{all} the destinations; whereas when one has multiple continuations, they can only return to one of them. Multiple destination arises when a destination for a pair gets filled with a hollow pair constructor:

\codehere{\newoperator
{\ottkw{fillWithPair}}{   \ottstype{\lfloor}\,\!_{\mydestm{   \ottsmode{1}  \hspace{-0.15ex}  \ottsmode{\nu}   } }  \ottstype{T} \ottstype{\otimes} \ottstype{U}  \ottstype{\rfloor}  \,_{\myfuntm{   \ottsmode{1}  \hspace{-0.15ex}  \ottsmode{\nu}   } }\!\ottstype{\multimap}\,  \ottstype{\lfloor}\,\!_{\mydestm{   \ottsmode{1}  \hspace{-0.15ex}  \ottsmode{\nu}   } } \ottstype{T} \ottstype{\rfloor}   \ottstype{\otimes}  \ottstype{\lfloor}\,\!_{\mydestm{   \ottsmode{1}  \hspace{-0.15ex}  \ottsmode{\nu}   } } \ottstype{U} \ottstype{\rfloor}  }
{\ottkw{fillWithPair}~\ottmv{d}}{\ottmv{d}  \triangleleft  \ottsctor{({,})}}}

After using $\ottkw{fillWithPair}$, both the first field \emph{and} the second field must be filled, using the destinations of type $ \ottstype{\lfloor}\,\!_{\mydestm{   \ottsmode{1}  \hspace{-0.15ex}  \ottsmode{\nu}   } } \ottstype{T} \ottstype{\rfloor} $ and $ \ottstype{\lfloor}\,\!_{\mydestm{   \ottsmode{1}  \hspace{-0.15ex}  \ottsmode{\nu}   } } \ottstype{U} \ottstype{\rfloor} $ respectively. The key remark here is that $\ottkw{fillWithPair}$ couldn't exist if we replaced destinations by continuations, as we couldn't use both returned continuations easily.

\paragraph{Structures with Holes}
It is crucial to note that while a destination is used to build a structure, the type of the structure being built might be different from the type of the destination that is being filled. A destination of type $ \ottstype{\lfloor}\,\!_{\mydestm{   \ottsmode{1}  \hspace{-0.15ex}  \ottsmode{\nu}   } } \ottstype{T} \ottstype{\rfloor} $ is a pointer to a yet-undefined part of a bigger structure. We say that such a structure has a hole of type $\ottstype{T}$; but the type of the structure itself isn't specified (and never appears in the signature of destination-filling functions). For instance, using $\ottkw{fillWithPair}$ only indicates that the structure being operated on has a hole of type $ \ottstype{T} \ottstype{\otimes} \ottstype{U} $ that is being written to.

Thus, we still need a type to tie the structure under construction --- left implicit by destination-filling primitives --- with the destinations representing its holes. To represent this, \destcalculus{} introduces a type $ \ottstype{S} \,\ottstype{\ltimes}\,  \ottstype{\lfloor}\,\!_{\mydestm{   \ottsmode{1}  \hspace{-0.15ex}  \ottsmode{\nu}   } } \ottstype{T} \ottstype{\rfloor}  $ for a structure of type $\ottstype{S}$ missing a value of type $\ottstype{T}$ to be complete. There can be several holes in $\ottstype{S}$ --- resulting in several destinations on the right hand side --- and as long as there remains holes in $\ottstype{S}$, it cannot be read. For instance, $ \ottstype{S} \,\ottstype{\ltimes}\,  \ottstype{(}   \ottstype{\lfloor}\,\!_{\mydestm{   \ottsmode{1}  \hspace{-0.15ex}  \ottsmode{\nu}   } } \ottstype{T} \ottstype{\rfloor}  \ottstype{\otimes}  \ottstype{\lfloor}\,\!_{\mydestm{   \ottsmode{1}  \hspace{-0.15ex}  \ottsmode{\nu}   } } \ottstype{U} \ottstype{\rfloor}   \ottstype{)}  $ represents a $\ottstype{S}$ that misses both a $\ottstype{T}$ and a $\ottstype{U}$ to be complete (thus to be readable). 

The general form $ \ottstype{S} \,\ottstype{\ltimes}\, \ottstype{T} $ is read “$\ottstype{S}$ ampar $\ottstype{T}$”. The name “ampar” stands for “asymmetric memory par”; we will explain how we came up with this type and name in \cref{ssec:ampar-motivation}. A similar connective is called $\ottstype{Incomplete}$ in~\cite{bagrel_destination-passing_2024}. For now, it's sufficient to observe that $ \ottstype{S} \,\ottstype{\ltimes}\,  \ottstype{\lfloor}\,\!_{\mydestm{   \ottsmode{1}  \hspace{-0.15ex}  \ottsmode{\nu}   } } \ottstype{T} \ottstype{\rfloor}  $ is akin to a “par” type $\ottstype{S \parr T^\perp}$ in linear logic; you can think of $ \ottstype{S} \,\ottstype{\ltimes}\,  \ottstype{\lfloor}\,\!_{\mydestm{   \ottsmode{1}  \hspace{-0.15ex}  \ottsmode{\nu}   } } \ottstype{T} \ottstype{\rfloor}  $ as a (linear) function from $\ottstype{T}$ to $\ottstype{S}$. That structures with holes could be seen as linear functions was first observed in~\cite{minamide_functional_1998}; we elaborate on the value of having a “par” type with access to first-class destinations, rather than just linear functions to represent structures with holes, in~\cref{sec:bft}.

Destinations always exist within the context of a structure with holes. A destination is both a witness of a hole present in the structure, and a handle to write to it. Crucially, destinations are otherwise ordinary values. To access the destinations of an ampar, \destcalculus{} provides a $\ottkw{upd}_{\ottkw{\ltimes} }$ construction, which lets us apply a function to the right-hand side of the ampar. It is in the body of $\ottkw{upd}_{\ottkw{\ltimes} }$ that functions operating on destinations can be called to update the structure:

\codehere{
  \newoperator
  {\ottkw{fillWithInl'}}{   \ottstype{S} \,\ottstype{\ltimes}\,  \ottstype{\lfloor}\,\!_{\mydestm{   \ottsmode{1}  \hspace{-0.15ex}  \ottsmode{\nu}   } }  \ottstype{T} \ottstype{\oplus} \ottstype{U}  \ottstype{\rfloor}   \,_{\myfuntm{   \ottsmode{1}  \hspace{-0.15ex}  \ottsmode{\nu}   } }\!\ottstype{\multimap}\, \ottstype{S}  \,\ottstype{\ltimes}\,  \ottstype{\lfloor}\,\!_{\mydestm{   \ottsmode{1}  \hspace{-0.15ex}  \ottsmode{\nu}   } } \ottstype{T} \ottstype{\rfloor}  }
  {\ottkw{fillWithInl'}~\ottmv{x}}{ \ottkw{upd}_{\ottkw{\ltimes} }\, \ottmv{x} ~\ottkw{with}~ \ottmv{d} \, \pmb{\mapsto}  \ottkw{fillWithInl}~ \ottmv{d}  }
  \newoperator
  {\ottkw{fillWithPair'}}{   \ottstype{S} \,\ottstype{\ltimes}\,  \ottstype{\lfloor}\,\!_{\mydestm{   \ottsmode{1}  \hspace{-0.15ex}  \ottsmode{\nu}   } }  \ottstype{T} \ottstype{\otimes} \ottstype{U}  \ottstype{\rfloor}   \,_{\myfuntm{   \ottsmode{1}  \hspace{-0.15ex}  \ottsmode{\nu}   } }\!\ottstype{\multimap}\, \ottstype{S}  \,\ottstype{\ltimes}\,  \ottstype{(}   \ottstype{\lfloor}\,\!_{\mydestm{   \ottsmode{1}  \hspace{-0.15ex}  \ottsmode{\nu}   } } \ottstype{T} \ottstype{\rfloor}  \ottstype{\otimes}  \ottstype{\lfloor}\,\!_{\mydestm{   \ottsmode{1}  \hspace{-0.15ex}  \ottsmode{\nu}   } } \ottstype{U} \ottstype{\rfloor}   \ottstype{)}  }
  {\ottkw{fillWithPair'}~\ottmv{x}}{ \ottkw{upd}_{\ottkw{\ltimes} }\, \ottmv{x} ~\ottkw{with}~ \ottmv{d} \, \pmb{\mapsto}  \ottkw{fillWithPair}~ \ottmv{d}  }
}

To tie this up, we need a way to introduce and to eliminate structures with holes. Structures with holes are introduced with $ \ottkw{new}_{\ottkw{\ltimes} } $ which creates a value of type $ \ottstype{T} \,\ottstype{\ltimes}\,  \ottstype{\lfloor}\,\!_{\mydestm{   \ottsmode{1}  \hspace{-0.15ex}  \ottsmode{\nu}   } } \ottstype{T} \ottstype{\rfloor}  $. $ \ottkw{new}_{\ottkw{\ltimes} } $ is a bit like the identity function: it is a hole (of type $\ottstype{T}$) that needs a value of type $\ottstype{T}$ to be a complete value of type $\ottstype{T}$. Memory-wise, it is an uninitialized block large enough to host a value of type $\ottstype{T}$, and a destination pointing to it. Conversely, structures with holes are eliminated with\footnote{As the name suggest, there is a more general elimination $\ottkw{from}_{\ottkw{\ltimes} }$. It will be discussed in~\cref{sec:syntax-type-system}.} $\ottkw{from}_{\ottkw{\ltimes} }' :   \ottstype{S} \,\ottstype{\ltimes}\,  \ottstype{1}   \,_{\myfuntm{   \ottsmode{1}  \hspace{-0.15ex}  \ottsmode{\nu}   } }\!\ottstype{\multimap}\, \ottstype{S} $: if all the destinations have been consumed and only unit remains on the right side, then $\ottstype{S}$ no longer has holes and thus is just a normal, complete structure.

Equipped with these, we can, for instance, derive traditional constructors from piecemeal filling. In fact, \destcalculus{} doesn't have primitive constructor forms, constructors in \destcalculus{} are syntactic sugar. We show here the definition of $ \ottsctor{Inl} $ and $ \ottsctor{({,})} $, but the other constructors are derived similarly. Operator $\patu$, present in second example, is used to chain operations returning unit type $ \ottstype{1} $.

\codehere{\newoperator
{ \ottsctor{Inl} }{  \ottstype{T} \,_{\myfuntm{   \ottsmode{1}  \hspace{-0.15ex}  \ottsmode{\nu}   } }\!\ottstype{\multimap}\, \ottstype{T}  \ottstype{\oplus} \ottstype{U} }
{ \ottsctor{Inl}\, \ottmv{x} }{ \ottkw{from}_{\ottkw{\ltimes} }'   (  \ottkw{upd}_{\ottkw{\ltimes} }\,  (   \ottkw{new}_{\ottkw{\ltimes} }  \pmb{:}{\scriptstyle    \ottstype{(}  \ottstype{T} \ottstype{\oplus} \ottstype{U}  \ottstype{)}  \,\ottstype{\ltimes}\,  \ottstype{\lfloor}\,\!_{\mydestm{   \ottsmode{1}  \hspace{-0.15ex}  \ottsmode{\nu}   } }  \ottstype{T} \ottstype{\oplus} \ottstype{U}  \ottstype{\rfloor}   }  )  ~\ottkw{with}~ \ottmv{d} \, \pmb{\mapsto}  \ottmv{d}  \triangleleft \, \ottsctor{Inl} \blacktriangleleft \ottmv{x}   )  }
\newoperator
{ \ottsctor{({,})} }{   \ottstype{T} \,_{\myfuntm{   \ottsmode{1}  \hspace{-0.15ex}  \ottsmode{\nu}   } }\!\ottstype{\multimap}\, \ottstype{U}  \,_{\myfuntm{   \ottsmode{1}  \hspace{-0.15ex}  \ottsmode{\nu}   } }\!\ottstype{\multimap}\, \ottstype{T}  \ottstype{\otimes} \ottstype{U} }
{ \ottsctor{(} \ottmv{x} \,\ottsctor{,}~ \ottmv{y} \ottsctor{)} }{\!\!\!\begin{array}[t]{l} \ottkw{from}_{\ottkw{\ltimes} }'   (  \ottkw{upd}_{\ottkw{\ltimes} }\,  (   \ottkw{new}_{\ottkw{\ltimes} }  \pmb{:}{\scriptstyle    \ottstype{(}  \ottstype{T} \ottstype{\otimes} \ottstype{U}  \ottstype{)}  \,\ottstype{\ltimes}\,  \ottstype{\lfloor}\,\!_{\mydestm{   \ottsmode{1}  \hspace{-0.15ex}  \ottsmode{\nu}   } }  \ottstype{T} \ottstype{\otimes} \ottstype{U}  \ottstype{\rfloor}   }  )  ~\ottkw{with}~ \ottmv{d} \, \pmb{\mapsto}   \mynewline   \myspace{3}       \ottkw{case}_{\mycasem{   \ottsmode{1}  \hspace{-0.15ex}  \ottsmode{\nu}   } }\,  ( \ottmv{d}  \triangleleft  \ottsctor{({,})} )  ~\ottkw{of}~\ottsctor{(} \ottmv{d_{{\mathrm{1}}}} \,\ottsctor{,}~ \ottmv{d_{{\mathrm{2}}}} \ottsctor{)} \pmb{\mapsto} \ottmv{d_{{\mathrm{1}}}}  \blacktriangleleft \ottmv{x}  \patu \ottmv{d_{{\mathrm{2}}}}  \blacktriangleleft \ottmv{y}     )  \end{array}}}

\paragraph{Memory Safety and Purity}
At this point, the reader may be forgiven for feeling distressed at all the talk of mutations and uninitialized memory. How is it consistent with our claim to be building a pure and memory-safe language? The answer is that it wouldn't be if we'd allow unrestricted use of destinations. Instead \destcalculus{} uses a linear type system to ensure that:

\begin{itemize}
\item destinations are written at least once, preventing examples like:

\codehere{\hspace*{-0.057\linewidth}\newoperator
  {\ottkw{forget}}{\ottstype{T}}
  {\ottkw{forget}}{ \ottkw{from}_{\ottkw{\ltimes} }'   (  \ottkw{upd}_{\ottkw{\ltimes} }\,  (   \ottkw{new}_{\ottkw{\ltimes} }  \pmb{:}{\scriptstyle   \ottstype{T} \,\ottstype{\ltimes}\,  \ottstype{\lfloor}\,\!_{\mydestm{   \ottsmode{1}  \hspace{-0.15ex}  \ottsmode{\nu}   } } \ottstype{T} \ottstype{\rfloor}   }  )  ~\ottkw{with}~ \ottmv{d} \, \pmb{\mapsto} \ottsctor{()}  )  }}

  where reading the result of $\ottkw{forget}$ would result in reading the location pointed to by a destination that we never used, in other words, reading uninitialized memory;\longshort{}{\\[1cm]} 

\item destinations are written at most once, preventing examples like:

\codehere{\hspace*{-0.057\linewidth}\newoperator
  {\ottkw{ambiguous1}}{ \ottstype{Bool} }
  {\ottkw{ambiguous1}}{ \ottkw{from}_{\ottkw{\ltimes} }'   (  \ottkw{upd}_{\ottkw{\ltimes} }\,  (   \ottkw{new}_{\ottkw{\ltimes} }  \pmb{:}{\scriptstyle    \ottstype{Bool}  \,\ottstype{\ltimes}\,  \ottstype{\lfloor}\,\!_{\mydestm{   \ottsmode{1}  \hspace{-0.15ex}  \ottsmode{\nu}   } }  \ottstype{Bool}  \ottstype{\rfloor}   }  )  ~\ottkw{with}~ \ottmv{d} \, \pmb{\mapsto}    \ottmv{d} \blacktriangleleft  \ottsctor{true}   \patu \ottmv{d}  \blacktriangleleft  \ottsctor{false}    )  }
  \hspace*{-0.057\linewidth}\newoperator
  {\ottkw{ambiguous2}}{ \ottstype{Bool} }
  {\ottkw{ambiguous2}}{ \ottkw{from}_{\ottkw{\ltimes} }'   (  \ottkw{upd}_{\ottkw{\ltimes} }\,  (   \ottkw{new}_{\ottkw{\ltimes} }  \pmb{:}{\scriptstyle    \ottstype{Bool}  \,\ottstype{\ltimes}\,  \ottstype{\lfloor}\,\!_{\mydestm{   \ottsmode{1}  \hspace{-0.15ex}  \ottsmode{\nu}   } }  \ottstype{Bool}  \ottstype{\rfloor}   }  )  ~\ottkw{with}~ \ottmv{d} \, \pmb{\mapsto}    \ottkw{let}~ \ottmv{x} \assigneq  (  \ottmv{d} \blacktriangleleft  \ottsctor{false}   )  ~\ottkw{in}~ \ottmv{d}  \blacktriangleleft  \ottsctor{true}   \patu \ottmv{x}   )  }}

  where $\ottkw{ambiguous1}$ would return $ \ottsctor{false} $ and $\ottkw{ambiguous2}$ would return $ \ottsctor{true} $ due to evaluation order, even though let-expansion should be valid in a pure language.
\end{itemize}

\subsection{Tail-Recursive Map}\label{ssec:map-tr}

Now that we have an intuition of how destinations work, let's see how they can be used to build usual data structures. For this section, we suppose that \destcalculus{} has equirecursive types and a fixed-point operator. These aren't part of the formal system of \cref{sec:syntax-type-system} but don't add any complication.

\paragraph{Linked Lists}

We define lists as a fixpoint, as usual: $ \ottstype{List}~ \ottstype{T}  \btriangleq   \ottstype{1}  \ottstype{\oplus}  \ottstype{(}  \ottstype{T} \ottstype{\otimes}  \ottstype{(}  \ottstype{List}~ \ottstype{T}  \ottstype{)}   \ottstype{)}  $. For convenience, we also define filling operators $\triangleleft\ottsctor{[]}$ and $\triangleleft\ottsctor{(::)}$:

\sidebysidecodehere{b}{0.50}{
\newoperator
  {\triangleleft\ottsctor{[]}}{  \ottstype{\lfloor}\,\!_{\mydestm{   \ottsmode{1}  \hspace{-0.15ex}  \ottsmode{\nu}   } }  \ottstype{List}~ \ottstype{T}  \ottstype{\rfloor}  \,_{\myfuntm{   \ottsmode{1}  \hspace{-0.15ex}  \ottsmode{\nu}   } }\!\ottstype{\multimap}\,  \ottstype{1}  }
  { \ottmv{d} \triangleleft\ottsctor{[]} }{\ottmv{d}  \triangleleft \, \ottsctor{Inl}  \triangleleft  \ottsctor{()}}
}{
\newoperator
  {\triangleleft\ottsctor{(::)}}{   \ottstype{\lfloor}\,\!_{\mydestm{   \ottsmode{1}  \hspace{-0.15ex}  \ottsmode{\nu}   } }  \ottstype{List}~ \ottstype{T}  \ottstype{\rfloor}  \,_{\myfuntm{   \ottsmode{1}  \hspace{-0.15ex}  \ottsmode{\nu}   } }\!\ottstype{\multimap}\,  \ottstype{\lfloor}\,\!_{\mydestm{   \ottsmode{1}  \hspace{-0.15ex}  \ottsmode{\nu}   } } \ottstype{T} \ottstype{\rfloor}   \ottstype{\otimes}  \ottstype{\lfloor}\,\!_{\mydestm{   \ottsmode{1}  \hspace{-0.15ex}  \ottsmode{\nu}   } }  \ottstype{List}~ \ottstype{T}  \ottstype{\rfloor}  }
  { \ottmv{d} \triangleleft\ottsctor{(::)} }{\ottmv{d}  \triangleleft \, \ottsctor{Inr}  \triangleleft  \ottsctor{({,})}}
}

Just like we did in~\cref{ssec:build-up-vocab} we can recover traditional constructors from filling operators, e.g.:

\codehere{\newoperator
{\ottsctor{(::)}}{  \ottstype{T} \ottstype{\otimes}  \ottstype{(}  \ottstype{List}~ \ottstype{T}  \ottstype{)}   \,_{\myfuntm{   \ottsmode{1}  \hspace{-0.15ex}  \ottsmode{\nu}   } }\!\ottstype{\multimap}\,  \ottstype{List}~ \ottstype{T}  }
{ \ottmv{x} \,\ottsctor{::}\, \ottmv{xs} }{\!\!\!\begin{array}[t]{l} \ottkw{from}_{\ottkw{\ltimes} }'   (  \ottkw{upd}_{\ottkw{\ltimes} }\,  (   \ottkw{new}_{\ottkw{\ltimes} }  \pmb{:}{\scriptstyle    \ottstype{(}  \ottstype{List}~ \ottstype{T}  \ottstype{)}  \,\ottstype{\ltimes}\,  \ottstype{\lfloor}\,\!_{\mydestm{   \ottsmode{1}  \hspace{-0.15ex}  \ottsmode{\nu}   } }  \ottstype{List}~ \ottstype{T}  \ottstype{\rfloor}   }  )  ~\ottkw{with}~ \ottmv{d} \, \pmb{\mapsto}   \mynewline   \myspace{3}       \ottkw{case}_{\mycasem{   \ottsmode{1}  \hspace{-0.15ex}  \ottsmode{\nu}   } }\,  (  \ottmv{d} \triangleleft\ottsctor{(::)}  )  ~\ottkw{of}~\ottsctor{(} \ottmv{dx} \,\ottsctor{,}~ \ottmv{dxs} \ottsctor{)} \pmb{\mapsto} \ottmv{dx}  \blacktriangleleft \ottmv{x}  \patu \ottmv{dxs}  \blacktriangleleft \ottmv{xs}     )  \end{array}}}

\paragraph{A Tail-Recursive Map Function}

List being ubiquitous in functional programming, the fact that the most natural way to write a map function on lists isn't tail recursive (hence consumes unbounded stack space), is unpleasant. Map can be made tail-recursive in two passes: first build the result list in reverse, then reverse it. But destinations let us avoid this two-pass process altogether, as they let us extend the tail of the result list directly.
We give a complete implementation in \cref{fig:impl-map-tr}.

The main function is $\ottkw{map'}$, it has type $
    \ottstype{(}  \ottstype{T} \,_{\myfuntm{   \ottsmode{1}  \hspace{-0.15ex}  \ottsmode{\nu}   } }\!\ottstype{\multimap}\, \ottstype{U}  \ottstype{)}  \,_{\myfuntm{   \ottsmode{\omega}  \hspace{-0.15ex}  \ottsmode{\infty}   } }\!\ottstype{\multimap}\,  \ottstype{List}~ \ottstype{T}   \,_{\myfuntm{   \ottsmode{1}  \hspace{-0.15ex}  \ottsmode{\nu}   } }\!\ottstype{\multimap}\,  \ottstype{\lfloor}\,\!_{\mydestm{   \ottsmode{1}  \hspace{-0.15ex}  \ottsmode{\nu}   } }  \ottstype{List}~ \ottstype{U}  \ottstype{\rfloor}   \,_{\myfuntm{   \ottsmode{1}  \hspace{-0.15ex}  \ottsmode{\nu}   } }\!\ottstype{\multimap}\,  \ottstype{1}  
$. That is, instead of returning a resulting list, it takes a destination as an input and fills it with the result. At each recursive call, $\ottkw{map'}$ creates a new hollow cons cell to fill the destination. A destination pointing to the tail of the new cons cell is also created, on which $\ottkw{map'}$ is called (tail) recursively. This is really the same algorithm that you could write to implement map on a mutable list in an imperative language. Nevertheless \destcalculus{} is a pure language with only immutable types.

To obtain the regular $\ottkw{map}$ function, all is left to do is to call $ \ottkw{new}_{\ottkw{\ltimes} } $ to create an initial destination, and $\ottkw{from}_{\ottkw{\ltimes} }'$, much like when we make constructors out of filling operators, like $\ottsctor{(::)}$ above.

\begin{codefig}{\caption{Tail-recursive map function on lists}\label{fig:impl-map-tr}}
\newtype{ \ottstype{List}~ \ottstype{T} }{  \ottstype{1}  \ottstype{\oplus}  \ottstype{(}  \ottstype{T} \ottstype{\otimes}  \ottstype{(}  \ottstype{List}~ \ottstype{T}  \ottstype{)}   \ottstype{)}  }
\newoperatorb
  {\ottkw{map'}}{    \ottstype{(}  \ottstype{T} \,_{\myfuntm{   \ottsmode{1}  \hspace{-0.15ex}  \ottsmode{\nu}   } }\!\ottstype{\multimap}\, \ottstype{U}  \ottstype{)}  \,_{\myfuntm{   \ottsmode{\omega}  \hspace{-0.15ex}  \ottsmode{\infty}   } }\!\ottstype{\multimap}\,  \ottstype{List}~ \ottstype{T}   \,_{\myfuntm{   \ottsmode{1}  \hspace{-0.15ex}  \ottsmode{\nu}   } }\!\ottstype{\multimap}\,  \ottstype{\lfloor}\,\!_{\mydestm{   \ottsmode{1}  \hspace{-0.15ex}  \ottsmode{\nu}   } }  \ottstype{List}~ \ottstype{U}  \ottstype{\rfloor}   \,_{\myfuntm{   \ottsmode{1}  \hspace{-0.15ex}  \ottsmode{\nu}   } }\!\ottstype{\multimap}\,  \ottstype{1}  }
  { \ottkw{map'}~ \ottmv{f} ~ \ottmv{l} ~ \ottmv{dl} }{\!\!\!\begin{array}[t]{l} \ottkw{case}_{\mycasem{   \ottsmode{1}  \hspace{-0.15ex}  \ottsmode{\nu}   } }\, \ottmv{l} ~\ottkw{of}~\{   \mynewline   \myspace{1}   \ottsctor{[]}   \pmb{\mapsto}  \ottmv{dl} \triangleleft\ottsctor{[]}  \,,~    \mynewline   \myspace{1}  \ottmv{x}  \,\ottsctor{::}\, \ottmv{xs}  \pmb{\mapsto}     \ottkw{case}_{\mycasem{   \ottsmode{1}  \hspace{-0.15ex}  \ottsmode{\nu}   } }\,  (  \ottmv{dl} \triangleleft\ottsctor{(::)}  )  ~\ottkw{of}~  \mynewline   \myspace{2}  \ottsctor{(} \ottmv{dx} \,\ottsctor{,}~ \ottmv{dxs} \ottsctor{)}\! \pmb{\mapsto} \ottmv{dx}  \blacktriangleleft \ottmv{f}  ~ \ottmv{x}  \patu  \ottkw{map'}~ \ottmv{f} ~ \ottmv{xs} ~ \ottmv{dxs}   \} \end{array}}
\newoperator
  {\ottkw{map}}{   \ottstype{(}  \ottstype{T} \,_{\myfuntm{   \ottsmode{1}  \hspace{-0.15ex}  \ottsmode{\nu}   } }\!\ottstype{\multimap}\, \ottstype{U}  \ottstype{)}  \,_{\myfuntm{   \ottsmode{\omega}  \hspace{-0.15ex}  \ottsmode{\infty}   } }\!\ottstype{\multimap}\,  \ottstype{List}~ \ottstype{T}   \,_{\myfuntm{   \ottsmode{1}  \hspace{-0.15ex}  \ottsmode{\nu}   } }\!\ottstype{\multimap}\,  \ottstype{List}~ \ottstype{U}  }
  { \ottkw{map}~ \ottmv{f} ~ \ottmv{l} }{ \ottkw{from}_{\ottkw{\ltimes} }'   (  \ottkw{upd}_{\ottkw{\ltimes} }\,  (   \ottkw{new}_{\ottkw{\ltimes} }  \pmb{:}{\scriptstyle    \ottstype{(}  \ottstype{List}~ \ottstype{U}  \ottstype{)}  \,\ottstype{\ltimes}\,  \ottstype{\lfloor}\,\!_{\mydestm{   \ottsmode{1}  \hspace{-0.15ex}  \ottsmode{\nu}   } }  \ottstype{List}~ \ottstype{U}  \ottstype{\rfloor}   }  )  ~\ottkw{with}~ \ottmv{dl} \, \pmb{\mapsto}  \ottkw{map'}~ \ottmv{f} ~ \ottmv{l} ~ \ottmv{dl}   )  }
\end{codefig}

\subsection{Functional Queues, with Destinations}\label{ssec:efficient-queue}

Implementations for a tail-recursive map are present in most previous work, from~\cite{minamide_functional_1998}, to recent work~\cite{bour_tmc_2021,leijen_trmc_2023,bagrel_destination-passing_2024}. Tail-recursive map doesn't need the full power of \destcalculus{}'s first-class destinations: it just needs a notion of structures with a (single) hole. In \cref{sec:bft}, we will build an example which fully uses first-class destinations, but first, we will need some more material.

\paragraph{Difference Lists}
\newcommand{\lstcat}{\mathop{+\!+}}
Just like in any language, iterated concatenation of lists
$((\ottmv{xs_{{\mathrm{1}}}} \lstcat \ottmv{xs_{{\mathrm{2}}}})\lstcat \ldots)\lstcat \ottmv{xs}_n$
is quadratic in \destcalculus{}. The usual solution to this is difference lists. The name difference lists covers many related implementations, but in pure functional languages, a difference list is usually represented as a function~\cite{hughes_dlist_1986}. A singleton difference list is $\lamnt{\ottmv{ys}}{  \ottsmode{1}  \hspace{-0.15ex}  \ottsmode{\nu}  }{ \ottmv{x} \,\ottsctor{::}\, \ottmv{ys} }$, and concatenation of difference lists is function composition. A difference list is turned into a list by applying it to the empty list. The consequence is that no matter how many compositions we have, each cons cell will be allocated a single time, making the iterated concatenation linear indeed.

However, each concatenation allocates a closure. If we're building a difference list from singletons and composition, there's roughly one composition per cons cell, so iterated composition effectively performs two traversals of the list. In \destcalculus{}, we can do better by representing a difference list as a list with a hole. A singleton difference list is $\ottmv{x} \ottsctor{::} \holesq$. Concatenation is filling the hole with another difference list, using operator $\mathop{\triangleleft\mycirc}$. The details are on the left of~\cref{fig:impl-dlist-queue}. The \destcalculus{} encoding for difference lists makes no superfluous traversal: concatenation is just an $O(1)$ in-place update.

\sidebysidecodefig{\caption{Difference list and queue implementation in equirecursive \destcalculus{}}\label{fig:impl-dlist-queue}}{t}{0.49}{
\newtype{ \ottstype{DList}~ \ottstype{T} }{  \ottstype{(}  \ottstype{List}~ \ottstype{T}  \ottstype{)}  \,\ottstype{\ltimes}\,  \ottstype{\lfloor}\,\!_{\mydestm{   \ottsmode{1}  \hspace{-0.15ex}  \ottsmode{\nu}   } }  \ottstype{List}~ \ottstype{T}  \ottstype{\rfloor}  }
\newoperatorb
  {\ottkw{append}}{   \ottstype{DList}~ \ottstype{T}  \,_{\myfuntm{   \ottsmode{1}  \hspace{-0.15ex}  \ottsmode{\nu}   } }\!\ottstype{\multimap}\, \ottstype{T}  \,_{\myfuntm{   \ottsmode{1}  \hspace{-0.15ex}  \ottsmode{\nu}   } }\!\ottstype{\multimap}\,  \ottstype{DList}~ \ottstype{T}  }
  { \ottmv{ys} ~\ottkw{append}~ \ottmv{y} }{\!\!\!\begin{array}[t]{l}   \ottkw{upd}_{\ottkw{\ltimes} }\, \ottmv{ys} ~\ottkw{with}~ \ottmv{dys} \, \pmb{\mapsto}  \ottkw{case}_{\mycasem{   \ottsmode{1}  \hspace{-0.15ex}  \ottsmode{\nu}   } }\,  (  \ottmv{dys} \triangleleft\ottsctor{(::)}  )  ~\ottkw{of}~  \mynewline   \myspace{1}  \ottsctor{(} \ottmv{dy} \,\ottsctor{,}~ \ottmv{dys'} \ottsctor{)}\! \pmb{\mapsto} \ottmv{dy}   \blacktriangleleft \ottmv{y}  \patu \ottmv{dys'} \end{array}}
\newoperator
  {\ottkw{concat}}{   \ottstype{DList}~ \ottstype{T}  \,_{\myfuntm{   \ottsmode{1}  \hspace{-0.15ex}  \ottsmode{\nu}   } }\!\ottstype{\multimap}\,  \ottstype{DList}~ \ottstype{T}   \,_{\myfuntm{   \ottsmode{1}  \hspace{-0.15ex}  \ottsmode{\nu}   } }\!\ottstype{\multimap}\,  \ottstype{DList}~ \ottstype{T}  }
  { \ottmv{ys} ~\ottkw{concat}~ \ottmv{ys'} }{  \ottkw{upd}_{\ottkw{\ltimes} }\, \ottmv{ys} ~\ottkw{with}~ \ottmv{d} \, \pmb{\mapsto} \ottmv{d}  \mathop{\triangleleft\mycirc} \ottmv{ys'} }
\newoperator
  {\ottkw{toList}}{  \ottstype{DList}~ \ottstype{T}  \,_{\myfuntm{   \ottsmode{1}  \hspace{-0.15ex}  \ottsmode{\nu}   } }\!\ottstype{\multimap}\,  \ottstype{List}~ \ottstype{T}  }
  { \ottkw{toList}~ \ottmv{ys} }{ \ottkw{from}_{\ottkw{\ltimes} }'   (  \ottkw{upd}_{\ottkw{\ltimes} }\, \ottmv{ys} ~\ottkw{with}~ \ottmv{d} \, \pmb{\mapsto}  \ottmv{d} \triangleleft\ottsctor{[]}   )  }
}{
\newtype{ \ottstype{Queue}~ \ottstype{T} }{  \ottstype{(}  \ottstype{List}~ \ottstype{T}  \ottstype{)}  \ottstype{\otimes}  \ottstype{(}  \ottstype{DList}~ \ottstype{T}  \ottstype{)}  }
\newoperator
  {\ottkw{singleton}}{ \ottstype{T} \,_{\myfuntm{   \ottsmode{1}  \hspace{-0.15ex}  \ottsmode{\nu}   } }\!\ottstype{\multimap}\,  \ottstype{Queue}~ \ottstype{T}  }
  { \ottkw{singleton}~ \ottmv{x} }{ \ottsctor{(}  \ottsctor{Inr}\,  (  \ottmv{x} \,\ottsctor{::}\,  \ottsctor{[]}   )   \,\ottsctor{,}~  (   \ottkw{new}_{\ottkw{\ltimes} }  \pmb{:}{\scriptstyle   \ottstype{DList}~ \ottstype{T}  }  )  \ottsctor{)} }
\newoperatorb
  {\ottkw{enqueue}}{   \ottstype{Queue}~ \ottstype{T}  \,_{\myfuntm{   \ottsmode{1}  \hspace{-0.15ex}  \ottsmode{\nu}   } }\!\ottstype{\multimap}\, \ottstype{T}  \,_{\myfuntm{   \ottsmode{1}  \hspace{-0.15ex}  \ottsmode{\nu}   } }\!\ottstype{\multimap}\,  \ottstype{Queue}~ \ottstype{T}  }
  { \ottmv{q} ~\ottkw{enqueue}~ \ottmv{y} }{ \ottkw{case}_{\mycasem{   \ottsmode{1}  \hspace{-0.15ex}  \ottsmode{\nu}   } }\, \ottmv{q} ~\ottkw{of}~\ottsctor{(} \ottmv{xs} \,\ottsctor{,}~ \ottmv{ys} \ottsctor{)} \pmb{\mapsto}  \ottsctor{(} \ottmv{xs} \,\ottsctor{,}~  \ottmv{ys} ~\ottkw{append}~ \ottmv{y}  \ottsctor{)}  }
\newoperatorb
  {\ottkw{dequeue}}{   \ottstype{Queue}~ \ottstype{T}  \,_{\myfuntm{   \ottsmode{1}  \hspace{-0.15ex}  \ottsmode{\nu}   } }\!\ottstype{\multimap}\,  \ottstype{1}   \ottstype{\oplus}  \ottstype{(}  \ottstype{T} \ottstype{\otimes}  \ottstype{(}  \ottstype{Queue}~ \ottstype{T}  \ottstype{)}   \ottstype{)}  }
  { \ottkw{dequeue}~ \ottmv{q} }{\!\!\!\begin{array}[t]{l} \ottkw{case}_{\mycasem{   \ottsmode{1}  \hspace{-0.15ex}  \ottsmode{\nu}   } }\, \ottmv{q} ~\ottkw{of}~\{   \mynewline   \myspace{1}   \ottsctor{(}  (  \ottmv{x} \,\ottsctor{::}\, \ottmv{xs}  )  \,\ottsctor{,}~ \ottmv{ys} \ottsctor{)}   \pmb{\mapsto}  \ottsctor{Inr}\,  \ottsctor{(} \ottmv{x} \,\ottsctor{,}~  \ottsctor{(} \ottmv{xs} \,\ottsctor{,}~ \ottmv{ys} \ottsctor{)}  \ottsctor{)}   \,,~   \mynewline   \myspace{1}   \ottsctor{(}  \ottsctor{[]}  \,\ottsctor{,}~ \ottmv{ys} \ottsctor{)}   \pmb{\mapsto}  \ottkw{case}_{\mycasem{   \ottsmode{1}  \hspace{-0.15ex}  \ottsmode{\nu}   } }\,  (  \ottkw{toList}~ \ottmv{ys}  )  ~\ottkw{of}~\{   \mynewline   \myspace{2}   \ottsctor{[]}   \pmb{\mapsto} \ottsctor{Inl} \, \ottsctor{()} \,,~    \mynewline   \myspace{2}  \ottmv{x}  \,\ottsctor{::}\, \ottmv{xs}  \pmb{\mapsto}  \ottsctor{Inr}\,  \ottsctor{(} \ottmv{x} \,\ottsctor{,}~  \ottsctor{(} \ottmv{xs} \,\ottsctor{,}~  (   \ottkw{new}_{\ottkw{\ltimes} }  \pmb{:}{\scriptstyle   \ottstype{DList}~ \ottstype{T}  }  )  \ottsctor{)}  \ottsctor{)}   \}  \} \end{array}}
}

\paragraph{Efficient Queue Using Difference Lists}
In an immutable functional language, a queue can be implemented as a pair of lists $ \ottsctor{(} \ottmv{front} \,\ottsctor{,}~ \ottmv{back} \ottsctor{)} $~\cite{hood_queue_1981}. $\ottmv{back}$ stores new elements in reverse order ($O(1)$ prepend). We pop elements from $\ottmv{front}$, except when it is empty, in which case we set the queue to $ \ottsctor{(}  \ottkw{reverse}~ \ottmv{back}  \,\ottsctor{,}~  \ottsctor{[]}  \ottsctor{)} $, and pop from the new front.

For their simple implementation, Hood-Melville queues are surprisingly efficient: the cost of the reverse operation is $O(1)$ amortized for a single-threaded use of the queue. Still, it would be better to get rid of this full traversal of the back list. Taking a step back, this $\ottmv{back}$ list that has to be reversed before it is accessed is really merely a representation of a list that can be extended from the back. And we already know an efficient implementation for this: difference lists.

So we can give an improved version of the simple functional queue using destinations. This implementation is presented on the right-hand side of~\cref{fig:impl-dlist-queue}. Note that contrary to an imperative programming language, we can't implement the queue as a single difference list: as mentioned earlier, our type system prevents us from reading the front elements of difference lists. Just like for the simple functional queue, we need a pair of one list that we can read from, and one that we can extend. Nevertheless this implementation of queues is both pure, as guaranteed by the \destcalculus{} type system, and nearly as efficient as what an imperative programming language would afford.

\section{Scope Escape of Destinations}\label{sec:scope-escape-dests}

In \cref{sec:working-with-dests}, we've been making an implicit assumption: establishing a linear discipline on destinations ensures that all destinations will eventually find their way to the left of a fill operator $\blacktriangleleft$ or $\triangleleft$, so that the associated holes get written to. This turns out to be slightly incomplete.

To see why, let's consider the type $ \ottstype{\lfloor}\,\!_{\mydestm{   \ottsmode{1}  \hspace{-0.15ex}  \ottsmode{\nu}   } }  \ottstype{\lfloor}\,\!_{\mydestm{   \ottsmode{1}  \hspace{-0.15ex}  \ottsmode{\nu}   } } \ottstype{T} \ottstype{\rfloor}  \ottstype{\rfloor} $: the type of a destination pointing to a hole where a destination is expected. Think of it as an equivalent of the pointer type $\ottstype{T*\!*}$ in the C language. Destinations are indeed ordinary values, so they can be stored in data structures, and before they get stored, holes stand in their place in the structure. For instance, if we have $\ottmv{d}\pmb{:} \ottstype{\lfloor}\,\!_{\mydestm{   \ottsmode{1}  \hspace{-0.15ex}  \ottsmode{\nu}   } } \ottstype{T} \ottstype{\rfloor} $ and $\ottmv{dd}\pmb{:} \ottstype{\lfloor}\,\!_{\mydestm{   \ottsmode{1}  \hspace{-0.15ex}  \ottsmode{\nu}   } }  \ottstype{\lfloor}\,\!_{\mydestm{   \ottsmode{1}  \hspace{-0.15ex}  \ottsmode{\nu}   } } \ottstype{T} \ottstype{\rfloor}  \ottstype{\rfloor} $, we can form $ \ottmv{dd} \blacktriangleleft \ottmv{d} $: $\ottmv{d}$ will be stored in the structure pointed to by $\ottmv{dd}$.

Should we count $\ottmv{d}$ as linearly used here? The alternatives don't seem promising:
\vspace{-0.04cm}\begin{itemize}
\item If we count this as a non-linear use of $\ottmv{d}$, then $ \ottmv{dd} \blacktriangleleft \ottmv{d} $ is rejected since destinations (represented here by $\ottmv{d}$) can only be used linearly. This choice is fairly limiting, as it would prevent us from storing destinations in structures with holes, as we do, crucially, in \cref{sec:bft}. Nonetheless, that's the option chosen in \cite{bagrel_destination-passing_2024}.
\item If we do not count this use of $\ottmv{d}$ at all, we can write $   \ottmv{dd} \blacktriangleleft \ottmv{d}  \patu \ottmv{d}  \blacktriangleleft \ottnt{v} $ so that $\ottmv{d}$ is both stored for later use \emph{and} filled immediately (resulting in the corresponding hole being potentially written to twice), which is unsound, as discussed in \cref{ssec:build-up-vocab}.
\end{itemize}\vspace{-0.04cm}
So linear use it is. But it creates a problem: there's no way, within our linear type system, to distinguish between ``a destination has been used on the left of a triangle so its corresponding hole has been filled'' and ``a destination has been stored and its hole still exists at the moment''. This oversight may allow us to read uninitialized memory!

Let's compare two examples. We assume a simple store semantics for now where structures with holes stay in the store until they are completed. We'll need the $\ottkw{alloc} \pmb{:}   \ottstype{(}   \ottstype{\lfloor}\,\!_{\mydestm{   \ottsmode{1}  \hspace{-0.15ex}  \ottsmode{\nu}   } } \ottstype{T} \ottstype{\rfloor}  \,_{\myfuntm{   \ottsmode{1}  \hspace{-0.15ex}  \ottsmode{\nu}   } }\!\ottstype{\multimap}\,  \ottstype{1}   \ottstype{)}  \,_{\myfuntm{   \ottsmode{1}  \hspace{-0.15ex}  \ottsmode{\nu}   } }\!\ottstype{\multimap}\, \ottstype{T} $ operator. The semantics of $\ottkw{alloc}$ is: allocate a structure with a single root hole in the store, call the supplied function with the destination to the root hole as an argument; when the function has consumed all destinations (so only unit remains), pop the structure from the store to obtain a complete $\ottstype{T}$.

In this snippet, structures with holes are given names $\ottnt{v}$ and $\ottnt{vd}$ in the store; holes are given names too and denoted by $ \ottshname{\hboxed{ \ottshname{h} } } $ and $ \ottshname{\hboxed{ \ottshname{hd} } } $, and concrete destinations are denoted by $ \ottshname{\destminus} \ottshname{h} $ and $ \ottshname{\destminus} \ottshname{hd} $.

When the building scope of $\ottnt{v} \pmb{:}  \ottstype{Bool} $ is parent to the one of $\ottnt{vd} \pmb{:}  \ottstype{\lfloor}\,\!_{\mydestm{   \ottsmode{1}  \hspace{-0.15ex}  \ottsmode{\nu}   } }  \ottstype{Bool}  \ottstype{\rfloor} $, everything works well because $\ottnt{vd}$, that contains destination pointing to $ \ottshname{\hboxed{ \ottshname{h} } } $, has to be consumed before $\ottnt{v}$ can be read:
\bgroup\setlength{\arraycolsep}{0.5ex}
\codehere{\!\!\!\begin{array}{crcl}
       & $\{ \}$ &|&  \ottkw{alloc}~  (  \lamnt{ \ottmv{d} }{   \ottsmode{1}  \hspace{-0.15ex}  \ottsmode{\nu}   }{   (   \ottkw{alloc}~  (  \lamnt{ \ottmv{dd} }{   \ottsmode{1}  \hspace{-0.15ex}  \ottsmode{\nu}   }{  \ottmv{dd} \blacktriangleleft \ottmv{d}  }  )   \pmb{:}{\scriptstyle   \ottstype{\lfloor}\,\!_{\mydestm{   \ottsmode{1}  \hspace{-0.15ex}  \ottsmode{\nu}   } }  \ottstype{Bool}  \ottstype{\rfloor}  }  )  \blacktriangleleft  \ottsctor{true}   }  )   \\
\to & \ottsym{\{}   \ottnt{v} \assigneq  \ottshname{\hboxed{ \ottshname{h} } }    \ottsym{\}} &|&    (   \ottkw{alloc}~  (  \lamnt{ \ottmv{dd} }{   \ottsmode{1}  \hspace{-0.15ex}  \ottsmode{\nu}   }{  \ottmv{dd} \blacktriangleleft  \ottshname{\destminus} \ottshname{h}   }  )   \pmb{:}{\scriptstyle   \ottstype{\lfloor}\,\!_{\mydestm{   \ottsmode{1}  \hspace{-0.15ex}  \ottsmode{\nu}   } }  \ottstype{Bool}  \ottstype{\rfloor}  }  )  \blacktriangleleft  \ottsctor{true}   \patu  \ottkw{deref}~ \ottnt{v}   \\
\to & \ottsym{\{}   \ottnt{v} \assigneq  \ottshname{\hboxed{ \ottshname{h} } }    \ottsym{,}   \ottnt{vd} \assigneq  \ottshname{\hboxed{ \ottshname{hd} } }    \ottsym{\}} &|&    (    \ottshname{\destminus} \ottshname{hd}  \blacktriangleleft  \ottshname{\destminus} \ottshname{h}   \patu  \ottkw{deref}~ \ottnt{vd}   )  \blacktriangleleft  \ottsctor{true}   \patu  \ottkw{deref}~ \ottnt{v}   \\
\to & \ottsym{\{}   \ottnt{v} \assigneq  \ottshname{\hboxed{ \ottshname{h} } }    \ottsym{,}   \ottnt{vd} \assigneq  \ottshname{\destminus} \ottshname{h}    \ottsym{\}} &|&    \ottkw{deref}~ \ottnt{vd}  \blacktriangleleft  \ottsctor{true}   \patu  \ottkw{deref}~ \ottnt{v}   \\
\to & \ottsym{\{}   \ottnt{v} \assigneq  \ottshname{\hboxed{ \ottshname{h} } }    \ottsym{\}} &|&    \ottshname{\destminus} \ottshname{h}  \blacktriangleleft  \ottsctor{true}   \patu  \ottkw{deref}~ \ottnt{v}   \\
\to & \ottsym{\{}   \ottnt{v} \assigneq  \ottsctor{true}    \ottsym{\}} &|&  \ottkw{deref}~ \ottnt{v}  \\
\to & $\{ \}$ &|&  \ottsctor{true} 
\end{array}}

However, when $\ottnt{vd}$'s scope is parent to $\ottnt{v}$'s, we can write a linearly typed yet unsound program, as we demonstrate in the following example.
\codehere{\!\!\!\begin{array}{crcl}
       & $\{ \}$ &|&  \ottkw{alloc}~  (  \lamnt{ \ottmv{dd} }{   \ottsmode{1}  \hspace{-0.15ex}  \ottsmode{\nu}   }{  \ottkw{case}_{\mycasem{   \ottsmode{1}  \hspace{-0.15ex}  \ottsmode{\nu}   } }\,  (   \ottkw{alloc}~  (  \lamnt{ \ottmv{d} }{   \ottsmode{1}  \hspace{-0.15ex}  \ottsmode{\nu}   }{  \ottmv{dd} \blacktriangleleft \ottmv{d}  }  )   \pmb{:}{\scriptstyle   \ottstype{Bool}  }  )  ~\ottkw{of}~\{  \ottsctor{true}  \pmb{\mapsto} \ottsctor{()} \,,~  \ottsctor{false}  \pmb{\mapsto} \ottsctor{()} \}  }  )   \\
\to & \ottsym{\{}   \ottnt{vd} \assigneq  \ottshname{\hboxed{ \ottshname{hd} } }    \ottsym{\}} &|&   \ottkw{case}_{\mycasem{   \ottsmode{1}  \hspace{-0.15ex}  \ottsmode{\nu}   } }\,  (   \ottkw{alloc}~  (  \lamnt{ \ottmv{d} }{   \ottsmode{1}  \hspace{-0.15ex}  \ottsmode{\nu}   }{   \ottshname{\destminus} \ottshname{hd}  \blacktriangleleft \ottmv{d}  }  )   \pmb{:}{\scriptstyle   \ottstype{Bool}  }  )  ~\ottkw{of}~\{  \ottsctor{true}  \pmb{\mapsto} \ottsctor{()} \,,~  \ottsctor{false}  \pmb{\mapsto} \ottsctor{()} \}  \patu  \ottkw{deref}~ \ottnt{vd}   \\
\to & \ottsym{\{}   \ottnt{vd} \assigneq  \ottshname{\hboxed{ \ottshname{hd} } }    \ottsym{,}   \ottnt{v} \assigneq  \ottshname{\hboxed{ \ottshname{h} } }    \ottsym{\}} &|&   \ottkw{case}_{\mycasem{   \ottsmode{1}  \hspace{-0.15ex}  \ottsmode{\nu}   } }\,  (    \ottshname{\destminus} \ottshname{hd}  \blacktriangleleft  \ottshname{\destminus} \ottshname{h}   \patu  \ottkw{deref}~ \ottnt{v}   )  ~\ottkw{of}~\{  \ottsctor{true}  \pmb{\mapsto} \ottsctor{()} \,,~  \ottsctor{false}  \pmb{\mapsto} \ottsctor{()} \}  \patu  \ottkw{deref}~ \ottnt{vd}   \\
\to & \ottsym{\{}   \ottnt{vd} \assigneq  \ottshname{\destminus} \ottshname{h}    \ottsym{,}   \ottnt{v} \assigneq  \ottshname{\hboxed{ \ottshname{h} } }    \ottsym{\}} &|&   \ottkw{case}_{\mycasem{   \ottsmode{1}  \hspace{-0.15ex}  \ottsmode{\nu}   } }\,  (  \ottkw{deref}~ \ottnt{v}  )  ~\ottkw{of}~\{  \ottsctor{true}  \pmb{\mapsto} \ottsctor{()} \,,~  \ottsctor{false}  \pmb{\mapsto} \ottsctor{()} \}  \patu  \ottkw{deref}~ \ottnt{vd}   \\
\to & \ottsym{\{}   \ottnt{vd} \assigneq  \ottshname{\destminus} \ottshname{h}    \ottsym{\}} &|&   \ottkw{case}_{\mycasem{   \ottsmode{1}  \hspace{-0.15ex}  \ottsmode{\nu}   } }\,  \ottshname{\hboxed{ \ottshname{h} } }  ~\ottkw{of}~\{  \ottsctor{true}  \pmb{\mapsto} \ottsctor{()} \,,~  \ottsctor{false}  \pmb{\mapsto} \ottsctor{()} \}  \patu  \ottkw{deref}~ \ottnt{vd}   \qquad\qquad \raisebox{-0.8ex}{\scalebox{0.35}{\bcbombe\bcbombe\bcbombe}}
\end{array}}\egroup
\noindent{}Here the expression $ \ottmv{dd} \blacktriangleleft \ottmv{d} $ results in $\ottmv{d}$ escaping its scope for the parent one, so $\ottnt{v}$ is just uninitialized memory (the hole $ \ottshname{\hboxed{ \ottshname{h} } } $) when we dereference it. This example must be rejected by our type system.

Again, using purely a linear type system, we can only reject this example if we also reject the first, sound example, as in~\cite{bagrel_destination-passing_2024}. In this case, the type $ \ottstype{\lfloor}\,\!_{\mydestm{   \ottsmode{1}  \hspace{-0.15ex}  \ottsmode{\nu}   } }  \ottstype{\lfloor}\,\!_{\mydestm{   \ottsmode{1}  \hspace{-0.15ex}  \ottsmode{\nu}   } } \ottstype{T} \ottstype{\rfloor}  \ottstype{\rfloor} $ becomes practically useless: such destinations can never be filled.

This isn't the direction we want to take: we really want to be able to store destinations in data structures with holes. So we want $\ottnt{t}$ in $ \ottmv{d} \blacktriangleleft \ottnt{t} $ to be allowed to be linear. Without further restrictions, it wouldn't be sound, so to address this, \destcalculus{} uses a system of ages to represent scopes. Ages are described in \cref{sec:syntax-type-system}.

\section{Breadth-First Tree Traversal}\label{sec:bft}

As a full-fledged example, which uses the full expressive power of \destcalculus{}, we borrow and improve on an example from~\cite{bagrel_destination-passing_2024}, breadth-first tree relabeling:
\emph{``
Given a tree, create a new one of the same shape, but with the values at the nodes replaced by the numbers $1\ldots|T|$ in breadth-first order.
''}

This isn't a very natural problem in functional programming, as breadth-first traversal implies that the order in which the structure must be built (left-to-right, top-to-bottom) is not the same as the structural order of a functional tree --- building the leaves first and going up to the root. So it usually requires fancy functional workarounds~\cite{okasaki_bfs_2000,jones_gibbons_linearbfs_93,gibbons_phases_2023}.

It's very tempting to implement this example in an efficient imperative-like fashion, where a queue drives the processing order, thanks to the power of destinations. For that, \citet{minamide_functional_1998}'s system where structures with holes are represented as linear functions is not enough. Destinations as first-class values are very much required.

\Cref{fig:impl-bfs} presents the \destcalculus{} implementation of the breadth-first tree traversal. The core idea is that we hold a queue of pairs, storing each an input subtree with (a destination to) its corresponding output subtree. When the element $ \ottsctor{(} \ottmv{tree} \,\ottsctor{,}~ \ottmv{dtree} \ottsctor{)} $ at the front of the queue has been processed, the children nodes of $\ottmv{tree}$ and children destinations of $\ottmv{dtree}$ are enqueued to be processed later. There, $ \ottstype{Tree}~ \ottstype{T} $ is defined unsurprisingly as $ \ottstype{Tree}~ \ottstype{T} \btriangleq   \ottstype{1}  \ottstype{\oplus}  \ottstype{(}  \ottstype{T} \ottstype{\otimes}  \ottstype{(}   \ottstype{(}  \ottstype{Tree}~ \ottstype{T}  \ottstype{)}  \ottstype{\otimes}  \ottstype{(}  \ottstype{Tree}~ \ottstype{T}  \ottstype{)}   \ottstype{)}   \ottstype{)}  $; we refer to the constructors of $ \ottstype{Tree}~ \ottstype{T} $ as $\ottsctor{Nil}$ and $\ottsctor{Node}$, defined in the obvious way. We also assume some encoding of the type $ \ottstype{Nat} $ of natural number. $ \ottstype{Queue}~ \ottstype{T} $ is the efficient queue type from \cref{ssec:efficient-queue}.

We implement the actual breadth-first relabeling $\ottkw{relabelDPS}$ as an instance of a more general breadth-first traversal function $\ottkw{mapAccumBFS}$, which applies any state-passing style transformation of labels in breadth-first order.

In $\ottkw{mapAccumBFS}$, we create a new destination $\ottmv{dtree}$ into which we will write the result of the traversal, then call $\ottkw{go}$. The $\ottkw{go}$ function is in destination-passing style, but what's remarkable is that $\ottkw{go}$ takes an unbounded number of destinations as arguments, since there are as many destinations as items in the queue. This is where we use the fact that destinations are ordinary values.

The implementation of \cref{fig:impl-bfs} is very close to the one found in~\cite{bagrel_destination-passing_2024}. The difference is that, because they can't store destinations in structures with holes (see the discussion in \cref{sec:scope-escape-dests}), their implementation can't use the efficient queue implementation from \cref{ssec:efficient-queue}. So they have to revert to using a Hood-Melville queue for breadth-first traversal.

However this improvement comes at a cost: we a introduce \emph{mode} system that combines linearity and age to make the system sound, hence the new \textcolor{modecolor}{fuchsia} annotations in the code. We'll describe modes in detail in \cref{sec:syntax-type-system}. In the meantime, $ \ottsmode{1} $ and $ \ottsmode{\omega} $ control linearity: we use $ \ottsmode{\omega} $ to mean that the state and function $\ottmv{f}$ can be used many times. On the other hand, $ \ottsmode{\infty} $ is an \emph{age} annotation; in particular, the associated argument cannot carry destinations. Arguments with no modes are otherwise linear and can capture destinations. We introduce the exponential modality $ \ottstype{!}_{ \ottsmode{m} } \ottstype{T} $ to reify mode $\ottsmode{m}$ in a type; this is useful to return several values having different modes from a function, like in $\ottmv{f}$. An exponential is rarely needed in an argument position, as we have $  \ottstype{(}  \ottstype{!}_{ \ottsmode{m} } \ottstype{T}  \ottstype{)}  \,_{\myfuntm{   \ottsmode{1}  \hspace{-0.15ex}  \ottsmode{\nu}   } }\!\ottstype{\multimap}\, \ottstype{U} \simeq \ottstype{T} \,_{\myfuntm{ \ottsmode{m} } }\!\ottstype{\multimap}\, \ottstype{U} $.

\begin{codefig}{\caption{Breadth-first tree traversal in destination-passing style}\label{fig:impl-bfs}}
\newoperator[~\mathsfbf{rec}]
{\ottkw{go}}{    \ottstype{(}    \ottstype{S} \,_{\myfuntm{   \ottsmode{\omega}  \hspace{-0.15ex}  \ottsmode{\infty}   } }\!\ottstype{\multimap}\, \ottstype{T_{{\mathrm{1}}}}  \,_{\myfuntm{   \ottsmode{1}  \hspace{-0.15ex}  \ottsmode{\nu}   } }\!\ottstype{\multimap}\,  \ottstype{(}  \ottstype{!}_{   \ottsmode{\omega}  \hspace{-0.15ex}  \ottsmode{\infty}   } \ottstype{S}  \ottstype{)}   \ottstype{\otimes} \ottstype{T_{{\mathrm{2}}}}  \ottstype{)}  \,_{\myfuntm{   \ottsmode{\omega}  \hspace{-0.15ex}  \ottsmode{\infty}   } }\!\ottstype{\multimap}\, \ottstype{S}  \,_{\myfuntm{   \ottsmode{\omega}  \hspace{-0.15ex}  \ottsmode{\infty}   } }\!\ottstype{\multimap}\,  \ottstype{Queue}~  \ottstype{(}   \ottstype{Tree}~ \ottstype{T_{{\mathrm{1}}}}  \ottstype{\otimes}  \ottstype{\lfloor}\,\!_{\mydestm{   \ottsmode{1}  \hspace{-0.15ex}  \ottsmode{\nu}   } }  \ottstype{Tree}~ \ottstype{T_{{\mathrm{2}}}}  \ottstype{\rfloor}   \ottstype{)}    \,_{\myfuntm{   \ottsmode{1}  \hspace{-0.15ex}  \ottsmode{\nu}   } }\!\ottstype{\multimap}\,  \ottstype{1}  }
{ \ottkw{go}~ \ottmv{f} ~ \ottmv{st} ~ \ottmv{q} }{\!\!\!\begin{array}[t]{l} \ottkw{case}_{\mycasem{   \ottsmode{1}  \hspace{-0.15ex}  \ottsmode{\nu}   } }\,  (  \ottkw{dequeue}~ \ottmv{q}  )  ~\ottkw{of}~\{   \mynewline   \myspace{1}  \ottsctor{Inl} \, \ottsctor{()}  \pmb{\mapsto} \ottsctor{()} \,,~   \mynewline   \myspace{1}   \ottsctor{Inr}\,  \ottsctor{(}  \ottsctor{(} \ottmv{tree} \,\ottsctor{,}~ \ottmv{dtree} \ottsctor{)}  \,\ottsctor{,}~ \ottmv{q'} \ottsctor{)}    \pmb{\mapsto}  \ottkw{case}_{\mycasem{   \ottsmode{1}  \hspace{-0.15ex}  \ottsmode{\nu}   } }\, \ottmv{tree} ~\ottkw{of}~\{   \mynewline   \myspace{2}   \ottsctor{Nil}   \pmb{\mapsto}   \ottmv{dtree} \triangleleft\ottsctor{Nil}  \patu  \ottkw{go}~ \ottmv{f} ~ \ottmv{st} ~ \ottmv{q'}   \,,~   \mynewline   \myspace{2}   \ottsctor{Node}~ \ottmv{x} ~ \ottmv{tl} ~ \ottmv{tr}   \pmb{\mapsto}   \ottkw{case}_{\mycasem{   \ottsmode{1}  \hspace{-0.15ex}  \ottsmode{\nu}   } }\,  (  \ottmv{dtree} \triangleleft\ottsctor{Node}  )  ~\ottkw{of}~   \mynewline   \myspace{3}   \ottsctor{(} \ottmv{dy} \,\ottsctor{,}~  \ottsctor{(} \ottmv{dtl} \,\ottsctor{,}~ \ottmv{dtr} \ottsctor{)}  \ottsctor{)}   \pmb{\mapsto}     \ottkw{case}_{\mycasem{   \ottsmode{1}  \hspace{-0.15ex}  \ottsmode{\nu}   } }\,  (    \ottmv{f} ~ \ottmv{st}   ~ \ottmv{x}  )  ~\ottkw{of}~   \mynewline   \myspace{4}   \ottsctor{(}  \expcons{   \ottsmode{\omega}  \hspace{-0.15ex}  \ottsmode{\infty}   } \ottmv{st'}  \,\ottsctor{,}~ \ottmv{y} \ottsctor{)}   \pmb{\mapsto}   \mynewline   \myspace{5}  \ottmv{dy}   \blacktriangleleft \ottmv{y}  \patu   \mynewline   \myspace{5}   \ottkw{go}~ \ottmv{f} ~ \ottmv{st'} ~  (   \ottmv{q'} ~\ottkw{enqueue}~  \ottsctor{(} \ottmv{tl} \,\ottsctor{,}~ \ottmv{dtl} \ottsctor{)}   ~\ottkw{enqueue}~  \ottsctor{(} \ottmv{tr} \,\ottsctor{,}~ \ottmv{dtr} \ottsctor{)}   )        \}  \} \end{array}}
\newoperator
{\ottkw{mapAccumBFS}}{    \ottstype{(}    \ottstype{S} \,_{\myfuntm{   \ottsmode{\omega}  \hspace{-0.15ex}  \ottsmode{\infty}   } }\!\ottstype{\multimap}\, \ottstype{T_{{\mathrm{1}}}}  \,_{\myfuntm{   \ottsmode{1}  \hspace{-0.15ex}  \ottsmode{\nu}   } }\!\ottstype{\multimap}\,  \ottstype{(}  \ottstype{!}_{   \ottsmode{\omega}  \hspace{-0.15ex}  \ottsmode{\infty}   } \ottstype{S}  \ottstype{)}   \ottstype{\otimes} \ottstype{T_{{\mathrm{2}}}}  \ottstype{)}  \,_{\myfuntm{   \ottsmode{\omega}  \hspace{-0.15ex}  \ottsmode{\infty}   } }\!\ottstype{\multimap}\, \ottstype{S}  \,_{\myfuntm{   \ottsmode{\omega}  \hspace{-0.15ex}  \ottsmode{\infty}   } }\!\ottstype{\multimap}\,  \ottstype{Tree}~ \ottstype{T_{{\mathrm{1}}}}   \,_{\myfuntm{   \ottsmode{1}  \hspace{-0.15ex}  \ottsmode{\infty}   } }\!\ottstype{\multimap}\,  \ottstype{Tree}~ \ottstype{T_{{\mathrm{2}}}}  }
{ \ottkw{mapAccumBFS}~ \ottmv{f} ~ \ottmv{st} ~ \ottmv{tree} }{\!\!\!\begin{array}[t]{l} \ottkw{from}_{\ottkw{\ltimes} }'   (  \ottkw{upd}_{\ottkw{\ltimes} }\,  (   \ottkw{new}_{\ottkw{\ltimes} }  \pmb{:}{\scriptstyle    \ottstype{(}  \ottstype{Tree}~ \ottstype{T_{{\mathrm{2}}}}  \ottstype{)}  \,\ottstype{\ltimes}\,  \ottstype{\lfloor}\,\!_{\mydestm{   \ottsmode{1}  \hspace{-0.15ex}  \ottsmode{\nu}   } }  \ottstype{Tree}~ \ottstype{T_{{\mathrm{2}}}}  \ottstype{\rfloor}   }  )  ~\ottkw{with}~ \ottmv{dtree} \, \pmb{\mapsto}   \mynewline   \myspace{3}    \ottkw{go}~ \ottmv{f} ~ \ottmv{st} ~  (  \ottkw{singleton}~  \ottsctor{(} \ottmv{tree} \,\ottsctor{,}~ \ottmv{dtree} \ottsctor{)}   )      )  \end{array}}
\newoperator
{\ottkw{relabelDPS}}{  \ottstype{Tree}~  \ottstype{1}   \,_{\myfuntm{   \ottsmode{1}  \hspace{-0.15ex}  \ottsmode{\infty}   } }\!\ottstype{\multimap}\,  \ottstype{Tree}~  \ottstype{Nat}   }
{ \ottkw{relabelDPS}~ \ottmv{tree} }{\!\!\!\begin{array}[t]{l} \ottkw{mapAccumBFS}~  (  \lamnt{ \ottmv{st} }{   \ottsmode{\omega}  \hspace{-0.15ex}  \ottsmode{\infty}   }{  \lamnt{ \ottmv{un} }{   \ottsmode{1}  \hspace{-0.15ex}  \ottsmode{\nu}   }{  \ottmv{un} \patu  \ottsctor{(}  \expcons{   \ottsmode{\omega}  \hspace{-0.15ex}  \ottsmode{\infty}   }  (  \ottkw{succ}~ \ottmv{st}  )   \,\ottsctor{,}~ \ottmv{st} \ottsctor{)}   }  }  )  ~  \ottsctor{1}  ~ \ottmv{tree} \end{array}}
\end{codefig}

\section{Type System}\label{sec:syntax-type-system}

\newcommand{\grammsep}{\hspace*{1.8ex}|\hspace*{1.8ex}}
\newcommand{\grammdef}{\mathrel{\raisebox{0.09ex}{$\mathop{:}$\hspace*{-0.1ex}$\mathop{:}$\hspace*{-0.1ex}}\shorteq}}
\newcommand{\pleq}{ \mathrel{\texttt{⥶} } ^{\scriptscriptstyle{\pmb{\mathsf{p}}}}}
\newcommand{\aleq}{ \mathrel{\texttt{⥶} } ^{\scriptscriptstyle{\pmb{\mathsf{a}}}}}

\begin{figure}[t]

\begin{minipage}{\linewidth}\small\textit{Core grammar of terms:}\end{minipage}

\smallskip

\hspace*{-0.05\linewidth}\begin{minipage}{1.2\linewidth}\codehere{\setlength{\arraycolsep}{0.6ex}\!\begin{array}{rrl}
\ottnt{t}, \ottnt{u} &\grammdef& \ottmv{x} \grammsep  \ottnt{t'} ~ \ottnt{t}  \grammsep  \ottnt{t} \patu \ottnt{t'}  \\
             &|\,&  \ottkw{case}_{\mycasem{ \ottsmode{m} } }\, \ottnt{t} ~\ottkw{of}~\{\ottsctor{Inl}\, \ottmv{x_{{\mathrm{1}}}} \pmb{\mapsto} \ottnt{u_{{\mathrm{1}}}} \,,~\ottsctor{Inr}\, \ottmv{x_{{\mathrm{2}}}} \pmb{\mapsto} \ottnt{u_{{\mathrm{2}}}} \}  \grammsep  \ottkw{case}_{\mycasem{ \ottsmode{m} } }\, \ottnt{t} ~\ottkw{of}~\ottsctor{(} \ottmv{x_{{\mathrm{1}}}} \,\ottsctor{,}~ \ottmv{x_{{\mathrm{2}}}} \ottsctor{)} \pmb{\mapsto} \ottnt{u}  \grammsep  \ottkw{case}_{\mycasem{ \ottsmode{m} } }\, \ottnt{t} ~\ottkw{of}~\expcons{ \ottsmode{n} } \ottmv{x} \, \pmb{\mapsto} \ottnt{u}  \\
             &|\,&  \ottkw{upd}_{\ottkw{\ltimes} }\, \ottnt{t} ~\ottkw{with}~ \ottmv{x} \, \pmb{\mapsto} \ottnt{t'}  \grammsep  \ottkw{to}_{\ottkw{\ltimes} }\, \ottnt{t}  \grammsep  \ottkw{from}_{\ottkw{\ltimes} }\, \ottnt{t}  \grammsep  \ottkw{new}_{\ottkw{\ltimes} }  \\
             &|\,& \ottnt{t}  \triangleleft  \ottsctor{()} \grammsep \ottnt{t}  \triangleleft \, \ottsctor{Inl} \grammsep \ottnt{t}  \triangleleft \, \ottsctor{Inr} \grammsep \ottnt{t}  \triangleleft  \ottsctor{({,})} \grammsep  \ottnt{t} \triangleleft \,\expcons{ \ottsmode{m} }  \grammsep  \ottnt{t} \triangleleft (\lamnt{ \ottmv{x} }{ \ottsmode{m} }{ \ottnt{u} })  \grammsep  \ottnt{t} \mathop{\triangleleft\mycirc} \ottnt{t'}  \grammsep  \ottnt{t} \blacktriangleleft \ottnt{t'} 
\end{array}}\end{minipage}

\bigskip

\begin{minipage}{\linewidth}\small\textit{Syntactic sugar for terms:}\end{minipage}

\smallskip

\hspace*{-0.05\linewidth}\begin{minipage}{1.2\linewidth}\sidebysidecodehere{t}{0.393}{
 \ottsctor{Inl}\, \ottnt{t}  \btriangleq \!\!\!\begin{array}[t]{l} \ottkw{from}_{\ottkw{\ltimes} }'   (  \ottkw{upd}_{\ottkw{\ltimes} }\,  \ottkw{new}_{\ottkw{\ltimes} }  ~\ottkw{with}~ \ottmv{d} \, \pmb{\mapsto}    \mynewline   \myspace{3}  \ottmv{d}   \triangleleft \, \ottsctor{Inl} \blacktriangleleft \ottnt{t}   )  \end{array}\\[\interdefskip]
 \ottsctor{Inr}\, \ottnt{t}  \btriangleq \!\!\!\begin{array}[t]{l} \ottkw{from}_{\ottkw{\ltimes} }'   (  \ottkw{upd}_{\ottkw{\ltimes} }\,  \ottkw{new}_{\ottkw{\ltimes} }  ~\ottkw{with}~ \ottmv{d} \, \pmb{\mapsto}    \mynewline   \myspace{3}  \ottmv{d}   \triangleleft \, \ottsctor{Inr} \blacktriangleleft \ottnt{t}   )  \end{array}\\[\interdefskip]
 \expcons{ \ottsmode{m} } \ottnt{t}  \btriangleq \!\!\!\begin{array}[t]{l} \ottkw{from}_{\ottkw{\ltimes} }'   (  \ottkw{upd}_{\ottkw{\ltimes} }\,  \ottkw{new}_{\ottkw{\ltimes} }  ~\ottkw{with}~ \ottmv{d} \, \pmb{\mapsto}     \mynewline   \myspace{3}  \ottmv{d}  \triangleleft \,\expcons{ \ottsmode{m} }  \blacktriangleleft \ottnt{t}   )  \end{array}\\[\interdefskip]
 \lamnt{ \ottmv{x} }{ \ottsmode{m} }{ \ottnt{u} }  \btriangleq \!\!\!\begin{array}[t]{l} \ottkw{from}_{\ottkw{\ltimes} }'   (  \ottkw{upd}_{\ottkw{\ltimes} }\,  \ottkw{new}_{\ottkw{\ltimes} }  ~\ottkw{with}~ \ottmv{d} \, \pmb{\mapsto}    \mynewline   \myspace{3}  \ottmv{d}  \triangleleft (\lamnt{ \ottmv{x} }{ \ottsmode{m} }{ \ottnt{u} })   )  \end{array}\hspace*{-0.4cm}
}{
 \ottkw{from}_{\ottkw{\ltimes} }'  \ottnt{t}  \btriangleq \\
\myspace{1}\!\!\!\begin{array}[t]{l} \ottkw{case}_{\mycasem{   \ottsmode{1}  \hspace{-0.15ex}  \ottsmode{\nu}   } }\,  (  \ottkw{from}_{\ottkw{\ltimes} }\,  (  \ottkw{upd}_{\ottkw{\ltimes} }\, \ottnt{t} ~\ottkw{with}~ \ottmv{un} \, \pmb{\mapsto}  \ottmv{un} \patu  \expcons{   \ottsmode{1}  \hspace{-0.15ex}  \ottsmode{\infty}   } \ottsctor{()}    )   )  ~\ottkw{of}~  \mynewline   \myspace{1}  \ottsctor{(} \ottmv{st} \,\ottsctor{,}~ \ottmv{ex} \ottsctor{)}\! \pmb{\mapsto}    \ottkw{case}_{\mycasem{   \ottsmode{1}  \hspace{-0.15ex}  \ottsmode{\nu}   } }\, \ottmv{ex} ~\ottkw{of}~  \mynewline   \myspace{2}  \expcons{   \ottsmode{1}  \hspace{-0.15ex}  \ottsmode{\infty}   } \ottmv{un} \pmb{\mapsto} \ottmv{un}  \patu \ottmv{st}   \end{array}\\[\interdefskip]
 \ottsctor{()}  \btriangleq \!\!\!\begin{array}[t]{l} \ottkw{from}_{\ottkw{\ltimes} }'   (  \ottkw{upd}_{\ottkw{\ltimes} }\,  \ottkw{new}_{\ottkw{\ltimes} }  ~\ottkw{with}~ \ottmv{d} \, \pmb{\mapsto} \ottmv{d}  \triangleleft  \ottsctor{()}  )  \end{array}\\[\interdefskip]
 \ottsctor{(} \ottnt{t_{{\mathrm{1}}}} \,\ottsctor{,}~ \ottnt{t_{{\mathrm{2}}}} \ottsctor{)}  \btriangleq \!\!\!\begin{array}[t]{l} \ottkw{from}_{\ottkw{\ltimes} }'   (  \ottkw{upd}_{\ottkw{\ltimes} }\,  \ottkw{new}_{\ottkw{\ltimes} }  ~\ottkw{with}~ \ottmv{d} \, \pmb{\mapsto}     \ottkw{case}_{\mycasem{   \ottsmode{1}  \hspace{-0.15ex}  \ottsmode{\nu}   } }\,  ( \ottmv{d}  \triangleleft  \ottsctor{({,})} )  ~\ottkw{of}~  \mynewline   \myspace{3}  \ottsctor{(} \ottmv{d_{{\mathrm{1}}}} \,\ottsctor{,}~ \ottmv{d_{{\mathrm{2}}}} \ottsctor{)}\! \pmb{\mapsto} \ottmv{d_{{\mathrm{1}}}}  \blacktriangleleft \ottnt{t_{{\mathrm{1}}}}  \patu \ottmv{d_{{\mathrm{2}}}}  \blacktriangleleft \ottnt{t_{{\mathrm{2}}}}   )  \end{array}
}\end{minipage}

\bgroup\renewcommand{\ottdruleTyXXtermXXVal}{}

\renewcommand\ottaltinferrule[4]{
  \inferrule*[narrower=0.3,lab=#1,#2]
    {#3}
    {#4}
}

\bigskip
\hrule
\bigskip

\begin{minipage}{\linewidth}\small\textit{Grammar of types, modes and contexts:}\end{minipage}

\smallskip

\hspace*{-0.05\linewidth}\begin{minipage}{\linewidth}\sidebysidecodehere{t}{0.58}{\setlength{\arraycolsep}{0.6ex}\!\begin{array}{rrl}
\ottstype{T}, \ottstype{U}, \ottstype{S} &\grammdef&  \ottstype{\lfloor}\,\!_{\mydestm{ \ottsmode{n} } } \ottstype{T} \ottstype{\rfloor}  \quad\quad\textit{(destination)} \\
                    &|\,&  \ottstype{S} \,\ottstype{\ltimes}\, \ottstype{T}  \hspace*{\widthof{$ \ottstype{\lfloor}\,\!_{\mydestm{ \ottsmode{n} } } \ottstype{T} \ottstype{\rfloor} $}-\widthof{$ \ottstype{U} \,\ottstype{\ltimes}\, \ottstype{T} $}}\quad\quad\textit{(ampar)} \\
                    &|\,&  \ottstype{1}  \grammsep  \ottstype{T_{{\mathrm{1}}}} \ottstype{\oplus} \ottstype{T_{{\mathrm{2}}}}  \grammsep  \ottstype{T_{{\mathrm{1}}}} \ottstype{\otimes} \ottstype{T_{{\mathrm{2}}}}  \grammsep  \ottstype{!}_{ \ottsmode{m} } \ottstype{T}  \grammsep  \ottstype{T} \,_{\myfuntm{ \ottsmode{m} } }\!\ottstype{\multimap}\, \ottstype{U}  \\
&&\\
\Gamma &\grammdef&  \smallbullet  \grammsep  \ottmv{x} :\!_{\! \ottsmode{m} } \ottstype{T}  \grammsep  \Gamma_{{\mathrm{1}}} ,~ \Gamma_{{\mathrm{2}}} 
\end{array}}{\setlength{\arraycolsep}{0.6ex}\!\begin{array}[b]{rrl}
\ottsmode{m}, \ottsmode{n} &\grammdef&  \ottsmode{p} \hspace{-0.15ex} \ottsmode{a}  \hspace*{\widthof{$ \ottstype{\lfloor}\,\!_{\mydestm{ \ottsmode{m} } } \ottstype{T} \ottstype{\rfloor} $}-\widthof{$ \ottsmode{p} \hspace{-0.15ex} \ottsmode{a} $}}\quad\textit{(pair of multiplicity and age)} \\
  \ottsmode{p} &\grammdef&  \ottsmode{1}  \grammsep  \ottsmode{\omega}  \\
  \ottsmode{a} &\grammdef&  \ottsmode{\uparrow}^{ \ottsmodee{k} }  \grammsep  \ottsmode{\infty} 
\end{array}\\[\interdefskip]
 \ottsmode{\nu}  \btriangleq  \ottsmode{\uparrow}^{  \ottsmodee{0}  }  \quad  \ottsmode{\uparrow}  \btriangleq  \ottsmode{\uparrow}^{  \ottsmodee{1}  } }\end{minipage}

\bigskip

\begin{minipage}{\linewidth}\small\textit{Ordering on modes:}\end{minipage}

{\small
\vspace*{-0.3cm}

\hfill $ \ottsmode{p} \hspace{-0.15ex} \ottsmode{a}   \mathrel{\texttt{⥶} }   \ottsmode{p'} \hspace{-0.15ex} \ottsmode{a'}  \Longleftrightarrow \ottsmode{p} \pleq \ottsmode{p'} \land~\ottsmode{a} \aleq \ottsmode{a'}$ \hfill \begin{tikzpicture}[baseline=(current bounding box.center), clip]
\node (omega) at (0,1) {$ \ottsmode{\omega} $};
\node (oneA) at (0,0) {$ \ottsmode{1} $};

\draw (omega) -- node[midway, below, rotate=90] {$\pleq$} (oneA);

\node (inf) at (4,1) {$ \ottsmode{\infty} $};
\node (zero) at (2,0) {$ \ottsmode{\uparrow}^{  \ottsmodee{0}  } $};
\node (oneB) at (3,0) {$ \ottsmode{\uparrow}^{  \ottsmodee{1}  } $};
\node (dots1) at (4,0) {\ldots};
\node (k) at (5,0) {$ \ottsmode{\uparrow}^{ \ottsmodee{k} } $};
\node (dots2) at (6,0) {\ldots};

\draw (inf) -- node[midway, above, rotate=30, inner sep=1pt] {$\aleq$} (zero);
\draw (inf) -- node[midway, below, rotate=45, inner sep=1pt] {$\aleq$} (oneB);
\draw (inf) -- node[midway, above, rotate=135, yscale=-1, inner sep=1pt] {$ \mathrel{\texttt{⥶} } ^{\reflectbox{$\scriptscriptstyle{\pmb{\mathsf{a}}}$}}$} (k);
\end{tikzpicture}\hfill\phantom{.}
}

\vspace*{-0.2cm}

\begin{minipage}{\linewidth}\small\textit{Operations on modes:}\end{minipage}

\smallskip

{\small
\hfill\begin{tabular}{|c||c|c|}\hline
$\ottsmode{+}$ & $ \ottsmode{1} $ & $ \ottsmode{\omega} $ \\\hhline{|=#=|=|}
$ \ottsmode{1} $        & $ \ottsmode{\omega} $ & $ \ottsmode{\omega} $ \\\hline
$ \ottsmode{\omega} $        & $ \ottsmode{\omega} $ & $ \ottsmode{\omega} $ \\\hline
\end{tabular}
\hfill
\begin{tabular}{|c||c|c|}\hline
$ \ottsmode{\hspace{-0.1ex}\cdot\hspace{-0.1ex} } $        & $ \ottsmode{1} $ & $ \ottsmode{\omega} $ \\\hhline{|=#=|=|}
$ \ottsmode{1} $        & $ \ottsmode{1} $ & $ \ottsmode{\omega} $ \\\hline
$ \ottsmode{\omega} $        & $ \ottsmode{\omega} $ & $ \ottsmode{\omega} $ \\\hline
\end{tabular}
\hfill
\vrule width 0.5pt 
\hfill
\begin{tabular}{|c||c|c|}\hline
$\ottsmode{+}$ & $ \ottsmode{\uparrow}^{ \ottsmodee{k} } $               & $ \ottsmode{\infty} $ \\\hhline{|=#=|=|}
$ \ottsmode{\uparrow}^{ \ottsmodee{j} } $     & $\text{if }\ottsmodee{k} = \ottsmodee{j}\text{ then } \ottsmode{\uparrow}^{ \ottsmodee{k} } \text{ else } \ottsmode{\infty} $ & $ \ottsmode{\infty} $ \\\hline
$ \ottsmode{\infty} $        & $ \ottsmode{\infty} $                   & $ \ottsmode{\infty} $ \\\hline
\end{tabular}
\hfill
\begin{tabular}{|c||c|c|}\hline
$ \ottsmode{\hspace{-0.1ex}\cdot\hspace{-0.1ex} } $        & $ \ottsmode{\uparrow}^{ \ottsmodee{k} } $    & $ \ottsmode{\infty} $ \\\hhline{|=#=|=|}
$ \ottsmode{\uparrow}^{ \ottsmodee{j} } $     & $ \ottsmode{\uparrow}^{  \ottsmodee{ \ottsmodee{k} + \ottsmodee{j} }  } $ & $ \ottsmode{\infty} $ \\\hline
$ \ottsmode{\infty} $        & $ \ottsmode{\infty} $       & $ \ottsmode{\infty} $ \\\hline
\end{tabular}\hfill\phantom{.}
\smallskip

\hfill\hspace*{-1.5cm}
$ \ottsmode{(}  \ottsmode{p} \hspace{-0.15ex} \ottsmode{a}  \ottsmode{)}   \ottsmode{\hspace{-0.1ex}\cdot\hspace{-0.1ex} }   \ottsmode{(}  \ottsmode{p'} \hspace{-0.15ex} \ottsmode{a'}  \ottsmode{)}  \btriangleq   \ottsmode{(} \ottsmode{p}  \ottsmode{\hspace{-0.1ex}\cdot\hspace{-0.1ex} }  \ottsmode{p'} \ottsmode{)}  \hspace{-0.15ex}  \ottsmode{(} \ottsmode{a}  \ottsmode{\hspace{-0.1ex}\cdot\hspace{-0.1ex} }  \ottsmode{a'} \ottsmode{)}  $
\hfill\hspace*{0.5cm}
$ \ottsmode{  \ottsmode{(}  \ottsmode{p} \hspace{-0.15ex} \ottsmode{a}  \ottsmode{)}  +  \ottsmode{(}  \ottsmode{p'} \hspace{-0.15ex} \ottsmode{a'}  \ottsmode{)}  }  \btriangleq   \ottsmode{(}  \ottsmode{ \ottsmode{p} + \ottsmode{p'} }  \ottsmode{)}  \hspace{-0.15ex}  \ottsmode{(}  \ottsmode{ \ottsmode{a} + \ottsmode{a'} }  \ottsmode{)}  $
\hfill\phantom{.}}

\bigskip

\begin{minipage}{\linewidth}\small\textit{Operations on typing contexts:}\end{minipage}

\smallskip

{\small
\bgroup
\renewcommand\tabcolsep{2pt}
\hfill\begin{tabular}[c]{rclcc@{\qquad}l}
  $\ottsmode{n}$ &$\cdot$& $ \smallbullet $ & $\btriangleq$ & $ \smallbullet $\\
  $\ottsmode{n}$ &$\cdot$& $ (   \ottmv{x} :\!_{\! \ottsmode{m} } \ottstype{T}  ,~ \Gamma  ) $ & $\btriangleq$ & $  (  \ottmv{x} :\!_{\! \ottsmode{n}  \ottsmode{\hspace{-0.1ex}\cdot\hspace{-0.1ex} }  \ottsmode{m} } \ottstype{T}  )  ,~  \ottsmode{n} \hspace{-0.3ex}\cdot\hspace{-0.3ex} \Gamma  $\\
\end{tabular}
\hfill\hspace*{-0.07cm}
\vrule width 0.5pt 
\hfill
\begin{tabular}[c]{rclcc@{\qquad}l}
  $ \smallbullet $ &$+$& $\Gamma$ & $\btriangleq$ & $\Gamma$\\
  $ (   \ottmv{x} :\!_{\! \ottsmode{m} } \ottstype{T}  ,~ \Gamma_{{\mathrm{1}}}  ) $ &$+$& $\Gamma_{{\mathrm{2}}}$ & $\btriangleq$ & $  \ottmv{x} :\!_{\! \ottsmode{m} } \ottstype{T}  ,~  ( \Gamma_{{\mathrm{1}}}  +  \Gamma_{{\mathrm{2}}} )  $ & \textrm{if $\ottmv{x}\notin\Gamma_{{\mathrm{2}}}$}\\
  $ (   \ottmv{x} :\!_{\! \ottsmode{m} } \ottstype{T}  ,~ \Gamma_{{\mathrm{1}}}  ) $ &$+$& $ (   \ottmv{x} :\!_{\! \ottsmode{m'} } \ottstype{T}  ,~ \Gamma_{{\mathrm{2}}}  ) $ & $\btriangleq$ & $  \ottmv{x} :\!_{\!  \ottsmode{ \ottsmode{m} + \ottsmode{m'} }  } \ottstype{T}  ,~  ( \Gamma_{{\mathrm{1}}}  +  \Gamma_{{\mathrm{2}}} )  $
\end{tabular}\hfill\phantom{.}
\egroup
}

\egroup

\caption{Terms, types and modes of \destcalculus{}}\label{fig:grammar}\label{fig:sterm}\label{fig:mul-age-tables}
\end{figure}

\destcalculus{} is a simply typed $\lambda$-calculus with unit ($ \ottstype{1} $), product ($\ottstype{\otimes}$) and sum ($\ottstype{\oplus}$) types. Its most salient features are the destination $ \ottstype{\lfloor}\,\!_{\mydestm{ \ottsmode{m} } } \ottstype{T} \ottstype{\rfloor} $ and ampar $ \ottstype{S} \,\ottstype{\ltimes}\, \ottstype{T} $ types which we've introduced in \cref{sec:working-with-dests,sec:scope-escape-dests,sec:bft}.

To ensure that destinations are used soundly, we need both to enforce the linearity of destination but also to prevent destinations from escaping their scope, as discussed in \cref{sec:scope-escape-dests}. To that effect, \destcalculus{} tracks the \emph{age} of destinations, that is how many nested scope have been open between the current expression and the scope from which a destination originates. We'll see in \cref{ssec:ty-term-val} that scopes are introduced by $ \ottkw{upd}_{\ottkw{\ltimes} }\, \ottnt{t} ~\ottkw{with}~ \ottmv{x} \, \pmb{\mapsto} \ottnt{t'} $. For instance, if we have a term $ \ottkw{upd}_{\ottkw{\ltimes} }\, \ottnt{t_{{\mathrm{1}}}} ~\ottkw{with}~ \ottmv{x_{{\mathrm{1}}}} \, \pmb{\mapsto}  \ottkw{upd}_{\ottkw{\ltimes} }\, \ottnt{t_{{\mathrm{2}}}} ~\ottkw{with}~ \ottmv{x_{{\mathrm{2}}}} \, \pmb{\mapsto}  \ottkw{upd}_{\ottkw{\ltimes} }\, \ottnt{t_{{\mathrm{3}}}} ~\ottkw{with}~ \ottmv{x_{{\mathrm{3}}}} \, \pmb{\mapsto} \ottmv{x_{{\mathrm{1}}}}   $, then we will say that the innermost occurrence of $\ottmv{x_{{\mathrm{1}}}}$ has age $ \ottsmode{\uparrow} ^{\ottsmodee{2}}$ because two nested $\ottkw{upd}_{\ottkw{\ltimes} }$ separate the definition and use site of $\ottmv{x_{{\mathrm{1}}}}$.

A natural idea, to track ages, is to introduce a modality $\ottstype{\uparrow}\ottstype{T}$ to mean ``a $\ottstype{T}$ in the previous scope''. Let's explore why this isn't quite going to work for us, hence why we need something more general.

A typical presentation of modal type theories is with a pair of context $\Gamma_{\uparrow}$~;~$\Gamma$ of bindings from the previous scope and the current scope respectively~\cite{pfenning_modal_1995}, and rules such as

\vspace*{-0.3cm}

$$
\begin{array}{cc}
  \begin{minipage}{0.3\linewidth}
    \begin{equation}
      \label{eq:1}
      \inferrule
      {\phantom{a}}
      {\Gamma_{\uparrow}~;~\Gamma,\ottmv{x}:\ottstype{T} \pmb{\vdash} \ottmv{x}\pmb{:}\ottstype{T}}
    \end{equation}
  \end{minipage}
  &
    \begin{minipage}{0.3\linewidth}
      \begin{equation}
        \label{eq:2}
        \inferrule
        { \smallbullet ~;~ \Gamma_{\uparrow} \pmb{\vdash} \ottnt{t}\pmb{:}\ottstype{T}}
        {\Gamma_{\uparrow}~;~\Gamma \pmb{\vdash} \ottnt{t}\pmb{:}\ottstype{\uparrow}\ottstype{T}}
      \end{equation}
    \end{minipage}
\end{array}
$$

The idea is that only variables from the current scope can be used to make a term for the current scope~\eqref{eq:1}, and to make a term at the previous scope, you need to make it only with variables from $\Gamma_{\uparrow}$~\eqref{eq:2}. But this can't be the whole story here. Indeed, there's no way to refer to variables from two scopes ago, and it would be unsound for \destcalculus{}, in the manner described in \cref{sec:scope-escape-dests}, to mash all the older scopes together in $\Gamma_{\uparrow}$. So, following this route, we'd need infinitely many contexts (and as many modalities, or more realistically a single graded modality), sequents would look like $\ldots~;~\Gamma_2~;~\Gamma_1~;~\Gamma_0 \pmb{\vdash} \ottnt{t}\pmb{:}\ottstype{T}$. Finicky, but manageable perhaps. But it's not all! We need yet another context: one for bindings which are ageless because they don't contain destinations, so that we can fill destinations with harmless values like $ \ottsctor{(}  \ottsctor{1}  \,\ottsctor{,}~  \ottsctor{2}  \ottsctor{)} $ that have been bound two or more scopes ago. More even: linear logic, famously, is also a modal logic (the modality is the exponential $\ottstype{!}\ottstype{T}$), so we'd need to double each of these contexts. This would be quite messy.

Fortunately, there's a way to simplify all this. First observe that having several contexts is the same as having a single context but with annotation on the bindings: instead of $\ottmv{x}:\ottstype{S}~;~\ottmv{y}:\ottstype{T}~;~\ottmv{z}:\ottstype{U} \pmb{\vdash} \ldots$, we can have $\ottmv{x}:_2 \ottstype{S}, \ottmv{y}:_1 \ottstype{T}, \ottmv{z}:_0 \ottstype{U} \pmb{\vdash} \ldots$ without adding or losing any information (note how the semicolons separating contexts are replaced by commas separating bindings). We'll call these annotations \emph{modes}. We build on the key insight, which seem to originate with~\cite{ghica_bounded_2014}, that equipping the set of modes with a particular algebraic structure is sufficient to express, algebraically, all the context manipulation that we need. They use a rig structure; for \destcalculus{} we don't need a $\ottsmode{0}$ element, but we'll need a partial order on modes. The motivation for this structure in~\cite{ghica_bounded_2014} is generating electronic circuits, the same approach has since been used, for instance, for Linear Haskell~\cite{bernardy_linear_2018} where it is used to support mode polymorphism. In both of those cases, this algebraic mode style is used in the context of linear types, but~\citet{bernardy_modality_2020} show that it can generalize to a variety of modalities.

For \destcalculus{}, we're following this approach: \destcalculus{} has a single modality $\ottstype{!}_{\ottsmode{m}}$ indexed by a mode. This will greatly simplify our context management (especially in the computer-mechanized proofs), but, just as interestingly, we can define the set of modes as being the Cartesian product of the set of multiplicities (which keep track of linearity) and of the set of ages. The algebraic structure carries over by a generic theorem on products. This means that we can define multiplicities and ages independently, and the combination is taken care of for free.
The syntax of \destcalculus{} terms is presented in~\cref{fig:grammar}, including the syntactic sugar that we've been using in \cref{sec:working-with-dests,sec:scope-escape-dests,sec:bft}.

\subsection{Modes and the Age Semiring}\label{ssec:age-control}

The precise algebraic structure that we'll be needing on modes is both a commutative additive semigroup $\ottsmode{+}$ and a multiplicative monoid $(\ottsmode{ \ottsmode{\hspace{-0.1ex}\cdot\hspace{-0.1ex} } }\,, \ottsmode{1})$, with the usual distributivity law $\ottsmode{n}  \ottsmode{\hspace{-0.1ex}\cdot\hspace{-0.1ex} }   \ottsmode{(}  \ottsmode{ \ottsmode{m_{{\mathrm{1}}}} + \ottsmode{m_{{\mathrm{2}}}} }  \ottsmode{)}  =  \ottsmode{ \ottsmode{n}  \ottsmode{\hspace{-0.1ex}\cdot\hspace{-0.1ex} }  \ottsmode{m_{{\mathrm{1}}}} + \ottsmode{n}  \ottsmode{\hspace{-0.1ex}\cdot\hspace{-0.1ex} }  \ottsmode{m_{{\mathrm{2}}}} } $. In addition we require a partial order $⥶$, such that $\ottsmode{+}$ and $\ottsmode{ \ottsmode{\hspace{-0.1ex}\cdot\hspace{-0.1ex} } }$ are order-preserving. In other words, we'll be dealing with ordered semirings\footnote{This literature is complicated by disputed terminology, where some prefer to use the term ``semiring'' when the additive semigroup has a zero. This terminology is arguably more popular, but leaves no term for the version without a zero. We'll follow the convention, in this article, that semirings with a zero are called ``rigs''.}. In the rest of the article, we'll just say ``semiring''. In practice, all our semirings will be commutative, and we won't be paying attention to the order of factors in mode multiplication.

Our mode semiring is, as promised, the product of a multiplicity semiring, to track linearity, and an age semiring, to prevent scope escape. The multiplicity semiring has elements $ \ottsmode{1} $ (linear) and $ \ottsmode{\omega} $ (unrestricted), it's the same semiring as in~\cite{qtt_2018} or~\cite{bernardy_linear_2018}. It's mostly unsurprising, the key is that that $ \ottsmode{  \ottsmode{1}  +  \ottsmode{1}  } = \ottsmode{\omega} $ will enforce that a linear variable can only be used once, the full description of the multiplicity semiring is given in~\cref{fig:mul-age-tables}.

Ages are more interesting. We write ages as $ \ottsmode{\uparrow}^{ \ottsmodee{k} } $ (with $\ottsmodee{k}$ a natural number), for ``defined $\ottsmodee{k}$ scopes ago''.
We also have an age $ \ottsmode{\infty} $ for variables that don't originate from a $ \ottkw{upd}_{\ottkw{\ltimes} }\, \ottnt{t} ~\ottkw{with}~ \ottmv{x} \, \pmb{\mapsto} \ottnt{t'} $ i.e. that aren't destinations, and can be freely used in and returned by any scope. The main role of age $ \ottsmode{\infty} $ is thus to act as a guarantee that a value doesn't contain destinations. Finally, we will write $ \ottsmode{\nu}  \btriangleq  \ottsmode{\uparrow}^{  \ottsmodee{0}  } $  (``now'') for the age of destinations that originate from the current scope; and $ \ottsmode{\uparrow}  \btriangleq  \ottsmode{\uparrow}^{  \ottsmodee{1}  } $.

The operations or order aren't the usual ones on natural numbers though. It is crucial that \destcalculus{} tracks the precise age of variables. Variables from $2$ scopes ago cannot be used as if they were from $1$ scope ago, or vice-versa. The ordering reflects this with finite ages being arranged in a flat order, with $ \ottsmode{\infty} $ being bigger than all of them. Multiplication of ages will reflect nesting of scope, as such, (finite) ages are multiplied by adding their numerical exponents $ \ottsmode{\uparrow}^{ \ottsmodee{k} }   \ottsmode{\hspace{-0.1ex}\cdot\hspace{-0.1ex} }   \ottsmode{\uparrow}^{ \ottsmodee{j} }  =  \ottsmode{\uparrow}^{  \ottsmodee{ \ottsmodee{k} + \ottsmodee{j} }  } $. In the typing rules, the most common form of scope nesting is opening a scope, which is represented by multiplying by $ \ottsmode{\uparrow} $ (that is, adding $1$ to the ages seen as a natural numbers). Finally $\ottsmode{+}$ is used to share a variable between two subterms, it's given by the least upper bound (for the age order above). The intuition here, is still precise age tracking: a variable must be at the same age in both subterms, or it can be $ \ottsmode{\infty} $, and assume whichever age it needs, including different ones in different subterms.

Bindings in the context are annotated by a mode. The insight of~\cite{ghica_bounded_2014} is that mode addition and multiplication by a mode (aka \emph{scaling}) lift to contexts pointwise, so we have all the tools we need to define a modal type system, including a sub-structural one like linear logic.

The operations and preorders on mode, contexts, etc. are presented in~\cref{fig:mul-age-tables}.
We will usually omit mode annotations when the mode is the multiplicative unit $  \ottsmode{1}  \hspace{-0.15ex}  \ottsmode{\nu}  $ of the semiring.

\subsection{Typing Rules}\label{ssec:ty-term-val}

The typing rules for \destcalculus{} are highly inspired from \citet{bernardy_modality_2020} and Linear Haskell \cite{bernardy_linear_2018}, and are detailed in~\cref{fig:ty-term-sterm}. In particular, we use the same algebraic approach on contexts for mode tracking. Per \cref{ssec:age-control}, a mode is a pair of a multiplicity and an age.

\begin{ottfig}{\caption{Typing rules of \destcalculus{}}\label{fig:ty-term-sterm}}\bgroup\renewcommand{\ottdruleTyXXtermXXVal}{}

\bgroup\SetPrefix{\CTyTerm\CSep}
\ottdefnTyXXterm{}
\egroup

\bigskip
\hrule
\bigskip

\renewcommand\ottaltinferrule[4]{
  \inferrule*[fraction={===},narrower=0.3,lab=#1,#2]
    {#3}
    {#4}
}
\bgroup\SetPrefix{\CTySTerm\CSep}
\ottdefnTyXXsterm{}\egroup
\egroup
\end{ottfig}

\cref{fig:ty-term-sterm} presents the typing rules, including rules for syntactic sugar forms. We'll now walk through the few peculiarities of the type system for terms.

Predicate $\mathtt{DisposableOnly}~\Gamma$ in rules \rref*{\CTyTerm\CSep\CVar}, \rref*{\CTyTerm\CSep\CNewA} and \rref*{\CTySTerm\CSep\CUnit} says that $\Gamma$ can only contain bindings with multiplicity $ \ottsmode{\omega} $, for which weakening is allowed in linear logic. We only need weakening in these three rules, as they are the only possible leaves of the typing tree.

Rule \rref*{\CTyTerm\CSep\CVar}, in addition to weakening, allows for coercion of the mode $\ottsmode{m}$ of the variable used, with ordering constraint $  \ottsmode{1}  \hspace{-0.15ex}  \ottsmode{\nu}    \mathrel{\texttt{⥶} }  \ottsmode{m}$ as defined in \cref{fig:mul-age-tables}. Notably, mode coercion still doesn't allow for a finite age to be changed to another, as $ \ottsmode{\uparrow}^{ \ottsmodee{j} } $ and $ \ottsmode{\uparrow}^{ \ottsmodee{k} } $ are not comparable w.r.t. $\aleq$ when $\ottsmodee{j}\neq\ottsmodee{k}$.

Rule \rref*{\CTyTerm\CSep\CPatU} is the elimination for unit, and is also used to chain fill operations.

Pattern-matching with rules \rref*{\CTyTerm\CSep\CApp}, \rref*{\CTyTerm\CSep\CPatS}, \rref*{\CTyTerm\CSep\CPatP} and \rref*{\CTyTerm\CSep\CPatE} is parametrized by a mode $\ottsmode{m}$ by which the typing context $\Gamma_{{\mathrm{1}}}$ of the scrutinee is multiplied. The variables which bind the subcomponents of the scrutinee then inherit this mode. In particular, this allows distributing the $\ottstype{!}_{\ottsmode{m}}$ modality over $\ottstype{\otimes}$, which is not part of Girard's intuitionistic linear logic, but is included in \cite{bernardy_linear_2018} and referred to as \emph{deep} modes in \cite{lorenzen_oxidizing_2024}.

\paragraph{Rules for Scoping}

As destinations always exist in the context of a structure with holes, and must stay in that context, we need a formal notion of \emph{scope}. Scopes are created by \rref*{\CTyTerm\CSep\CUpdA}, as destinations are only ever accessed through $\ottkw{upd}_{\ottkw{\ltimes} }$. More precisely, $ \ottkw{upd}_{\ottkw{\ltimes} }\, \ottnt{t} ~\ottkw{with}~ \ottmv{x} \, \pmb{\mapsto} \ottnt{t'} $ creates a new scope which spans over $\ottnt{t'}$. In that scope, $\ottmv{x}$ has age $ \ottsmode{\nu} $ (now), and the ages of the existing bindings in $\Gamma_{{\mathrm{2}}}$ are multiplied by $ \ottsmode{\uparrow} $ (i.e. we add $1$ to ages seen as a numbers). That is represented by $\ottnt{t'}$ typing in $    \ottsmode{1}  \hspace{-0.15ex}  \ottsmode{\uparrow}   \hspace{-0.3ex}\cdot\hspace{-0.3ex} \Gamma_{{\mathrm{2}}}  ,~  \ottmv{x} :\!_{\!   \ottsmode{1}  \hspace{-0.15ex}  \ottsmode{\nu}   } \ottstype{T}  $ while the parent term $ \ottkw{upd}_{\ottkw{\ltimes} }\, \ottnt{t} ~\ottkw{with}~ \ottmv{x} \, \pmb{\mapsto} \ottnt{t'} $ types in unscaled contexts $\Gamma_{{\mathrm{1}}}  +  \Gamma_{{\mathrm{2}}}$. This difference of age between $\ottmv{x}$ --- introduced by $\ottkw{upd}_{\ottkw{\ltimes} }$, containing destinations --- and $\Gamma_{{\mathrm{2}}}$ lets us see what originates from older scopes. Specifically, distinguishing the age of destinations is crucial when typing filling primitives to avoid the pitfalls of \cref{sec:scope-escape-dests}.

\begin{figure}[t]
  \scalebox{0.85}{\begin{tikzpicture}
	\begin{pgfonlayer}{nodelayer}
		\node [style=code] (173) at (11, 7.5) {\scz{$\mathsfbf{upd_{\ltimes} }$}};
		\node [style=none] (174) at (13.25, 8) {};
		\node [style=none] (175) at (13, 7.5) {};
		\node [style=none] (176) at (13.25, 7) {};
		\node [style=code] (181) at (17, 7.25) {\scz{$\!\mathsf{\bbcomma}$}};
		\node [style=none] (182) at (17.5, 7.875) {};
		\node [style=none] (183) at (21.875, 7.875) {};
		\node [style=none] (184) at (17.5, 7.125) {};
		\node [style=none] (185) at (21.875, 7.125) {};
		\node [style=none] (186) at (22, 8) {};
		\node [style=none] (187) at (22.25, 7.5) {};
		\node [style=none] (188) at (22, 7) {};
		\node [style=code] (189) at (22.25, 7.5) {\scz{$\mathsfbf{with}$}};
		\node [style=code] (190) at (24.25, 7.5) {\sci{$\mathit{d_1}\,\pmb{\mapsto}$}};
		\node [style=none] (191) at (24.25, 8.625) {};
		\node [style=none] (192) at (24.25, 6.625) {};
		\node [style=none] (193) at (11, 6.625) {};
		\node [style=code] (194) at (13.25, 4.5) {\sci{$\mathit{d_{11}}\!\blacktriangleleft$}};
		\node [style=none] (195) at (15.5, 3.75) {};
		\node [style=none] (196) at (15.25, 3.25) {};
		\node [style=none] (197) at (15.5, 2.75) {};
		\node [style=code] (202) at (17.5, 3) {\sci{$\!\mathsf{\bbcomma}$}};
		\node [style=none] (203) at (18.125, 3.625) {};
		\node [style=none] (204) at (19.875, 3.625) {};
		\node [style=none] (205) at (18.125, 2.875) {};
		\node [style=none] (206) at (19.875, 2.875) {};
		\node [style=none] (207) at (20, 3.75) {};
		\node [style=none] (208) at (20.25, 3.25) {};
		\node [style=none] (209) at (20, 2.75) {};
		\node [style=code] (210) at (13.25, 3.25) {\sci{$\mathsfbf{upd_{\ltimes} }$}};
		\node [style=code] (211) at (20.25, 3.25) {\sci{$\mathsfbf{with}$}};
		\node [style=code] (212) at (22.5, 3.25) {\scii{$\mathit{d_2}\,\pmb{\mapsto}$}};
		\node [style=none] (213) at (22.25, 4.375) {};
		\node [style=none] (214) at (22.25, 2.375) {};
		\node [style=none] (215) at (11, 2.375) {};
		\node [style=none] (220) at (15.25, 4.875) {};
		\node [style=none] (221) at (16.125, 4.875) {};
		\node [style=none] (222) at (15.25, 4.125) {};
		\node [style=none] (223) at (16.125, 4.125) {};
		\node [style=code] (224) at (16, 4.5) {\sci{$\patu$}};
		\node [style=code] (225) at (15, 4.5) {\scz{$~\mathit{x_0}$}};
		\node [style=none] (226) at (15.225, 7) {};
		\node [style=none] (227) at (15.75, 5) {};
		\node [style=none] (228) at (16.375, 2.75) {};
		\node [style=none] (229) at (23, 3.75) {};
		\node [style=none] (230) at (24.825, 8) {};
		\node [style=none] (231) at (30.5, 3.875) {};
		\node [style=none] (233) at (37.5, 8.625) {};
		\node [style=none] (234) at (37.5, 4.375) {};
		\node [style=none] (235) at (30.5, 8.625) {};
		\node [style=none] (236) at (30.5, 4.375) {};
		\node [style=none] (239) at (30.5, 7.875) {};
		\node [style=none] (240) at (28.5, 7.875) {};
		\node [style=none] (241) at (28.5, 8.625) {};
		\node [style=none] (242) at (30.5, 3.625) {};
		\node [style=none] (243) at (28.5, 4.375) {};
		\node [style=none] (244) at (28.5, 3.625) {};
		\node [style=scopename] (245) at (28.5, 8.25) {\sci{scope 1}};
		\node [style=scopename] (246) at (28.5, 4) {\scii{scope 2}};
		\node [style=smallannot] (247) at (23.6, 9.175) {\sci{stands for}};
		\node [style=smallannot] (248) at (21.825, 4.825) {\scii{stands for}};
		\node [style=smallannot] (249) at (14.475, 6.85) {\scz{fill}};
		\node [style=code] (250) at (32.25, 8) {\scz{$\mathit{x_0}\,:\,\mathsf{1\hspace{-0.15ex}\!\uparrow}$}};
		\node [style=code] (251) at (32.25, 7.25) {\sci{$\mathit{d_{1}}\,:\,\mathsf{1\hspace{-0.15ex}\nu}$}};
		\node [style=code] (252) at (32.25, 0.75) {\scii{$\mathit{d_2}\,:\,\mathsf{1\hspace{-0.15ex}\nu}$}};
		\node [style=none] (253) at (16.925, 2.025) {};
		\node [style=none] (254) at (17.15, 2.75) {};
		\node [style=smallannot] (255) at (16.45, 2.55) {\sci{fill}};
		\node [style=none] (256) at (15.75, 7.5) {\scz{$(\fbox{$\mathit{h_{11}}$},\fbox{$\mathit{h_{12}}$})$}};
		\node [style=none] (257) at (13.75, 7.5) {\scz{$\mathsf{Inl}$}};
		\node [style=none] (258) at (13.125, 8.5) {\scz{$\mathsfbf{let}~\mathit{x_0}=\mathsf{()}~\mathsfbf{in}$}};
		\node [style=code] (259) at (32.25, 9.5) {\scz{$\mathit{x_0}\,:\,\mathsf{1\hspace{-0.15ex}\nu}$}};
		\node [style=none] (260) at (19.625, 7.5) {\sci{$(\to\!h_{11}, \to\!h_{12})$}};
		\node [style=none] (261) at (16.75, 5.75) {\sci{$\mathsfbf{case}~\mathit{d_1}~\mathsfbf{of}~\mathsf{(}\mathit{d_{11}},\mathit{d_{12}}\mathsf{)}\pmb{\mapsto}$}};
		\node [style=none] (262) at (16, 3.25) {\sci{$\mathsf{Inr}$}};
		\node [style=none] (263) at (17, 3.25) {\sci{\fbox{$\mathit{h_2}$}}};
		\node [style=none] (264) at (19, 3.25) {\scii{$\to\!\mathit{h_2}$}};
		\node [style=code] (265) at (14.75, 1.5) {\scii{$\mathit{d_{2}}\!\blacktriangleleft$}};
		\node [style=none] (266) at (16.5, 1.875) {};
		\node [style=none] (267) at (17.5, 1.875) {};
		\node [style=none] (268) at (16.5, 1.125) {};
		\node [style=none] (269) at (17.5, 1.125) {};
		\node [style=code] (270) at (17.375, 1.5) {\scii{$\patu$}};
		\node [style=code] (271) at (16.125, 1.5) {\sci{$~\mathit{d_{12}}$}};
		\node [style=none] (272) at (30.5, 9.875) {};
		\node [style=none] (273) at (30.5, 9.125) {};
		\node [style=none] (274) at (28.5, 9.125) {};
		\node [style=none] (275) at (28.5, 9.875) {};
		\node [style=scopename] (276) at (28.5, 9.5) {\scz{scope 0}};
		\node [style=code] (277) at (32.25, 6.5) {\sci{$\mathit{d_{11}}\,:\,\mathsf{1\hspace{-0.15ex}\nu}$}};
		\node [style=code] (278) at (32.25, 5.75) {\sci{$\mathit{d_{12}}\,:\,\mathsf{1\hspace{-0.15ex}\nu}$}};
		\node [style=code] (279) at (32.25, 1.5) {\sci{$\mathit{d_{12}}\,:\,\mathsf{1\hspace{-0.15ex}\!\uparrow}$}};
		\node [style=code] (280) at (32.25, 3.75) {\sczf{$\mathit{x_0}\,:\,\mathsf{1\hspace{-0.15ex}\!\uparrow^2}$}};
		\node [style=code] (281) at (32.25, 3) {\scif{$\mathit{d_{1}}\,:\,\mathsf{1\hspace{-0.15ex}\!\uparrow}$}};
		\node [style=code] (282) at (32.25, 2.25) {\scif{$\mathit{d_{11}}\,:\,\mathsf{1\hspace{-0.15ex}\!\uparrow}$}};
		\node [style=none] (283) at (30.5, 0.25) {};
		\node [style=none] (284) at (36.25, 3.75) {\sczf{\small (used)}};
		\node [style=none] (285) at (36.25, 3) {\scif{\small (used)}};
		\node [style=none] (286) at (36.25, 2.25) {\scif{\small (used)}};
		\node [style=none] (287) at (21.25, 8.025) {};
		\node [style=none] (288) at (19.425, 3.725) {};
	\end{pgfonlayer}
	\begin{pgfonlayer}{edgelayer}
		\draw [style=s0s] (174.center) to (175.center);
		\draw [style=s0s] (175.center) to (176.center);
		\draw [style=s0s] (186.center) to (187.center);
		\draw [style=s0s] (187.center) to (188.center);
		\draw [style=s1d, bend left=15] (182.center) to (183.center);
		\draw [style=s1d, bend right=15] (182.center) to (184.center);
		\draw [style=s1d, bend right=15] (184.center) to (185.center);
		\draw [style=s1d, bend left=15] (183.center) to (185.center);
		\draw [style=s1s] (195.center) to (196.center);
		\draw [style=s1s] (196.center) to (197.center);
		\draw [style=s1s] (207.center) to (208.center);
		\draw [style=s1s] (208.center) to (209.center);
		\draw [style=s2d, bend left=15] (203.center) to (204.center);
		\draw [style=s2d, bend left=15] (204.center) to (206.center);
		\draw [style=s2d, bend left=15] (206.center) to (205.center);
		\draw [style=s2d, bend left=15] (205.center) to (203.center);
		\draw [style=s0d, bend left=15] (220.center) to (221.center);
		\draw [style=s0d, in=105, out=-105] (220.center) to (222.center);
		\draw [style=s0d, bend right=15] (222.center) to (223.center);
		\draw [style=s0d, bend right=15] (223.center) to (221.center);
		\draw [style=a0, in=-90, out=90, looseness=1.25] (227.center) to (226.center);
		\draw [style=a2, in=75, out=105] (229.center) to (288.center);
		\draw [style=a1, in=75, out=105] (230.center) to (287.center);
		\draw [style=s2s] (243.center) to (244.center);
		\draw [style=s2s] (244.center) to (242.center);
		\draw [style=s1s] (241.center) to (240.center);
		\draw [style=s1s] (240.center) to (239.center);
		\draw [style=a1, in=-90, out=90] (253.center) to (254.center);
		\draw [style=s1d, bend left=15] (266.center) to (267.center);
		\draw [style=s1d, in=105, out=-105, looseness=1.25] (266.center) to (268.center);
		\draw [style=s1d, bend right=15] (268.center) to (269.center);
		\draw [style=s1d, in=285, out=75, looseness=1.25] (269.center) to (267.center);
		\draw [style=s1s] (235.center) to (239.center);
		\draw [style=s2s] (236.center) to (242.center);
		\draw [style=s0s] (275.center) to (274.center);
		\draw [style=s0s] (274.center) to (273.center);
		\draw [style=s0s] (272.center) to (273.center);
		\draw [style=s0s] (275.center) to (272.center);
		\draw [style=s1d] (193.center) to (192.center);
		\draw [style=s1d] (191.center) to (192.center);
		\draw [style=s1d] (191.center) to (235.center);
		\draw [style=s1d] (235.center) to (233.center);
		\draw [style=s2d] (215.center) to (214.center);
		\draw [style=s2d] (214.center) to (213.center);
		\draw [style=s2d] (213.center) to (236.center);
		\draw [style=s2d] (236.center) to (234.center);
		\draw [style=s0d] (273.center) to (235.center);
		\draw [style=s1d] (239.center) to (236.center);
		\draw [style=s2d] (242.center) to (283.center);
	\end{pgfonlayer}
\end{tikzpicture}}
  \vspace*{-0.5cm}
  \caption{Scope rules for $\ottkw{upd}_{\ottkw{\ltimes} }$ in \destcalculus{}}
  \label{fig:scope-rules}
\end{figure}

\Cref{fig:scope-rules} illustrates scopes introduced by $\ottkw{upd}_{\ottkw{\ltimes} }$, and how the typing rules for $\ottkw{upd}_{\ottkw{\ltimes} }$ and $\blacktriangleleft$ interact. Anticipating \cref{ssec:runtime-values}, ampar values are pairs with a structure with holes on the left, and destinations on the right. With $\ottkw{upd}_{\ottkw{\ltimes} }$ we enter a new scope where the destinations are accessible, but the structure with holes remains in the outer scope. As a result, when filling a destination with \rref*{\CTyTerm\CSep\CFillLeaf}, for instance $ \ottmv{d_{{\mathrm{11}}}} \blacktriangleleft \ottmv{x_{{\mathrm{0}}}} $ in~\cref{fig:scope-rules}, we type $\ottmv{d_{{\mathrm{11}}}}$ in the new scope, while we type $\ottmv{x_{{\mathrm{0}}}}$ in the outer scope, as it’s being moved to the structure with holes on the left of the ampar, which lives in the outer scope too. This is the opposite of the scaling that $\ottkw{upd}_{\ottkw{\ltimes} }$ does: while $\ottkw{upd}_{\ottkw{\ltimes} }$ creates a new scope for its body, operator $\blacktriangleleft$, and similarly, $\triangleleft\mycirc$ and $ \triangleleft (\lamnt{\ottmv{x}}{\ottsmode{m}}{\ottnt{u}})$\footnote{We chose the form $ \triangleleft (\lamnt{\ottmv{x}}{\ottsmode{m}}{\ottnt{u}})$ for function creation so that any data can be built through piecemeal destination filling}, transfer their right operand to the outer scope. In other words, the right-hand side of $ \blacktriangleleft $ or $ \triangleleft $ is an enclave for the parent scope.

When using $\ottkw{from}_{\ottkw{\ltimes} }'$ (rule \rref*{\CTySTerm\CSep\CFromA'}), the left of an ampar is extracted to the current scope: this is the fundamental reason why the left of an ampar has to ``take place'' in the current scope. We know the structure is complete and can be extracted because the right-hand side is of type unit ($ \ottstype{1} $), and thus no destination on the right-hand side means no hole can remain on the left. $\ottkw{from}_{\ottkw{\ltimes} }'$ is implemented in terms of $\ottkw{from}_{\ottkw{\ltimes} }$ in~\cref{fig:sterm} to keep the core calculus tidier (and limit the number of typing rules, evaluation contexts, etc), but it can be implemented much more efficiently in a real-world implementation.

When an ampar is eliminated with the more general $\ottkw{from}_{\ottkw{\ltimes} }$ in rule \rref*{\CTyTerm\CSep\CFromA} however, we extract both sides of the ampar to the current scope, even though the right-hand side is normally in a different scope. This is only safe to do because the right-hand side is required to have type $ \ottstype{!}_{   \ottsmode{1}  \hspace{-0.15ex}  \ottsmode{\infty}   } \ottstype{T} $, which means it is scope-insensitive: it can't contain any scope-controlled resource. This also ensures that the right-hand side cannot contain destinations, so the structure is ready to be read.

In \rref*{\CTyTerm\CSep\CToA}, on the other hand, there is no need to bother with scopes: the operator $\ottkw{to}_{\ottkw{\ltimes} }$ embeds an already completed structure in an ampar whose left side is the structure (that continues to type in the current scope), and right-hand side is unit.

The remaining operators $ \triangleleft  \ottsctor{()} ,  \triangleleft  \ottsctor{Inl} ,  \triangleleft  \ottsctor{Inr} ,  \triangleleft \,\expcons{\ottsmode{m}},  \triangleleft  \ottsctor{({,})} $ from rules \IfFancyRuleNames{of the form \textsc{\CTyTerm\CSep$\ottstype{\lfloor}~\ottstype{\rfloor}$E}}{\textsc{Ty-term-Fill$*$}} are the other destination-filling primitives. They write a hollow constructor to the hole pointed by the destination operand, and return the potential new destinations that are created for new holes in the hollow constructor (or unit if there is none).

\section{Operational Semantics}\label{sec:ectxs-sem}

Before we define the operational semantics of \destcalculus{} we need to introduce a few more concepts. We'll need commands $ \ottnt{E} \biggerbrackl  \ottnt{t}  \biggerbrackr $, they're described in \cref{ssec:ty-ectxs-cmd}; and we'll need runtime values (we'll often just say \emph{values}), described in \cref{ssec:runtime-values}. Indeed, the terms of \destcalculus{} lack any way to represent destinations or holes, or really any kind of value (for instance $\ottsctor{Inl} \, \ottsctor{()}$ has been, so far, just syntactic sugar for a term $ \ottkw{from}_{\ottkw{\ltimes} }'   (  \ottkw{upd}_{\ottkw{\ltimes} }\,  \ottkw{new}_{\ottkw{\ltimes} }  ~\ottkw{with}~ \ottmv{d} \, \pmb{\mapsto}  \ldots   )  $). It's a peculiarity of \destcalculus{} that values (in particular, data constructors) only exist during the reduction; usually they are part of the term syntax of functional languages. We also extend the type system to cover commands and values, so as to be able to state and prove type safety theorems.

\subsection{Runtime Values and New Typing Context Forms}\label{ssec:runtime-values}

\begin{ottfig}{\caption{Runtime values and new typing context forms}\label{fig:grammar-val}\label{fig:ty-val}}
\hspace*{-0.1\linewidth}\begin{minipage}{\linewidth}\sidebysidecodehere{c}{0.39}{\begin{minipage}{\linewidth}
\begin{minipage}{\linewidth}\small\textit{Grammar extended with values:}\end{minipage}

\bigskip

$\setlength{\arraycolsep}{0.6ex}\newlength\myskip\setlength{\myskip}{2.38ex}\!\begin{array}{rrl}
  \ottnt{t}, \ottnt{u} &\grammdef& \ldots \grammsep \ottnt{v} \\
  \\
       \ottnt{v} &\grammdef&  \ottshname{\hboxed{ \ottshname{h} } }  \hspace*{\widthof{$ _{ \ottshname{H} \!}\ottsctor{\langle} \ottnt{v_{{\mathrm{2}}}} \,\ottsctor{\bbcomma}~ \ottnt{v_{{\mathrm{1}}}} \ottsctor{\rangle} $}-\widthof{$ \ottshname{\hboxed{ \ottshname{h} } } $}}\quad\quad\textit{(hole)} \\
             &|\,&  \ottshname{\destminus} \ottshname{h}  \hspace*{\widthof{$ _{ \ottshname{H} \!}\ottsctor{\langle} \ottnt{v_{{\mathrm{2}}}} \,\ottsctor{\bbcomma}~ \ottnt{v_{{\mathrm{1}}}} \ottsctor{\rangle} $}-\widthof{$ \ottshname{\destminus} \ottshname{h} $}}\quad\quad\textit{(destination)} \\
             &|\,&  _{ \ottshname{H} \!}\ottsctor{\langle} \ottnt{v_{{\mathrm{2}}}} \,\ottsctor{\bbcomma}~ \ottnt{v_{{\mathrm{1}}}} \ottsctor{\rangle}  \quad\quad\textit{(ampar value)} \\
             &|\,&  \ottsctor{()}  \grammsep  \lamvnt{ \ottmv{x} }{ \ottsmode{m} }{ \ottnt{u} }  \grammsep \ottsctor{Inl} \, \ottnt{v} \\
             &|\,& \ottsctor{Inr} \, \ottnt{v} \grammsep  \expcons{ \ottsmode{m} } \ottnt{v}  \grammsep  \ottsctor{(} \ottnt{v_{{\mathrm{1}}}} \,\ottsctor{,}~ \ottnt{v_{{\mathrm{2}}}} \ottsctor{)}  \\
\end{array}$

\bigskip\bigskip
\hrule
\bigskip

\begin{minipage}{\linewidth}\small\textit{Typing values as terms:}\end{minipage}

\smallskip

\[\bgroup\SetPrefix{\CTyTerm\CSep}
  \drule{Ty-term-Val}\egroup
\]

\bigskip

\end{minipage}}{\hspace*{-0.1cm}\begin{minipage}{\linewidth}

\begin{minipage}{\linewidth}\small\textit{Extended grammar of typing contexts:}\end{minipage}

\smallskip

$\setlength{\myskip}{2.38ex}\!\begin{array}{rrlcccc}
\Delta &\grammdef&  \smallbullet  \grammsep  \ottshname{\destminus} \ottshname{h} :\!_{\! \ottsmode{m} }\ottstype{\lfloor}\,\!_{ \ottsmode{n} } \ottstype{T} \ottstype{\rfloor}  &|&  \Delta_{{\mathrm{1}}} ,~ \Delta_{{\mathrm{2}}}  \\
\Gamma &\grammdef&  \smallbullet  \grammsep  \ottshname{\destminus} \ottshname{h} :\!_{\! \ottsmode{m} }\ottstype{\lfloor}\,\!_{ \ottsmode{n} } \ottstype{T} \ottstype{\rfloor}  \grammsep  \ottmv{x} :\!_{\! \ottsmode{m} } \ottstype{T}  &|&  \Gamma_{{\mathrm{1}}} ,~ \Gamma_{{\mathrm{2}}}  \\
\hskip \myskip \Theta &\grammdef&  \smallbullet  \grammsep  \ottshname{\destminus} \ottshname{h} :\!_{\! \ottsmode{m} }\ottstype{\lfloor}\,\!_{ \ottsmode{n} } \ottstype{T} \ottstype{\rfloor}  \grammsep  \ottshname{\hboxed{ \ottshname{h} } }:\!_{\! \ottsmode{n} } \ottstype{T}  &|&  \Theta_{{\mathrm{1}}} ,~ \Theta_{{\mathrm{2}}}  
\end{array}$

\bigskip

\begin{minipage}{\linewidth}\small\textit{Operations extended to new typing context forms:}\end{minipage}

\smallskip

{\small
\bgroup
\renewcommand\tabcolsep{2pt}
\begin{tabular}[c]{rclcc}
  $\ottsmode{n'}$ &$\cdot$& $ (   \ottshname{\hboxed{ \ottshname{h} } }:\!_{\! \ottsmode{n} } \ottstype{T}  ,~ \Theta  ) $ & $\btriangleq$ & $  (  \ottshname{\hboxed{ \ottshname{h} } }:\!_{\! \ottsmode{n'}  \ottsmode{\hspace{-0.1ex}\cdot\hspace{-0.1ex} }  \ottsmode{n} } \ottstype{T}  )  ,~  \ottsmode{n'} \hspace{-0.3ex}\cdot\hspace{-0.3ex} \Theta  $\\
  $\ottsmode{n'}$ &$\cdot$& $ (   \ottshname{\destminus} \ottshname{h} :\!_{\! \ottsmode{m} }\ottstype{\lfloor}\,\!_{ \ottsmode{n} } \ottstype{T} \ottstype{\rfloor}  ,~ \Gamma  ) $ & $\btriangleq$ & $  (  \ottshname{\destminus} \ottshname{h} :\!_{\! \ottsmode{n'}  \ottsmode{\hspace{-0.1ex}\cdot\hspace{-0.1ex} }  \ottsmode{m} }\ottstype{\lfloor}\,\!_{ \ottsmode{n} } \ottstype{T} \ottstype{\rfloor}  )  ,~  \ottsmode{n'} \hspace{-0.3ex}\cdot\hspace{-0.3ex} \Gamma  $ ~$\phantom{.}^{\dagger}$\\
\end{tabular}

\bigskip

\begin{tabular}[c]{rclcc@{\quad}l}
  $ (   \ottshname{\hboxed{ \ottshname{h} } }:\!_{\! \ottsmode{n} } \ottstype{T}  ,~ \Theta_{{\mathrm{1}}}  ) $ &$+$& $\Theta_{{\mathrm{2}}}$ & $\btriangleq$ & $  \ottshname{\hboxed{ \ottshname{h} } }:\!_{\! \ottsmode{n} } \ottstype{T}  ,~  ( \Theta_{{\mathrm{1}}}  +  \Theta_{{\mathrm{2}}} )  $ & \textrm{if $\ottshname{h}\notin\Theta_{{\mathrm{2}}}$}\\
  $ (   \ottshname{\hboxed{ \ottshname{h} } }:\!_{\! \ottsmode{n} } \ottstype{T}  ,~ \Theta_{{\mathrm{1}}}  ) $ &$+$& $ (   \ottshname{\hboxed{ \ottshname{h} } }:\!_{\! \ottsmode{n'} } \ottstype{T}  ,~ \Theta_{{\mathrm{2}}}  ) $ & $\btriangleq$ & $  \ottshname{\hboxed{ \ottshname{h} } }:\!_{\!  \ottsmode{ \ottsmode{n} + \ottsmode{n'} }  } \ottstype{T}  ,~  ( \Theta_{{\mathrm{1}}}  +  \Theta_{{\mathrm{2}}} )  $\\
\end{tabular}

\smallskip 

\begin{tabular}[c]{rcl}
  $ (   \ottshname{\destminus} \ottshname{h} :\!_{\! \ottsmode{m} }\ottstype{\lfloor}\,\!_{ \ottsmode{n} } \ottstype{T} \ottstype{\rfloor}  ,~ \Gamma_{{\mathrm{1}}}  ) $ &$+$& $\Gamma_{{\mathrm{2}}}$  $\btriangleq$  $  \ottshname{\destminus} \ottshname{h} :\!_{\! \ottsmode{m} }\ottstype{\lfloor}\,\!_{ \ottsmode{n} } \ottstype{T} \ottstype{\rfloor}  ,~  ( \Gamma_{{\mathrm{1}}}  +  \Gamma_{{\mathrm{2}}} )  $  \quad\textrm{if $\ottshname{h}\notin\Gamma_{{\mathrm{2}}}$}~$\phantom{.}^{\dagger}$\\
  $ (   \ottshname{\destminus} \ottshname{h} :\!_{\! \ottsmode{m} }\ottstype{\lfloor}\,\!_{ \ottsmode{n} } \ottstype{T} \ottstype{\rfloor}  ,~ \Gamma_{{\mathrm{1}}}  ) $ &$+$& $ (   \ottshname{\destminus} \ottshname{h} :\!_{\! \ottsmode{m'} }\ottstype{\lfloor}\,\!_{ \ottsmode{n} } \ottstype{T} \ottstype{\rfloor}  ,~ \Gamma_{{\mathrm{2}}}  ) $  $\btriangleq$  $  \ottshname{\destminus} \ottshname{h} :\!_{\!  \ottsmode{ \ottsmode{m} + \ottsmode{m'} }  }\ottstype{\lfloor}\,\!_{ \ottsmode{n} } \ottstype{T} \ottstype{\rfloor}  ,~  ( \Gamma_{{\mathrm{1}}}  +  \Gamma_{{\mathrm{2}}} )  $$\phantom{.}^{\dagger}$\\
\end{tabular}
\egroup

\smallskip

\smallskip

\begin{tabular}[c]{rcl}
  $ \ottshname{\destminus^{\scriptscriptstyle\text{-}1} }  (  \smallbullet  )  $ &$\btriangleq$&  $ \smallbullet $\\
  $ \ottshname{\destminus^{\scriptscriptstyle\text{-}1} }  (   \ottshname{\destminus} \ottshname{h} :\!_{\!   \ottsmode{1}  \hspace{-0.15ex}  \ottsmode{\nu}   }\ottstype{\lfloor}\,\!_{ \ottsmode{n} } \ottstype{T} \ottstype{\rfloor}  ,~ \Delta  )  $ &$\btriangleq$& $  (  \ottshname{\hboxed{ \ottshname{h} } }:\!_{\! \ottsmode{n} } \ottstype{T}  )  ,~  \ottshname{\destminus^{\scriptscriptstyle\text{-}1} }  ( \Delta )   $ \\
\end{tabular}

\begin{center}$\phantom{.}^{\dagger}$ : \textit{same rule is also true for $\Theta$ or $\Delta$ replacing $\Gamma$}\end{center}

}\end{minipage}}
\end{minipage}

\bigskip
\hrule
\bigskip

  \bgroup\SetPrefix{\CTyVal\CSep}
  \ottdefnTyXXval{}\egroup
\end{ottfig}

The syntax of runtime values is given in \cref{fig:grammar-val}. It features constructors for all of our basic types, as well as functions (note that in $ \lamvnt{ \ottmv{x} }{ \ottsmode{m} }{ \ottnt{u} } $, $\ottnt{u}$ is a term, not a value). The more interesting values are holes $ \ottshname{\hboxed{ \ottshname{h} } } $, destinations $ \ottshname{\destminus} \ottshname{h} $, and ampars $ _{ \ottshname{H} \!}\ottsctor{\langle} \ottnt{v_{{\mathrm{2}}}} \,\ottsctor{\bbcomma}~ \ottnt{v_{{\mathrm{1}}}} \ottsctor{\rangle} $, which we'll describe in the rest of the section. In order for the operational semantics to use substitution, which requires substituting variables with values, we also extend the syntax of terms to include values through rule \rref*{\CTyTerm\CSep\CVal}.

Destinations and holes are two faces of the same coin, as seen in~\cref{ssec:build-up-vocab}, and we must ensure that throughout the reduction, a destination always points to a hole, and a hole is always the target of exactly one destination. Thus, the new idea of our system is to feature \emph{hole bindings} $ \ottshname{\hboxed{ \ottshname{h} } }:\!_{\! \ottsmode{n} } \ottstype{T} $ and \emph{destination bindings} $ \ottshname{\destminus} \ottshname{h} :\!_{\! \ottsmode{m} }\ottstype{\lfloor}\,\!_{ \ottsmode{n} } \ottstype{T} \ottstype{\rfloor} $ in typing contexts in addition to the usual variable bindings $ \ottmv{x} :\!_{\! \ottsmode{m} } \ottstype{T} $. In both cases, we call $\ottshname{h}$ a \emph{hole name}. By definition, a context $\Theta$ can contain both destination bindings and hole bindings, but \emph{not a destination binding and a hole binding for the same hole name}.

We extend our previous context operations $+$ and $\cdot$ to act on the new binding forms, as described in \cref{fig:grammar-val}. Context addition is still very partial; for instance, $ (  \ottshname{\hboxed{ \ottshname{h} } }:\!_{\! \ottsmode{n} } \ottstype{T}  )   +   (  \ottshname{\destminus} \ottshname{h} :\!_{\! \ottsmode{m} }\ottstype{\lfloor}\,\!_{ \ottsmode{n'} } \ottstype{T} \ottstype{\rfloor}  ) $ is not defined, as $\ottshname{h}$ is present on both sides but with different binding forms.

One of the main goals of \destcalculus{} is to ensure that a hole value is never read. The type system maintains this invariant by simply not allowing any hole bindings in the context when typing terms (see \cref{fig:grammar-val} for the different type of contexts used in the typing judgment). In fact, the only place where holes are introduced, is the left-hand side $\ottnt{v_{{\mathrm{2}}}}$ in an ampar $ _{ \ottshname{H} \!}\ottsctor{\langle} \ottnt{v_{{\mathrm{2}}}} \,\ottsctor{\bbcomma}~ \ottnt{v_{{\mathrm{1}}}} \ottsctor{\rangle} $, in \rref*{\CTyVal\CSep\CAmpar}.

Specifically, holes come from the operator $\ottshname{\destminus^{\scriptscriptstyle\text{-}1} }$, which represents the matching hole bindings for a set of destination bindings. It's a partial, pointwise operation on typing contexts $\Delta$, as defined in \cref{fig:grammar-val}.
Note that $ \ottshname{\destminus^{\scriptscriptstyle\text{-}1} } \Delta $ is undefined if any destination binding in $\Delta$ has a mode other than $  \ottsmode{1}  \hspace{-0.15ex}  \ottsmode{\nu}  $.

Furthermore, in \rref*{\CTyVal\CSep\CAmpar}, the holes $ \ottshname{\destminus^{\scriptscriptstyle\text{-}1} } \Delta_{{\mathrm{3}}} $ and the corresponding destinations $\Delta_{{\mathrm{3}}}$ are bound together and consequently removed from the ampar's typing context: this is how we ensure that, indeed, there's one destination per hole and one hole per destination. That being said, both sides of the ampar may also contain stored destinations from other scopes, represented by $   \ottsmode{1}  \hspace{-0.15ex}  \ottsmode{\uparrow}   \hspace{-0.3ex}\cdot\hspace{-0.3ex} \Delta_{{\mathrm{1}}} $ and $\Delta_{{\mathrm{2}}}$ in the respective typing contexts of $\ottnt{v_{{\mathrm{1}}}}$ and $\ottnt{v_{{\mathrm{2}}}}$.

Rule \rref*{\CTyVal\CSep\CHole} indicates that a hole must have mode $  \ottsmode{1}  \hspace{-0.15ex}  \ottsmode{\nu}  $ in typing context to be well-typed; in particular mode coercion is not allowed here, and neither is weakening. Only when a hole is behind an exponential, that mode can change to some arbitrary mode $\ottsmode{n}$. The mode of a hole constrains which values can be written to it, e.g. in $ \ottshname{\hboxed{ \ottshname{h} } }:\!_{\! \ottsmode{n} } \ottstype{T}   \!\!\pmb{\phantom{a}^{\scriptscriptstyle \mathrm{v} }\!\!\vdash}\,   \expcons{ \ottsmode{n} }  \ottshname{\hboxed{ \ottshname{h} } }    \pmb{:}   \ottstype{!}_{ \ottsmode{n} } \ottstype{T} $, only a value with mode $\ottsmode{n}$ (more precisely, a value typed in a context of the form $ \ottsmode{n} \hspace{-0.3ex}\cdot\hspace{-0.3ex} \Theta $) can be written to $ \ottshname{\hboxed{ \ottshname{h} } } $.

Surprisingly, in \rref*{\CTyVal\CSep\CDest}, we see that a destination can be typed with any mode $\ottsmode{m}$ coercible to $  \ottsmode{1}  \hspace{-0.15ex}  \ottsmode{\nu}  $. We did this to mimic the rule \rref*{\CTyTerm\CSep\CVar} and make the general modal substitution lemma expressible for \destcalculus{}\footnote{Generally, in modal systems, if $  \ottmv{x} :\!_{\! \ottsmode{m} } \ottstype{T}  ,~ \Gamma   \,\pmb{\vdash}\,  \ottnt{u}  \pmb{:}  \ottstype{U}$ and $\Delta  \,\pmb{\vdash}\,  \ottnt{v}  \pmb{:}  \ottstype{T}$ then $  \ottsmode{m} \hspace{-0.3ex}\cdot\hspace{-0.3ex} \Delta  ,~ \Gamma   \,\pmb{\vdash}\,   \ottnt{u} [  \ottmv{x}  \assigneq  \ottnt{v}  ]   \pmb{:}  \ottstype{U}$~\cite{bernardy_modality_2020}.\\We have $ \ottmv{x} :\!_{\!   \ottsmode{\omega}  \hspace{-0.15ex}  \ottsmode{\infty}   }  \ottstype{\lfloor}\,\!_{\mydestm{ \ottsmode{n} } } \ottstype{T} \ottstype{\rfloor}    \,\pmb{\vdash}\,  \ottsctor{()}  \pmb{:}   \ottstype{1} $ and $ \ottshname{\destminus} \ottshname{h} :\!_{\!   \ottsmode{1}  \hspace{-0.15ex}  \ottsmode{\nu}   }\ottstype{\lfloor}\,\!_{ \ottsmode{n} } \ottstype{T} \ottstype{\rfloor}   \,\pmb{\vdash}\,   \ottshname{\destminus} \ottshname{h}   \pmb{:}   \ottstype{\lfloor}\,\!_{\mydestm{ \ottsmode{n} } } \ottstype{T} \ottstype{\rfloor} $ so $   \ottsmode{\omega}  \hspace{-0.15ex}  \ottsmode{\infty}   \hspace{-0.3ex}\cdot\hspace{-0.3ex}  (  \ottshname{\destminus} \ottshname{h} :\!_{\!   \ottsmode{1}  \hspace{-0.15ex}  \ottsmode{\nu}   }\ottstype{\lfloor}\,\!_{ \ottsmode{n} } \ottstype{T} \ottstype{\rfloor}  )    \,\pmb{\vdash}\,   \ottsctor{()} [  \ottmv{x}  \assigneq   \ottshname{\destminus} \ottshname{h}   ]   \pmb{:}   \ottstype{1} $ should be valid.}. We formally proved however that throughout a well-typed closed program, $\ottsmode{m}$ will never be of multiplicity $ \ottsmode{\omega} $ or age $ \ottsmode{\infty} $ --- a destination is always linear and of finite age --- so mode coercion is never actually used; and we used this result during the formal proof of the substitution lemma to make it quite easier. The other mode $\ottsmode{n}$, appearing in \rref*{\CTyVal\CSep\CDest}, is not the mode of the destination binding; instead it is part of the type $ \ottstype{\lfloor}\,\!_{\mydestm{ \ottsmode{n} } } \ottstype{T} \ottstype{\rfloor} $ and corresponds to the mode of values that we can write to the corresponding $ \ottshname{\hboxed{ \ottshname{h} } } $; so for it no coercion can take place.

\paragraph{Other Salient Points}
We don't distinguish values with holes from fully-defined values at the syntactic level: instead types prevent holes from being read. In particular, while values are typed in contexts $\Theta$ allowing both destination and hole bindings, when using a value as a term in \rref*{\CTyTerm\CSep\CVal}, it's only allowed to have free destinations, but no free holes.

Notice, also, that values can't have free variables, since contexts $\Theta$ only contain hole and destination bindings, no variable binding. That values are closed is a standard feature of denotational semantics or abstract machine semantics. This is true even for function values (\rref*{\CTyVal\CSep\CFun}), which, also is prevented from containing free holes. Since a function's body is unevaluated, it's unclear what it'd mean for a function to contain holes; at the very least it'd complicate our system a lot, and we are unaware of any benefit supporting free holes in functions could bring.

One might wonder how we can represent a curried function $  \lamnt{ \ottmv{x} }{   \ottsmode{1}  \hspace{-0.15ex}  \ottsmode{\nu}   }{  \lamnt{ \ottmv{y} }{   \ottsmode{1}  \hspace{-0.15ex}  \ottsmode{\nu}   }{ \ottmv{x} }  }  ~\ottkw{concat}~ \ottmv{y} $ at the value level, as the inner abstraction captures the free variable $\ottmv{x}$. The answer is that such a function, at value level, is encoded as $ \lamvnt{ \ottmv{x} }{   \ottsmode{1}  \hspace{-0.15ex}  \ottsmode{\nu}   }{  \ottkw{from}_{\ottkw{\ltimes} }'   (  \ottkw{upd}_{\ottkw{\ltimes} }\,  \ottkw{new}_{\ottkw{\ltimes} }  ~\ottkw{with}~ \ottmv{d} \, \pmb{\mapsto}  \ottmv{d} \triangleleft (\lamnt{ \ottmv{y} }{   \ottsmode{1}  \hspace{-0.15ex}  \ottsmode{\nu}   }{  \ottmv{x} ~\ottkw{concat}~ \ottmv{y}  })   )   } $, where the inner closure is not yet in value form. As the form $ \ottmv{d} \triangleleft (\lamnt{ \ottmv{y} }{   \ottsmode{1}  \hspace{-0.15ex}  \ottsmode{\nu}   }{  \ottmv{x} ~\ottkw{concat}~ \ottmv{y}  }) $ is part of term syntax, it's allowed to have free variable $\ottmv{x}$.

\subsection{Evaluation Contexts and Commands}\label{ssec:ectxs}\label{ssec:ty-ectxs-cmd}

The semantics of \destcalculus{} is given using small-step reductions on a pair $ \ottnt{E} \biggerbrackl  \ottnt{t}  \biggerbrackr $ of an evaluation context $\ottnt{E}$, and an (extended) term $\ottnt{t}$ under focus. We call such a pair $ \ottnt{E} \biggerbrackl  \ottnt{t}  \biggerbrackr $ a \emph{command}, borrowing the terminology from~\citet{herbelin_curien_2000}.

The grammar of evaluation contexts is given in~\cref{fig:grammar-ty-ectxs}. An evaluation context $\ottnt{E}$ is the composition of an arbitrary number of focusing components $\ottnt{e}$. We chose to represent evaluation contexts syntactically, taking inspiration from \citet{felleisen_calculi_1987} and subsequent \cite{danvy_refocusing_2004,biernacka_syntactic_2007}. The intuition here is that destination filling only require a very tame notion of state. So tame, in fact, that we can simply represent writing to a hole by a substitution in the evaluation context, instead of using more heavy store semantics. With this choice, focusing and defocusing steps are made explicit in the semantics, resulting in a verbose but simpler proof. It is also easier to derive an abstract machine for the language, should one want to do that.

Consequently, $ \ottnt{E} \biggerbrackl  \ottnt{t}  \biggerbrackr $ is formally a pair (although we use the notation usually reserved for one-hole contexts, to make rules look more familiar). It's important to keep in mind that won't always have a corresponding term (for instance, when $\ottnt{E}$ contains open ampar focusing components).

\begin{ottfig}[p]{\caption{Evaluation contexts and their typing rules}\label{fig:grammar-ty-ectxs}}{\setlength{\arraycolsep}{1ex}
\hspace*{-0.05\linewidth}\begin{minipage}{\linewidth}\codehere{\hspace*{-0.3cm}\begin{minipage}{\linewidth}
\begin{minipage}{\linewidth}\small\textit{Grammar of evaluation contexts:}\end{minipage}

\smallskip

$\!\begin{array}{rrl}
\ottnt{e} &\grammdef&  \ottnt{t'} ~ \raisebox{0.075em}{$\scriptstyle []$}  \grammsep  \raisebox{0.075em}{$\scriptstyle []$} ~ \ottnt{v}  \grammsep  \raisebox{0.075em}{$\scriptstyle []$} \patu \ottnt{u}  \\
      &|\,&  \ottkw{case}_{ \ottsmode{m} }~ \raisebox{0.075em}{$\scriptstyle []$} ~\ottkw{of}~\{\ottsctor{Inl}\, \ottmv{x_{{\mathrm{1}}}} \pmb{\mapsto} \ottnt{u_{{\mathrm{1}}}} \,,~\ottsctor{Inr}\, \ottmv{x_{{\mathrm{2}}}} \pmb{\mapsto} \ottnt{u_{{\mathrm{2}}}} \}  \grammsep  \ottkw{case}_{ \ottsmode{m} }~ \raisebox{0.075em}{$\scriptstyle []$} ~\ottkw{of}~\ottsctor{(} \ottmv{x_{{\mathrm{1}}}} \,\ottsctor{,}~ \ottmv{x_{{\mathrm{2}}}} \ottsctor{)} \pmb{\mapsto} \ottnt{u}  \grammsep  \ottkw{case}_{ \ottsmode{m} }~ \raisebox{0.075em}{$\scriptstyle []$} ~\ottkw{of}~\expcons{ \ottsmode{n} } \ottmv{x} \pmb{\mapsto} \ottnt{u}  \\
      &|\,&  \ottkw{upd}_{\ottkw{\ltimes} }\, \raisebox{0.075em}{$\scriptstyle []$} ~\ottkw{with}~ \ottmv{x} \, \pmb{\mapsto} \ottnt{t'}  \grammsep  \ottkw{to}_{\ottkw{\ltimes} }\, \raisebox{0.075em}{$\scriptstyle []$}  \grammsep  \ottkw{from}_{\ottkw{\ltimes} }\, \raisebox{0.075em}{$\scriptstyle []$}  \grammsep  \raisebox{0.075em}{$\scriptstyle []$} \mathop{\triangleleft\mycirc} \ottnt{t'}  \grammsep  \ottnt{v} \mathop{\triangleleft\mycirc} \raisebox{0.075em}{$\scriptstyle []$}  \grammsep  \raisebox{0.075em}{$\scriptstyle []$} \blacktriangleleft \ottnt{t'}  \grammsep  \ottnt{v} \blacktriangleleft \raisebox{0.075em}{$\scriptstyle []$}  \\
      &|\,& \raisebox{0.075em}{$\scriptstyle []$}  \triangleleft  \ottsctor{()} \grammsep \raisebox{0.075em}{$\scriptstyle []$}  \triangleleft \, \ottsctor{Inl} \grammsep \raisebox{0.075em}{$\scriptstyle []$}  \triangleleft \, \ottsctor{Inr} \grammsep \raisebox{0.075em}{$\scriptstyle []$}  \triangleleft  \ottsctor{({,})} \grammsep  \raisebox{0.075em}{$\scriptstyle []$} \triangleleft \,\expcons{ \ottsmode{m} }  \grammsep  \raisebox{0.075em}{$\scriptstyle []$} \triangleleft (\lamnt{ \ottmv{x} }{ \ottsmode{m} }{ \ottnt{u} })  \\
      &|\,&  ^{\text{op}\!}_{ \ottshname{H} \!}\ottsctor{\langle} \ottnt{v_{{\mathrm{2}}}} \,\ottsctor{\bbcomma}~ \raisebox{0.075em}{$\scriptstyle []$} \ottsctor{\rangle}  \quad\quad\textit{(open ampar focusing component)} \\
\ottnt{E} &\grammdef&  \raisebox{0.075em}{$\scriptstyle []$}  \grammsep  \ottnt{E} \hspace*{0.4em}\circ\hspace*{0.4em} \ottnt{e}  
\end{array}$\end{minipage}}\end{minipage}

\bigskip
\hrule
\bigskip

\begin{augmentwidth}{1.2cm}
\bgroup\SetPrefix{\CTyEctxs\CSep}
\ottdefnTyXXectxs{}\egroup
\end{augmentwidth}

}\end{ottfig}

Focusing components are all directly derived from the term syntax, except for the \emph{open ampar} component $ ^{\text{op}\!}_{ \ottshname{H} \!}\ottsctor{\langle} \ottnt{v_{{\mathrm{2}}}} \,\ottsctor{\bbcomma}~ \raisebox{0.075em}{$\scriptstyle []$} \ottsctor{\rangle} $. This focusing component indicates that an ampar is currently being processed by $\ottkw{upd}_{\ottkw{\ltimes} }$, with its left-hand side $\ottnt{v_{{\mathrm{2}}}}$ (the structure being built) being attached to the open ampar focusing component, while its right-hand side (containing destinations) is either in subsequent focusing components, or in the term under focus. Ampars being open during the evaluation of $\ottkw{upd}_{\ottkw{\ltimes} }$'s body and closed back afterwards is counterpart to the notion of scopes in typing rules.

Evaluation contexts are typed in a context $\Delta$ that can only contain destination bindings. As we will later see in rule \rref*{\CTyCmd} of \cref{fig:sem}, $\Delta$ is exactly the typing context that the term $\ottnt{t}$ has to use to form a valid $ \ottnt{E} \biggerbrackl  \ottnt{t}  \biggerbrackr $. In other words, while $\Gamma  \,\pmb{\vdash}\,  \ottnt{t}  \pmb{:}  \ottstype{T}$ \emph{requires} the bindings of $\Gamma$, judgment $ \Delta \,\pmb{\dashv}\, \ottnt{E} \pmb{:} \ottstype{T} \ottstype{\rightarrowtail} \ottstype{U_{{\mathrm{0}}}} $ \emph{provides} the bindings of $\Delta$. Typing rules for evaluation contexts are given in~\cref{fig:grammar-ty-ectxs}.

An evaluation context has a context type $\ottstype{T}\ottstype{\rightarrowtail}\ottstype{U_{{\mathrm{0}}}}$. The meaning of $\ottnt{E} \pmb{:}  \ottstype{T}\ottstype{\rightarrowtail}\ottstype{U_{{\mathrm{0}}}}$ is that given $\ottnt{t} \pmb{:} \ottstype{T}$, $ \ottnt{E} \biggerbrackl  \ottnt{t}  \biggerbrackr $ returns a value of type $\ottstype{U_{{\mathrm{0}}}}$. Composing an evaluation context $\ottnt{E} \pmb{:} \ottstype{T}\ottstype{\rightarrowtail}\ottstype{U_{{\mathrm{0}}}}$ with a new focusing component never affects the type $\ottstype{U_{{\mathrm{0}}}}$ of the future command; only the type $\ottstype{T}$ of the focus is altered.

All typing rules for evaluation contexts can be derived systematically from the ones for the corresponding term (except for the rule \rref*{\CTyEctxs\CSep\COpenAmpar} that is a truly new form). Let's take the rule \rref*{\CTyEctxs\CSep\CPatP} as an example:

\medskip

\sidebysidecodehere{t}{0.45}{
\drule{Ty-term-PatP}
}{
\drule{Ty-ectxs-PatP}
}

\medskip

\begin{itemize}
  \item the typing context $  \ottsmode{m} \hspace{-0.3ex}\cdot\hspace{-0.3ex} \Delta_{{\mathrm{1}}}  ,~ \Delta_{{\mathrm{2}}} $ in the premise for $\ottnt{E}$ corresponds to $ \ottsmode{m} \hspace{-0.3ex}\cdot\hspace{-0.3ex} \Gamma_{{\mathrm{1}}}   +  \Gamma_{{\mathrm{2}}}$ in the conclusion of \rref*{\CTyTerm\CSep\CPatP};
  \item the typing context $  \Delta_{{\mathrm{2}}} ,~  \ottmv{x_{{\mathrm{1}}}} :\!_{\! \ottsmode{m} } \ottstype{T_{{\mathrm{1}}}}   ,~  \ottmv{x_{{\mathrm{2}}}} :\!_{\! \ottsmode{m} } \ottstype{T_{{\mathrm{2}}}}  $ in the premise for term $\ottnt{u}$ corresponds to the typing context $  \Gamma_{{\mathrm{2}}} ,~  \ottmv{x_{{\mathrm{1}}}} :\!_{\! \ottsmode{m} } \ottstype{T_{{\mathrm{1}}}}   ,~  \ottmv{x_{{\mathrm{2}}}} :\!_{\! \ottsmode{m} } \ottstype{T_{{\mathrm{2}}}}  $ for the same term in \rref*{\CTyTerm\CSep\CPatP};
  \item the typing context $\Delta_{{\mathrm{1}}}$ in the conclusion for $ \ottnt{E} \hspace*{0.4em}\circ\hspace*{0.4em}   \ottkw{case}_{ \ottsmode{m} }~ \raisebox{0.075em}{$\scriptstyle []$} ~\ottkw{of}~\ottsctor{(} \ottmv{x_{{\mathrm{1}}}} \,\ottsctor{,}~ \ottmv{x_{{\mathrm{2}}}} \ottsctor{)} \pmb{\mapsto} \ottnt{u}   $ corresponds to the typing context $\Gamma_{{\mathrm{1}}}$ in the premise for $\ottnt{t}$ in \rref*{\CTyTerm\CSep\CPatP} (the term $\ottnt{t}$ is located where the focus $ \raisebox{0.075em}{$\scriptstyle []$} $ is in \rref*{\CTyEctxs\CSep\CPatP}).
\end{itemize}

We think of the typing rule for an evaluation context as a rotation of the typing rule for the associated term, where the typing contexts of one subterm and the conclusion are swapped, and the typing contexts of the other potential subterms are kept unchanged (with the difference that typing contexts for evaluation contexts are of shape $\Delta$ instead of $\Gamma$).

\subsection{Small-Step Semantics}\label{ssec:sem}

We equip \destcalculus{} with small-step semantics. There are three sorts of semantic rules:
\begin{itemize}
  \item focus rules, where we focus on a subterm of a term, by pushing a corresponding focusing component on the stack $\ottnt{E}$;
  \item unfocus rules, where the term under focus is in fact a value, and thus we pop a focusing component from the stack $\ottnt{E}$ and transform it back to the corresponding term so that a redex appears (or so that another focus/unfocus rule can be triggered);
  \item reduction rules, where the actual computation logic takes place.
\end{itemize}


\newlength{\tempwidth}
\newcommand{\ifnonempty}[2]{%
  \settowidth{\tempwidth}{#1}%
  \ifthenelse{\lengthtest{\tempwidth < 1ex}}{}{#2}
}

\bgroup
\renewcommand\arraystretch{1.4}
\renewcommand\ottaltinferrule[4]{
  \ensuremath{#4} \ifnonempty{\ensuremath{#3}}{\quad\textit{when}\quad\ensuremath{#3}} \\
}
\bgroup\SetPrefix{\CRed\CSep}
Here the focus, unfocus, and reduction rules for \textsc{PatP}:
{\small\[\begin{array}{ll}\drule{PatP-Focus}
\drule{PatP-Unfocus}
\drule{PatP-Red}
\end{array}\]}
\egroup\egroup

Rules are triggered in a purely deterministic fashion; once a subterm is a value, it cannot be focused on again. Focusing and defocusing rules are entirely mechanical ---they are just a matter of pushing or popping a focusing component on the stack--- so we only present the set of reduction rules for the system in~\cref{fig:sem}\longshort{, but the whole system is included in the annex (\cref{fig:sem-full1,fig:sem-full2})}{ (the rest of the system can be recovered very easily)}.

\begin{ottfig}{\caption{Small-step semantics}\label{fig:sem}}

\hspace*{-0.1\linewidth}\begin{minipage}{\linewidth}\sidebysidecodehere{t}{0.50}{\begin{minipage}{\linewidth}
\bgroup\SetPrefix{\CTyCmd}
\ottdefnTy{}\egroup
\end{minipage}}{\begin{minipage}{\linewidth}
\hspace*{0.5cm}\begin{minipage}{\linewidth}\small\textit{Name set shift and conditional name shift:}\end{minipage}

\bigskip\bigskip

\hspace*{0.5cm}$\!\begin{array}{rcl}
\ottshname{H}  \pluseq  \ottshname{h'} &\btriangleq& \{  \ottshname{h} \hspace*{-.2ex}\ottshname{+}\hspace*{-.2ex} \ottshname{h'} ~|~\ottshname{h}\in \ottshname{H} \}\\
 \ottshname{h} \ottshname{[} \ottshname{H} \pluseq \ottshname{h'} \ottshname{]}  &\btriangleq& \left\{\begin{array}{ll} \ottshname{h} \hspace*{-.2ex}\ottshname{+}\hspace*{-.2ex} \ottshname{h'}  & \text{if}~\ottshname{h}\in\ottshname{H}\\\ottshname{h} & \text{otherwise}\end{array}\right.\end{array}$
\end{minipage}}\end{minipage}

\bigskip
\hrule
\bigskip

\begin{minipage}{\linewidth}\small\textit{Special substitution for open ampars:}\end{minipage}

\smallskip

\hfill$\!\begin{array}{rcll}
    \biggerparenl  \ottnt{E} \hspace*{0.4em}\circ\hspace*{0.4em}  ^{\text{op}\!}_{  \ottsym{\{}  \ottshname{h}  \ottsym{\}} \ottshname{\sqcup}\, \ottshname{H}  \!}\ottsctor{\langle} \ottnt{v_{{\mathrm{2}}}} \,\ottsctor{\bbcomma}~ \raisebox{0.075em}{$\scriptstyle []$} \ottsctor{\rangle}   \biggerparenr  \hspace*{0.11em}\llparenthesis  \ottshname{h} \assigneq_{ \ottshname{H'} }\, \ottnt{v'} \hspace*{0.11em}\rrparenthesis  &=&  \ottnt{E} \hspace*{0.4em}\circ\hspace*{0.4em}  ^{\text{op}\!}_{  \ottshname{H} \ottshname{\sqcup}\, \ottshname{H'}  \!}\ottsctor{\langle}  \ottnt{v_{{\mathrm{2}}}} \hspace*{0.11em}\llparenthesis  \ottshname{h} \assigneq_{ \ottshname{H'} }\, \ottnt{v'} \hspace*{0.11em}\rrparenthesis  \,\ottsctor{\bbcomma}~ \raisebox{0.075em}{$\scriptstyle []$} \ottsctor{\rangle}   &\\
    \biggerparenl  \ottnt{E} \hspace*{0.4em}\circ\hspace*{0.4em} \ottnt{e}  \biggerparenr  \hspace*{0.11em}\llparenthesis  \ottshname{h} \assigneq_{ \ottshname{H'} }\, \ottnt{v'} \hspace*{0.11em}\rrparenthesis  &=&   \ottnt{E} \hspace*{0.11em}\llparenthesis  \ottshname{h} \assigneq_{ \ottshname{H'} }\, \ottnt{v'} \hspace*{0.11em}\rrparenthesis  \hspace*{0.4em}\circ\hspace*{0.4em} \ottnt{e} &\text{if $\ottshname{h} \notin \ottnt{e}$}
\end{array}$\hfill\phantom{.}

\bigskip
\hrule
\bigskip

\bgroup
\renewcommand\arraystretch{1.4}
\renewcommand\ottaltinferrule[4]{
  \ensuremath{#4} & \text{\textsc{#1}} \\
}
\makeatletter
\renewenvironment{drulepar}[3][\relax]
  {\ifx#1\relax\else\def\ottalt@rulesection@prefix{#1-}\fi
  \drulesectionhead{#2}{#3}$\!\!\!\array{ll}}
  {\endarray$}
\makeatother
\bgroup\SetPrefix{\CRed\CSep}
\drules{$ \ottnt{E} ~\big[\, \ottnt{t} \,\big]~ ~\longrightarrow~ ~ \ottnt{E}' ~\big[\, \ottnt{t'} \,\big] $}{Small-step evaluation of commands}{%
App-Red,
PatU-Red,
PatL-Red,
PatR-Red,
PatP-Red,
PatE-Red,
ToA-Red,
FromA-Red,
NewA-Red,
FillU-Red,
FillF-Red,
FillL-Red,
FillR-Red,
FillE-Red,
FillP-Red,
FillComp-Red,
FillLeaf-Red,
Ampar-Open,
Ampar-Close}
\egroup\egroup
\vspace*{-0.5cm}
\[
\text{\textit{where}}\quad\left\{\begin{array}{rcl}
\ottshname{h'} &=&   \ottshname{\mathsfbf{max}(}   \ottshname{\mathsfbf{hnames}(} \ottnt{E} \ottshname{)}  \ottshname{\cup}\, \ottsym{\{}  \ottshname{h}  \ottsym{\}}  \ottshname{)}  \hspace*{-.2ex}\ottshname{+}\hspace*{-.2ex}  \textcolor{hnamecolor}{\mathtt{1} }  \\
\ottshname{h''} &=&   \ottshname{\mathsfbf{max}(}  \ottshname{H} \ottshname{\cup}\,  \ottshname{(}   \ottshname{\mathsfbf{hnames}(} \ottnt{E} \ottshname{)}  \ottshname{\cup}\, \ottsym{\{}  \ottshname{h}  \ottsym{\}}  \ottshname{)}   \ottshname{)}  \hspace*{-.2ex}\ottshname{+}\hspace*{-.2ex}  \textcolor{hnamecolor}{\mathtt{1} }   \\
\ottshname{h'''} &=&   \ottshname{\mathsfbf{max}(}  \ottshname{H} \ottshname{\cup}\,  \ottshname{\mathsfbf{hnames}(} \ottnt{E} \ottshname{)}   \ottshname{)}  \hspace*{-.2ex}\ottshname{+}\hspace*{-.2ex}  \textcolor{hnamecolor}{\mathtt{1} }  
\end{array}\right.\]
\end{ottfig}

Reduction rules for function application, pattern-matching, $\ottkw{to}_{\ottkw{\ltimes} }$ and $\ottkw{from}_{\ottkw{\ltimes} }$ are straightforward.

We introduce a special substitution $ \ottnt{E} \hspace*{0.11em}\llparenthesis  \ottshname{h} \assigneq_{ \ottshname{H} }\, \ottnt{v} \hspace*{0.11em}\rrparenthesis $ that is used to update structures under construction, that are attached to open ampar focusing components in the stack. Such a substitution is triggered when a destination $ \ottshname{\destminus} \ottshname{h} $ is filled in the term under focus, typically in destination-filling primitives reductions, and results in the value $\ottnt{v}$ being written to hole $ \ottshname{\hboxed{ \ottshname{h} } } $. The value $\ottnt{v}$ may contain holes itself (e.g. when the hollow constructor $\ottsctor{Inl} \,  \ottshname{\hboxed{  \ottshname{h'} \hspace*{-.2ex}\ottshname{+}\hspace*{-.2ex}  \textcolor{hnamecolor}{\mathtt{1} }   } } $ is being written to the hole $ \ottshname{\hboxed{ \ottshname{h} } } $ in \rref*{\CFillL\CSep\CRed}), hence the set $\ottshname{H}$ tracks the potential hole names introduced by value $\ottnt{v}$, and is used to update the hole name set of the corresponding (open) ampar. Proper definition of $ \ottnt{E} \hspace*{0.11em}\llparenthesis  \ottshname{h} \assigneq_{ \ottshname{H} }\, \ottnt{v} \hspace*{0.11em}\rrparenthesis $ is given in \cref{fig:sem}.

\rref*{\CFillU\CSep\CRed} and \rref*{\CFillF\CSep\CRed} do not create any new hole; they only write a value to an existing one. On the other hand, rules \rref*{\CFillL\CSep\CRed}, \rref*{\CFillR\CSep\CRed}, \rref*{\CFillE\CSep\CRed} and \rref*{\CFillP\CSep\CRed} all write a hollow constructor to the hole $ \ottshname{\hboxed{ \ottshname{h} } } $ that contains new holes. Thus, we need to generate fresh names for these new holes, and also return a destination for each new hole with a matching name.

The substitution $ \ottnt{E} \hspace*{0.11em}\llparenthesis  \ottshname{h} \assigneq_{ \ottshname{H} }\, \ottnt{v} \hspace*{0.11em}\rrparenthesis $ should only be performed if $\ottshname{h}$ is a globally unique name; otherwise we break the promise of a write-once memory model. To this effect, we allow name shadowing while an ampar is closed, but as soon as an ampar is open, it should have globally unique hole names. This restriction is enforced in rule \rref*{\CTyEctxs\CSep\COpenAmpar} by premise $ \ottshname{\mathsfbf{hnames}(} \ottnt{E} \ottshname{)}  ~\mathtt{\#\#}~  \ottshname{\mathsfbf{hnames}(} \Delta_{{\mathrm{3}}} \ottshname{)} $, requiring hole name sets from $\ottnt{E}$ and $\Delta_{{\mathrm{3}}}$ to be disjoint when an open ampar focusing component is created during reduction of $\ottkw{upd}_{\ottkw{\ltimes} }$. Likewise, any hollow constructor written to a hole should have globally unique hole names. We assume that hole names are natural numbers for simplicity's sake.

To obtain globally fresh names, in the premises of the corresponding rules, we first set\\ $\ottshname{h'} =   \ottshname{\mathsfbf{max}(}   \ottshname{\mathsfbf{hnames}(} \ottnt{E} \ottshname{)}  \ottshname{\cup}\, \ottsym{\{}  \ottshname{h}  \ottsym{\}}  \ottshname{)}  \hspace*{-.2ex}\ottshname{+}\hspace*{-.2ex}  \textcolor{hnamecolor}{\mathtt{1} }  $ or similar definitions for $\ottshname{h''}$ and $\ottshname{h'''}$ (see in \cref{fig:sem}) to find the next unused name. Then we use either the \emph{shifted set} $\ottshname{H}  \pluseq  \ottshname{h'}$ or the \emph{conditional shift operator} $ \ottshname{h} \ottshname{[} \ottshname{H} \pluseq \ottshname{h'} \ottshname{]} $ as defined in \cref{fig:sem} to replace all names or just specific one with fresh unused names.
We extend \emph{conditional shift} $\smallbullet\ottshname{[}\ottshname{H}  \pluseq  \ottshname{h'}\ottshname{]}$ to arbitrary values, terms, and typing contexts in the obvious way (keeping in mind that $ _{ \ottshname{H'} \!}\ottsctor{\langle} \ottnt{v_{{\mathrm{2}}}} \,\ottsctor{\bbcomma}~ \ottnt{v_{{\mathrm{1}}}} \ottsctor{\rangle} $ binds the names in $\ottshname{H'}$).

Rules \rref*{\CAmpar\CSep\COpen} and \rref*{\CAmpar\CSep\CClose} dictate how and when a closed ampar (a value) is converted to an open ampar (a focusing component) and vice-versa, and they make use of the shifting strategy we've just introduced. With \rref*{\CAmpar\CSep\COpen}, the hole names bound by the ampar gets renamed to fresh ones, and the left-hand side gets attached to the focusing component $ ^{\text{op}\!}_{ \ottshname{H}  \pluseq  \ottshname{h'} \!}\ottsctor{\langle}  \ottnt{v_{{\mathrm{2}}}} \ottshname{[} \ottshname{H} \pluseq \ottshname{h'''} \ottshname{]}  \,\ottsctor{\bbcomma}~ \raisebox{0.075em}{$\scriptstyle []$} \ottsctor{\rangle} $ while the right-hand side (containing destinations) is substituted in the body of the $\ottkw{upd}_{\ottkw{\ltimes} }$ statement (which becomes the new term under focus). The rule \rref*{\CAmpar\CSep\CClose} triggers when the body of a $\ottkw{upd}_{\ottkw{\ltimes} }$ statement has reduced to a value. In that case, we can close the ampar, by popping the focusing component from the stack $\ottnt{E}$ and merging back with $\ottnt{v_{{\mathrm{2}}}}$ to form a closed ampar again.

In rule \rref*{\CFillComp\CSep\CRed}, we write the left-hand side $\ottnt{v_{{\mathrm{2}}}}$ of a closed ampar $ _{ \ottshname{H} \!}\ottsctor{\langle} \ottnt{v_{{\mathrm{2}}}} \,\ottsctor{\bbcomma}~ \ottnt{v_{{\mathrm{1}}}} \ottsctor{\rangle} $ to a hole $ \ottshname{\hboxed{ \ottshname{h} } } $ that is part of a structure with holes somewhere inside $\ottnt{E}$. This results in the composition of two structures with holes. Because we dissociate $\ottnt{v_{{\mathrm{2}}}}$ and $\ottnt{v_{{\mathrm{1}}}}$ that were previously bound together by the ampar connective ($\ottnt{v_{{\mathrm{2}}}}$ is merged with another structure, while $\ottnt{v_{{\mathrm{1}}}}$ becomes the new focus), their hole names are no longer bound, so we need to make them globally unique, as we do when an ampar is opened with $\ottkw{upd}_{\ottkw{\ltimes} }$. This renaming is carried out by the conditional shift $ \ottnt{v_{{\mathrm{2}}}} \ottshname{[} \ottshname{H} \pluseq \ottshname{h''} \ottshname{]} $ and $ \ottnt{v_{{\mathrm{1}}}} \ottshname{[} \ottshname{H} \pluseq \ottshname{h''} \ottshname{]} $.

\paragraph{Type Safety} With the semantics now defined, we can state the usual type safety theorems:

\begin{theorem}[Type preservation]\label{thm:preservation}
  If $ \,\pmb{\vdash}\, \ottnt{E} \big[\, \ottnt{t} \,\big] \pmb{:} \ottstype{T} $ and $ \ottnt{E} ~\big[\, \ottnt{t} \,\big]~ ~\longrightarrow~ ~ \ottnt{E}' ~\big[\, \ottnt{t'} \,\big] $ then $ \,\pmb{\vdash}\, \ottnt{E}' \big[\, \ottnt{t'} \,\big] \pmb{:} \ottstype{T} $.
\end{theorem}

\begin{theorem}[Progress]\label{thm:progress}
  If $ \,\pmb{\vdash}\, \ottnt{E} \big[\, \ottnt{t} \,\big] \pmb{:} \ottstype{T} $ and $\forall \ottnt{v},  \ottnt{E} \biggerbrackl  \ottnt{t}  \biggerbrackr  \neq  \raisebox{0.075em}{$\scriptstyle []$} \biggerbrackl  \ottnt{v}  \biggerbrackr $ then $\exists \ottnt{E}', \ottnt{t'}.~ \ottnt{E} ~\big[\, \ottnt{t} \,\big]~ ~\longrightarrow~ ~ \ottnt{E}' ~\big[\, \ottnt{t'} \,\big] $.
\end{theorem}

A command of the form $ \raisebox{0.075em}{$\scriptstyle []$} \biggerbrackl  \ottnt{v}  \biggerbrackr $ cannot be reduced further, as it only contains a fully determined value, and no pending computation. This it is the stopping point of the reduction, and any well-typed command eventually reaches this form.

\section{Formal Proof of Type Safety}\label{sec:formal-proof}

\longshort{

We've proved type preservation and progress theorems with the \href{https://coq.inria.fr/}{Coq proof assistant}. The artifact containing the machine-verified proofs is available at \url{https://doi.org/10.5281/zenodo.14982363}. One can check if a newer version is available using \url{https://doi.org/10.5281/zenodo.14534422} or \url{https://github.com/tweag/destination-calculus}.

Turning to a proof assistant was a pragmatic choice: typing context handling in \destcalculus{} can be quite finicky, and it was hard, without computer assistance, to make sure that we hadn't made mistakes in our proofs. The version of \destcalculus{} that we've proved is written in Ott~\cite{sewell_ott_2007}, the same Ott file is used as a source for this article, making sure that we've proved the same system as we're presenting; though some visual simplification is applied by a script to produce the version in the article.

}{

We've proved type preservation and progress theorems with the Coq proof assistant. Turning to a proof assistant was a pragmatic choice: typing context handling in \destcalculus{} can be quite finicky, and it was hard, without computer assistance, to make sure that we hadn't made mistakes in our proofs. The version of \destcalculus{} that we've proved is written in Ott~\cite{sewell_ott_2007}, the same Ott file is used as a source for this article, making sure that we've proved the same system as we're presenting; though some visual simplification is applied by a script to produce the version in the article.

}

Most of the proof was done by an author with little prior experience with Coq. This goes to show that Coq is reasonably approachable even for non-trivial development. The proof is about 7000 lines long, and contains nearly 500 lemmas. Many of the cases of the type preservation and progress lemmas are similar. To handle such repetitive cases, the use of a large-language-model based autocompletion system has proven quite effective.

The proofs aren't particularly elegant. For instance, we don't have any abstract formalization of semirings: it was more expedient to brute-force the properties we needed by hand. We've observed up to 232 simultaneous goals, but a computer makes short work of this: it was solved by a single call to the \verb|congruence| tactic. Nevertheless there are a few points of interest.

First, we represent contexts as finite-domain functions, rather than as syntactic lists. This works much better when defining sums of context. There are a handful of finite-function libraries in the ecosystem, but we needed finite dependent functions (because the type of binders depend on whether we're binding a variable name or a hole name). This didn't exist, but for our limited purpose, it ended up not being too costly rolling our own (about 1000 lines of proofs). The underlying data type is actual functions: this was simpler to develop, but in exchange equality gets more complex than with a bespoke data type.

Secondly, Addition of context is partial since we can only add two binding of the same name if they also have the same type. Instead of representing addition as a binary function to an optional context, we represent addition as a total function to contexts, but we change contexts to allow faulty bindings on some names. This works well better for our Ott-written rules, at the cost of needing well-formedness preconditions in the premises of typing rules as well as some lemmas.

Finally, to simplify equalities mostly, we assumed a few axioms: functional extensionality, classical logic, and indefinite description:

\begin{verbatim}
Axiom constructive_indefinite_description :
    forall (A : Type) (P : A->Prop), (exists x, P x) -> { x : A | P x }.
\end{verbatim}
This isn't particularly elegant: we could have avoided some of these axioms at the price of more complex development. But for the sake of this article, we decided to favor expediency over elegance.




\section{Implementation of \destcalculus{} Using in-Place Memory Mutations}\label{sec:implementation}

The formal language presented in~\cref{sec:syntax-type-system,sec:ectxs-sem} is not meant to be implemented as-is.

First, \destcalculus{} doesn't have recursion, this would have obscured the presentation of the system. However, adding a standard form of recursion doesn't create any complication.

Secondly, ampars are not managed linearly in \destcalculus{}; only destinations are. That is to say that an ampar can be wrapped in an exponential, e.g. $ \expcons{   \ottsmode{\omega}  \hspace{-0.15ex}  \ottsmode{\nu}   }  _{ \ottsym{\{}  \ottshname{h}  \ottsym{\}} \!}\ottsctor{\langle}    \ottsctor{0}  \,\ottsctor{::}\,  \ottshname{\hboxed{ \ottshname{h} } }    \,\ottsctor{\bbcomma}~  \ottshname{\destminus} \ottshname{h}  \ottsctor{\rangle}  $ (representing a difference list $0 \ottsctor{::} \holesq$ that can be used non-linearly), and then used twice, each time in a different way:

\begin{minipage}{0.50\linewidth}\codehere{ \ottkw{case}_{\mycasem{   \ottsmode{1}  \hspace{-0.15ex}  \ottsmode{\nu}   } }\,   \expcons{   \ottsmode{\omega}  \hspace{-0.15ex}  \ottsmode{\nu}   }  _{ \ottsym{\{}  \ottshname{h}  \ottsym{\}} \!}\ottsctor{\langle}   \ottsctor{0}  \,\ottsctor{::}\,  \ottshname{\hboxed{ \ottshname{h} } }   \,\ottsctor{\bbcomma}~  \ottshname{\destminus} \ottshname{h}  \ottsctor{\rangle}    ~\ottkw{of}~\expcons{   \ottsmode{\omega}  \hspace{-0.15ex}  \ottsmode{\nu}   } \ottmv{x} \, \pmb{\mapsto}   \mynewline   \myspace{1}   \ottkw{let}~ \ottmv{x_{{\mathrm{1}}}} \assigneq  \ottmv{x} ~\ottkw{append}~  \ottsctor{1}   ~\ottkw{in}  \mynewline   \myspace{1}   \ottkw{let}~ \ottmv{x_{{\mathrm{2}}}} \assigneq  \ottmv{x} ~\ottkw{append}~  \ottsctor{2}   ~\ottkw{in}  \mynewline   \myspace{2}   \ottkw{toList}~  (  \ottmv{x_{{\mathrm{1}}}} ~\ottkw{concat}~ \ottmv{x_{{\mathrm{2}}}}  )      }\end{minipage}$ ~\longrightarrow^*~ \quad   \ottsctor{0}  \,\ottsctor{::}\,    \ottsctor{1}  \,\ottsctor{::}\,    \ottsctor{0}  \,\ottsctor{::}\,    \ottsctor{2}  \,\ottsctor{::}\,  \ottsctor{[]}         $\\[\interdefskip]

It may seem counter-intuitive at first, but this program is valid and safe in \destcalculus{}. Thanks to the renaming discipline we detailed in~\cref{ssec:sem}, every time an ampar is operated over with $\ottkw{upd}_{\ottkw{\ltimes} }$, its hole names are renamed to fresh ones. One way we can support this is to allocate a fresh copy of $\ottmv{x}$ every time we call $\ottkw{append}$ (which is implemented in terms of $\ottkw{upd}_{\ottkw{\ltimes} }$), in a copy-on-write fashion. This way filling destinations is still implemented as mutation.

However, this is a long way from the efficient implementation promised in \cref{sec:working-with-dests}. Copy-on-write can be optimized using fully-in-place functional programming ~\cite{lorenzen_fp_2023}, where, thanks to reference counting, we don't need to perform a copy when the difference list isn't aliased.

An alternative is to refine the linear type system further in order to guarantee that ampars are unique and avoid copy-on-write altogether. We held back from doing that in the formalization of \destcalculus{} as it obfuscates the presentation of the system without adding much in return.

To make ampars linear, we follow a recipe proposed by~\citet{spiwack_linearly_2022} and introduce a new type $ \ottstype{Token} $, together with primitives $\ottkw{dup}$ and $\ottkw{drop}$. We also switch $ \ottkw{new}_{\ottkw{\ltimes} } $ for $ \ottkw{new}_{\ottkw{\ltimes IP} } $:

\codehere{\phantom{a}\!\!\!\!\!\!\begin{array}[t]{l}%
\ottkw{dup} ~\pmb{:}~    \ottstype{Token}  \,_{\myfuntm{   \ottsmode{1}  \hspace{-0.15ex}  \ottsmode{\nu}   } }\!\ottstype{\multimap}\,  \ottstype{Token}   \ottstype{\otimes}  \ottstype{Token}  \\[\interdefskip]
\ottkw{drop} ~\pmb{:}~   \ottstype{Token}  \,_{\myfuntm{   \ottsmode{1}  \hspace{-0.15ex}  \ottsmode{\nu}   } }\!\ottstype{\multimap}\,  \ottstype{1}  \\[\interdefskip]
 \ottkw{new}_{\ottkw{\ltimes IP} }  ~\pmb{:}~    \ottstype{Token}  \,_{\myfuntm{   \ottsmode{1}  \hspace{-0.15ex}  \ottsmode{\nu}   } }\!\ottstype{\multimap}\, \ottstype{T}  \,\ottstype{\ltimes}\,  \ottstype{\lfloor}\,\!_{\mydestm{   \ottsmode{1}  \hspace{-0.15ex}  \ottsmode{\nu}   } } \ottstype{T} \ottstype{\rfloor}  
\end{array}}

For the in-place system to work, we consider that a linear root token variable, $\ottmv{tok_{{\mathrm{0}}}}$, is available to a program. ``Closed'' programs can now typecheck in the non-empty context $\{ \ottmv{tok_{{\mathrm{0}}}} :\!_{\!   \ottsmode{1}  \hspace{-0.15ex}  \ottsmode{\infty}   }  \ottstype{Token}  \}$. $\ottmv{tok_{{\mathrm{0}}}}$ can be used to create new tokens $\ottmv{tok_{\ottmv{k}}}$ via $\ottkw{dup}$, but each of these tokens still has to be used linearly.

Ampar produced by $ \ottkw{new}_{\ottkw{\ltimes IP} } $ have a linear dependency on a variable $\ottmv{tok_{\ottmv{k}}}$. If an ampar produced by $  \ottkw{new}_{\ottkw{\ltimes IP} }  ~ \ottmv{tok_{\ottmv{k}}} $ were to be used twice in a block $\ottnt{t}$, then $\ottnt{t}$ would require a typing context $\{ \ottmv{tok_{\ottmv{k}}} :\!_{\!   \ottsmode{\omega}  \hspace{-0.15ex}  \ottsmode{\nu}   }  \ottstype{Token}  \}$, that itself would require $\ottmv{tok_{{\mathrm{0}}}}$ to have multiplicity $ \ottsmode{\omega} $ too. Thus the program would be rejected.

An alternative to having a linear root token variable is to add a primitive function\\$\ottkw{withToken} ~\pmb{:}~   \ottstype{(}   \ottstype{Token}  \,_{\myfuntm{   \ottsmode{1}  \hspace{-0.15ex}  \ottsmode{\infty}   } }\!\ottstype{\multimap}\,  \ottstype{!}_{   \ottsmode{\omega}  \hspace{-0.15ex}  \ottsmode{\infty}   } \ottstype{T}   \ottstype{)}  \,_{\myfuntm{   \ottsmode{1}  \hspace{-0.15ex}  \ottsmode{\nu}   } }\!\ottstype{\multimap}\,  \ottstype{!}_{   \ottsmode{\omega}  \hspace{-0.15ex}  \ottsmode{\infty}   } \ottstype{T}  $ that fulfill the same goal.

Now that ampars are managed linearly, we can change the allocation and renaming mechanisms:
\begin{itemize}
  \item the hole name for a new ampar is chosen fresh right from the start (this corresponds to a new heap allocation);
  \item adding a new hollow constructor still require freshness for its hole names (this corresponds to a new heap allocation too);
  \item Using $\ottkw{upd}_{\ottkw{\ltimes} }$ over an ampar and filling destinations or composing two ampars using $\triangleleft\mycirc$ no longer require any renaming: we have the guarantee that the all the names involved are globally fresh, and can only be used once, so we can do in-place memory updates.
\end{itemize}

\destcalculus{} extended with $ \ottstype{Token} $s and $ \ottkw{new}_{\ottkw{\ltimes IP} } $ is in fact very close to the implementation described in~\cite{bagrel_destination-passing_2024}. Our claim of efficiency is thus based on the results published in the latter and also \cite{lorenzen_searchtree_2024,bour_tmc_2021}, as an hypothetical implementation of \destcalculus{} would mostly resort to the same memory operations -- that is, in-place updates in functional settings.

\paragraph{From Purely Linked Structures to More Efficient Memory Forms}

In \destcalculus{} we only have binary product in sum types. However, it's very straightforward to extend the language and implement destination-based building for n-ary sums of n-ary products, with constructors for each variant having multiple fields directly, instead of each field needing an extra indirection as in the binary sum of products $  \ottstype{1}  \ottstype{\oplus}  \ottstype{(}  \ottstype{S} \ottstype{\otimes}  \ottstype{(}  \ottstype{T} \ottstype{\otimes} \ottstype{U}  \ottstype{)}   \ottstype{)}  $. This is, in fact, already implemented in \cite{bagrel_destination-passing_2024} without any issues. However, it's probably better for field's values to still be represented by pointers.

Indeed, composition of incomplete structures relies on the idea that destinations pointing to holes of a structure $\ottnt{v}$ will still be valid if $\ottnt{v}$ get assigned to a field $\ottmv{f}$ of a bigger structure $\ottnt{v'}$. That's true indeed if just the address of $\ottnt{v}$ is written to $\ottnt{v'}\!.\ottmv{f}$. However, if $\ottnt{v}$ is moved into $\ottnt{v'}$ completly (i.e. if $\ottmv{f}$ is an in-place/unpacked field), then the pointers representing destinations of $\ottnt{v}$ are now invalid.

Our early experiments around DPS support for unpacked fields seem to indicate that we would need two classes of destinations, one supporting composition (for indirected fields) and one disallowing it (for unpacked fields).

\section{Related Work}\label{sec:related-work}

\subsection{Destination-Passing Style for Efficient Memory Management}\label{ssec:shaikhha-dps}

\citet{shaikhha_destination-passing_2017} present a destination-based intermediate language for a functional array programming language, with destination-specific optimizations, that boasts near-C performance.

This is the most comprehensive evidence to date of the benefits of destination-passing style for performance in functional languages, although their work is on array programming, while this article focuses on linked data structures. They can therefore benefit from optimizations that are perhaps less valuable for us, such as allocating one contiguous memory chunk for several arrays.

The main difference between their work and ours is that their language is solely an intermediate language: it would be unsound to program in it manually. We, on the other hand, are proposing a type system to make it sound for the programmer to program directly with destinations.

We see these two aspects as complementing each other: good compiler optimizations are important to alleviate the burden from the programmer and allow high-level abstraction; having the possibility to use destinations in code affords the programmer more control, should they need it.

\subsection{Tail Modulo Constructor}\label{ssec:tmc}

Another example of destinations in a compiler's optimizer is~\cite{bour_tmc_2021}. It's meant to address the perennial problem that the map function on linked lists isn't tail-recursive, hence consumes stack space. The observation is that there's a systematic transformation of functions where the only recursive call is under a constructor to a destination-passing tail-recursive implementation.

Here again, there's no destination in user land, only in the intermediate representation. However, there is a programmatic interface: the programmer annotates a function like
\begin{verbatim}
let[@tail_mod_cons] rec map =
\end{verbatim}
to ask the compiler to perform the translation. The compiler will then throw an error if it can't. This way, contrary to the optimizations in~\cite{shaikhha_destination-passing_2017}, it is entirely predictable.

This has been available in OCaml since version 4.14. This is the one example we know of of destinations built in a production-grade compiler. Our \destcalculus{} makes it possible to express the result tail-modulo-constructor in a typed language. It can be used to write programs directly in that style,  or it could serve as a typed target language for an automatic transformation. On the flip-side, tail modulo constructor is too weak to handle our difference lists or breadth-first traversal examples.

\subsection{A Functional Representation of Data Structures With a Hole}\label{ssec:minamide}\label{ssec:ampar-motivation}

The idea of using linear types as a foundation of a functional calculus in which incomplete data structures can exist and be composed as first class values dates back to~\cite{minamide_functional_1998}. Our system is strongly inspired by theirs. In~\cite{minamide_functional_1998}, a first-class structure with a hole is called a \emph{hole abstraction}. Hole abstractions are represented by a special kind of linear functions with bespoke restrictions. As with any function, we can't pattern-match on their output (or pass it to another function) until they have been applied; but they also have the restriction that we cannot pattern-match on their argument ---the \emph{hole variable}--- as that one can only be used directly as argument of data constructors, or of other hole abstractions. The type of hole abstractions, $\ottstype{(\ottstype{T}, \ottstype{S}) hfun}$ is thus a weak form of linear function type $\ottstype{\ottstype{T} \multimap \ottstype{S}}$.

In~\cite{minamide_functional_1998}, it's only ever possible to represent structures with a single hole. But this is a rather superficial restriction. The author doesn't comment on this, but we believe that this restriction only exists for convenience of the exposition: the language is lowered to a language without function abstraction and where composition is performed by combinators. While it's easy to write a combinator for single-argument-function composition, it's cumbersome to write combinators for functions with multiple arguments. But having multiple-hole data structures wouldn't have changed their system in any profound way.

The more important difference is that while their system is based on a type of linear functions, ours is based on the linear logic's ``par'' type. In classical linear logic, linear implication $\ottstype{\ottstype{T} \multimap \ottstype{S}}$ is reinterpreted as $\ottstype{\ottstype{S}\mathop{\parr}\ottstype{T}^{\perp}}$. We, likewise, reinterpret $\ottstype{(\ottstype{T}, \ottstype{S}) hfun}$ as $ \ottstype{S} \,\ottstype{\ltimes}\,  \ottstype{\lfloor}\,\!_{\mydestm{   \ottsmode{1}  \hspace{-0.15ex}  \ottsmode{\nu}   } } \ottstype{T} \ottstype{\rfloor}  $ (a sort of weak ``par'').

A key consequence is that destinations ---as first-class representations of holes--- appear naturally in \destcalculus{}, while \cite{minamide_functional_1998} doesn't have them. This means that using~\cite{minamide_functional_1998}, or the more recent but similarly expressive system from~\cite{lorenzen_searchtree_2024}, one can implement the examples with difference lists and queues from~\cref{ssec:efficient-queue}, but couldn't do our breadth-first traversal example from~\cref{sec:bft}, since it requires to be able to store destinations in a structure.

Nevertheless, we still retain the main restrictions that \citet{minamide_functional_1998} places on hole abstractions. For instance, we can't pattern-match on $\ottstype{S}$ in (unapplied) $\ottstype{(\ottstype{T}, \ottstype{S}) hfun}$; so in \destcalculus{}, we can't act directly on the left-hand side $\ottstype{S}$ of $ \ottstype{S} \,\ottstype{\ltimes}\, \ottstype{T} $, only on the right-hand side $\ottstype{T}$. Similarly, hole variables can only be used as arguments of constructors or hole abstractions; it's reflected in \destcalculus{} by the fact that the only way to act on destinations is via fill operations, with either hollow constructors or another ampar.

The ability to manipulate destinations, and in particular, store them, does come at a cost though: the system needs this additional notion of ages to ensure that destinations are used soundly. On the other hand, our system is strictly more general, in that \citet{minamide_functional_1998}'s system can be embedded in \destcalculus{}, and if one stays in this fragment, we're never confronted with ages.

\subsection{Destination-Passing Style Programming: A Haskell Implementation}\label{ssec:dps-haskell}

The system introduced in \cite{bagrel_destination-passing_2024} is very similar to \destcalculus{}: it has a destination type, and a \emph{par}-like construct (called $\ottstype{Incomplete}$), where only the right-hand side can be modified; together these elements give extra expressiveness to the language compared to~\cite{minamide_functional_1998}.

In that system, $ \ottmv{d} \blacktriangleleft \ottnt{t} $ requires $\ottnt{t}$ to be unrestricted, while in \destcalculus{}, $\ottnt{t}$ can be linear. The consequence is that in~\cite{bagrel_destination-passing_2024}, destinations can be stored in data structures but not in data structures with holes; so in a breadth-first search algorithm like in~\cref{sec:bft}, the queue has to be built using normal constructors, and cannot use destination-filling primitives. Therefore both normal constructors and DPS primitives must coexist, while in \destcalculus{}, only DPS primitives are required to bootstrap the system, as we later derive normal constructors from them. In exchange, just linearity is enough to make \cite{bagrel_destination-passing_2024} system safe.

A more profound difference is that this previous work describe a practical implementation of destination passing for an existing functional language, while here we present a more theoretical framework that is meant to justify safety of DPS implementations (such as \cite{bagrel_destination-passing_2024} itself).

\subsection{Semi-Axiomatic Sequent Calculus}\label{ssec:sax}

In~\cite{deyoung_sax_2020} constructors return to a destination rather than allocating memory. It is very unlike the other systems described in this section in that it's completely founded in the Curry-Howard isomorphism. Specifically it gives an interpretation of a sequent calculus which mixes Gentzen-style deduction rules and Hilbert-style axioms. As a consequence, the \emph{par} connective is completely symmetric, and, unlike our $ \ottstype{\lfloor}\,\!_{\mydestm{   \ottsmode{1}  \hspace{-0.15ex}  \ottsmode{\nu}   } } \ottstype{T} \ottstype{\rfloor} $ type, their dualization connective is involutive.

The cost of this elegance is that computations may try to pattern-match on a hole, in which case they must wait for the hole to be filled. So the semantics of holes is that of a future or a promise. In turns this requires the semantics of their calculus to be fully concurrent, which is a very different point in the design space.

\subsection{Rust Lifetimes}\label{ssec:rust-lifetimes}

Rust uses a system of lifetimes (see e.g. \cite{pearce_lifetime_2021}) to ensure that borrows don't live longer than what they reference. It plays a similar role as our system of ages.

Rust lifetimes are symbolic. Borrows and moves generate constraints (inequalities of the form $\alpha\leqslant\beta$) on the symbolic lifetimes. For instance, that the lifetime of a reference is larger than the lifetime of any structure the reference is stored in. Without such constraints, Rust would have similar problems to those of \cref{sec:scope-escape-dests}. The borrow checker then checks that the constraints are solvable. This contrasts with \destcalculus{} where ages are set explicitly, with no analysis needed.

Another difference between the two systems is that \destcalculus{}'s ages (and modes in general) are relative. An explicit modality $\ottstype{!}_{  \ottsmode{1}  \hspace{-0.15ex}  \ottsmode{\uparrow}^{ \ottsmodee{k} }  }$ must be used when a part has an age different than its parent, and means that the part is $\ottsmodee{k}$ scope older than the parent. On the other hand, Rust's lifetimes are absolute, the lifetime of a part is tracked independently of the lifetime of its parent.

\subsection{Oxidizing OCaml}

\citet{lorenzen_oxidizing_2024} present an extension of the OCaml type system to support modes. Their modes are split along three different ``axes'', among which affinity and locality are comparable to our multiplicities and ages.
Like our multiplicities, there are two modes for affinity \verb|once| and \verb|many|, though in~\cite{lorenzen_oxidizing_2024}, \verb|once| supports weakening, whereas \destcalculus{}'s $ \ottsmode{1} $ multiplicity is properly linear (proper linearity matters for destination lest we end up reading uninitialized memory).

Locality tracks scope. There are two locality modes, \verb|local| (doesn't escape the current scope) and \verb|global| (can escape the current scope). The authors present their locality mode as a drastic simplification of Rust's lifetime system, which nevertheless fits their need.

However, such a simplified system would be a bit too weak to track the scope of destinations. The observation is that if destinations from two nested scopes are given the same mode, then we can't safely do anything with them, as it would be enough to reproduce the counterexamples of \cref{sec:scope-escape-dests}. So in order to type the breadth-first traversal example of \cref{sec:bft}, where destinations are stored in a structure, we need at least $ \ottsmode{\nu} $ (for the current scope), $ \ottsmode{\uparrow} $ (for the previous scope exactly), plus at least one extra mode for the rest of the scopes (destinations of this generic age cannot be safely used). It turns out that such systems with finitely many ages are incredibly easy to get wrong, and it was in fact much simpler to design a system with infinitely many exact ages.

\section{Conclusion and Future Work}\label{sec:conclusion}

Using a system of ages in addition to linearity, \destcalculus{} is a purely functional calculus which supports destinations in a very flexible way. It subsumes existing calculi from the literature for destination passing, allowing both composition of data structures with holes and storing destinations in data structures. Data structures are allowed to have multiple holes, and destinations can be stored in data structures that, themselves, have holes. The latter is the main reason to introduce ages and is key to \destcalculus{}'s flexibility.

We don't anticipate that a system of ages like \destcalculus{} will actually be used in a programming language: it's unlikely that destinations are so central to the design of a programming language that it's worth baking them so deeply in the type system. Perhaps a compiler that makes heavy use of destinations in its optimizer could use \destcalculus{} as a typed intermediate representation. But, more realistically, our expectation is that \destcalculus{} can be used as a theoretical framework to analyze destination-passing systems: if an API can be defined in \destcalculus{} then it's sound.

In fact, we plan to use this very strategy to design an API for destination passing in Haskell, leveraging only the existing linear types, but retaining the possibility of storing destinations in data structures with holes.

\clearpage{}

\longshort{}{

\section*{Data-Availability Statement}

We have submitted for artifact evaluation the formalization of the \destcalculus{} as described in \cref{sec:syntax-type-system,sec:ectxs-sem}, using the \href{https://coq.inria.fr/}{Coq proof assistant} with some classical axioms. The main reproducible results are the machine-verified proofs of type-safety \cref{thm:preservation,thm:progress} for \destcalculus{}.

The version submitted with the article is available at \url{https://doi.org/10.5281/zenodo.14982363}. One can check if a newer version is available using \url{https://doi.org/10.5281/zenodo.14534422} or \url{https://github.com/tweag/destination-calculus}.
}

\longshort{

\clearpage{}
\appendix

\section{Full Reduction Rules for \destcalculus{}}\label{apx:full-reduction-rules}

\renewenvironment{rulesection}[3][\relax]
  {\trivlist\item
   \ifx#1\relax\else\def\ottalt@rulesection@prefix{#1-}\fi
   \nopagebreak[4]%
   \noindent}
  {\endtrivlist}
\renewcommand\ottaltinferrule[4]{
  \inferrule*[narrower=0.3,right=#1,#2]
    {#3}
    {#4}
}
\bgroup\SetPrefix{\CRed\CSep}
\begin{ottfig}[h]{\caption{Full reduction rules for \destcalculus{} (part 1)}\label{fig:sem-full1}}\begin{augmentwidth}{3cm}
\drules{$ \ottnt{E} ~\big[\, \ottnt{t} \,\big]~ ~\longrightarrow~ ~ \ottnt{E}' ~\big[\, \ottnt{t'} \,\big] $}{Small-step evaluation of commands}{%
App-FocusOne,
App-UnfocusOne,
App-FocusTwo,
App-UnfocusTwo,
App-Red,
PatU-Focus,
PatU-Unfocus,
PatU-Red,
PatS-Focus,
PatS-Unfocus,
PatL-Red,
PatR-Red,
PatP-Focus,
PatP-Unfocus,
PatP-Red,
PatE-Focus,
PatE-Unfocus,
PatE-Red,
UpdA-Focus,
UpdA-Unfocus,
Ampar-Open,
Ampar-Close,
ToA-Focus,
ToA-Unfocus,
ToA-Red,
NewA-Red
}
\end{augmentwidth}\end{ottfig}

\begin{ottfig}[h]{\caption{Full reduction rules for \destcalculus{} (part 2)}\label{fig:sem-full2}}\begin{augmentwidth}{3cm}
\drules{}{}{%
FromA-Focus,
FromA-Unfocus,
FromA-Red,
FillU-Focus,
FillU-Unfocus,
FillU-Red,
FillL-Focus,
FillL-Unfocus,
FillL-Red,
FillR-Focus,
FillR-Unfocus,
FillR-Red,
FillE-Focus,
FillE-Unfocus,
FillE-Red,
FillP-Focus,
FillP-Unfocus,
FillP-Red,
FillF-Focus,
FillF-Unfocus,
FillF-Red,
FillComp-FocusOne,
FillComp-UnfocusOne,
FillComp-FocusTwo,
FillComp-UnfocusTwo,
FillComp-Red,
FillLeaf-FocusOne,
FillLeaf-UnfocusOne,
FillLeaf-FocusTwo,
FillLeaf-UnfocusTwo,
FillLeaf-Red
}
\end{augmentwidth}\end{ottfig}
\egroup

}{}

\clearpage{}

\bibliographystyle{ACM-Reference-Format}
\bibliography{bibliography}{}

\end{document}
